\documentclass[twocolumn,amssymb,nobibnotes, aps, prc,
superscriptaddress, nobalancelastpage, floatfix, nofootinbib]{revtex4}

 \usepackage{graphicx}
\usepackage{amsmath} \usepackage{braket} \usepackage{epsfig}
\usepackage{upgreek}
\usepackage{textcomp}
\usepackage[free-standing-units]{siunitx}
\usepackage{xcolor}
\usepackage[section]{placeins}
\usepackage{multirow}
\usepackage{bm}
\usepackage{longtable}
\usepackage{listings}
\newcolumntype{N}{>{\centering\arraybackslash}m{1.3in}}
\newcolumntype{M}{>{\centering\arraybackslash}m{1.0in}}
\newcolumntype{G}{>{\centering\arraybackslash}m{0.5in}}
\newcolumntype{R}{>{\raggedleft\arraybackslash}m{0.4in}}

\newcommand{\agNat}{\ensuremath{^{\text{nat}}}A\lowercase{g}}
\newcommand{\agOneHundredSeven}{\ensuremath{^{107}}A\lowercase{g}}
\newcommand{\alNat}{\ensuremath{^{\text{nat}}}A\lowercase{l}}
\newcommand{\alTwentySeven}{\ensuremath{^{27}}A\lowercase{l}}
\newcommand{\arNat}{\ensuremath{^{\text{nat}}}A\lowercase{r}}
\newcommand{\arForty}{\ensuremath{^{40}}A\lowercase{r}}
\newcommand{\asNat}{\ensuremath{^{\text{nat}}}A\lowercase{s}}
\newcommand{\auOneHundredNinetySeven}{\ensuremath{^{197}}A\lowercase{u}}

\newcommand{\baNat}{\ensuremath{^{\text{nat}}}B\lowercase{a}}
\newcommand{\baOneHundredThirtyTwo}{\ensuremath{^{132}}B\lowercase{a}}
\newcommand{\baOneHundredThirtyFour}{\ensuremath{^{134}}B\lowercase{a}}
\newcommand{\baOneHundredThirtySix}{\ensuremath{^{136}}B\lowercase{a}}
\newcommand{\baOneHundredThirtyEight}{\ensuremath{^{138}}B\lowercase{a}}

\newcommand{\biTwoHundredNine}{\ensuremath{^{209}}B\lowercase{i}}

\newcommand{\caNat}{\ensuremath{^{\text{nat}}}C\lowercase{a}}
\newcommand{\caForty}{\ensuremath{^{40}}C\lowercase{a}}
\newcommand{\caFortyFour}{\ensuremath{^{44}}C\lowercase{a}}
\newcommand{\caFortyEight}{\ensuremath{^{48}}C\lowercase{a}}
\newcommand{\cdNat}{\ensuremath{^{\text{nat}}}C\lowercase{d}}
\newcommand{\cdOneHundredSix}{\ensuremath{^{106}}C\lowercase{d}}
\newcommand{\cdOneHundredEight}{\ensuremath{^{108}}C\lowercase{d}}
\newcommand{\cdOneHundredTen}{\ensuremath{^{110}}C\lowercase{d}}
\newcommand{\cdOneHundredTwelve}{\ensuremath{^{112}}C\lowercase{d}}
\newcommand{\cdOneHundredFourteen}{\ensuremath{^{114}}C\lowercase{d}}
\newcommand{\cdOneHundredSixteen}{\ensuremath{^{116}}C\lowercase{d}}
\newcommand{\ceNat}{\ensuremath{^{\text{nat}}}C\lowercase{e}}
\newcommand{\clNat}{\ensuremath{^{\text{nat}}}C\lowercase{l}}
\newcommand{\coFiftyNine}{\ensuremath{^{59}}C\lowercase{o}}
\newcommand{\crNat}{\ensuremath{^{\text{nat}}}C\lowercase{r}}
\newcommand{\crFiftyTwo}{\ensuremath{^{52}}C\lowercase{r}}
\newcommand{\cuNat}{\ensuremath{^{\text{nat}}}C\lowercase{u}}
\newcommand{\cuSixtyThree}{\ensuremath{^{63}}C\lowercase{u}}
\newcommand{\cuSixtyFive}{\ensuremath{^{65}}C\lowercase{u}}

\newcommand{\feNat}{\ensuremath{^{\text{nat}}}F\lowercase{e}}
\newcommand{\feFiftyFour}{\ensuremath{^{54}}F\lowercase{e}}
\newcommand{\feFiftySix}{\ensuremath{^{56}}F\lowercase{e}}

\newcommand{\gdNat}{\ensuremath{^{\text{nat}}}G\lowercase{d}}

\newcommand{\geNat}{\ensuremath{^{\text{nat}}}G\lowercase{e}}
\newcommand{\geSeventyTwo}{\ensuremath{^{72}}G\lowercase{e}}
\newcommand{\geSeventyFour}{\ensuremath{^{74}}G\lowercase{e}}

\newcommand{\hgNat}{\ensuremath{^{\text{nat}}}H\lowercase{g}}
\newcommand{\hoOneHundredSixtyFive}{\ensuremath{^{165}}H\lowercase{g}}

\newcommand{\iOneHundredTwentySeven}{\ensuremath{^{127}}I}
\newcommand{\inNat}{\ensuremath{^{\text{nat}}}I\lowercase{n}}

\newcommand{\kNat}{\ensuremath{^{\text{nat}}}K}

\newcommand{\laNat}{\ensuremath{^{\text{nat}}}L\lowercase{a}}

\newcommand{\mgNat}{\ensuremath{^{\text{nat}}}M\lowercase{g}}
\newcommand{\mgTwentyFour}{\ensuremath{^{24}}M\lowercase{g}}
\newcommand{\mnFiftyFive}{\ensuremath{^{55}}M\lowercase{n}}

\newcommand{\moNat}{\ensuremath{^{\text{nat}}}M\lowercase{o}}
\newcommand{\moNinetyTwo}{\ensuremath{^{92}}M\lowercase{o}}
\newcommand{\moNinetyFive}{\ensuremath{^{95}}M\lowercase{o}}
\newcommand{\moNinetySix}{\ensuremath{^{96}}M\lowercase{o}}
\newcommand{\moNinetyEight}{\ensuremath{^{98}}M\lowercase{o}}
\newcommand{\moOneHundred}{\ensuremath{^{100}}M\lowercase{o}}

\newcommand{\nbNinetyThree}{\ensuremath{^{93}}N\lowercase{b}}
\newcommand{\ndNat}{\ensuremath{^{\text{nat}}}N\lowercase{d}}
\newcommand{\ndOneHundredFortyTwo}{\ensuremath{^{142}}N\lowercase{d}}
\newcommand{\ndOneHundredFortyFour}{\ensuremath{^{144}}N\lowercase{d}}
\newcommand{\niNat}{\ensuremath{^{\text{nat}}}N\lowercase{i}}
\newcommand{\niFiftyEight}{\ensuremath{^{58}}N\lowercase{i}}
\newcommand{\niSixty}{\ensuremath{^{60}}N\lowercase{i}}
\newcommand{\niSixtyTwo}{\ensuremath{^{62}}N\lowercase{i}}
\newcommand{\niSixtyFour}{\ensuremath{^{64}}N\lowercase{i}}

\newcommand{\pThirtyOne}{\ensuremath{^{31}}P}
\newcommand{\pbTwoHundredFour}{\ensuremath{^{204}}P\lowercase{b}}
\newcommand{\pbTwoHundredSix}{\ensuremath{^{206}}P\lowercase{b}}
\newcommand{\pbTwoHundredEight}{\ensuremath{^{208}}P\lowercase{b}}
\newcommand{\pbNat}{\ensuremath{^{\text{nat}}}P\lowercase{b}}
\newcommand{\pdNat}{\ensuremath{^{\text{nat}}}P\lowercase{d}}
\newcommand{\prOneHundredFortyOne}{\ensuremath{^{141}}P\lowercase{r}}

\newcommand{\reNat}{\ensuremath{^{\text{nat}}}R\lowercase{e}}

\newcommand{\rhOneHundredThree}{\ensuremath{^{103}}R\lowercase{h}}

\newcommand{\sNat}{\ensuremath{^{\text{nat}}}S}
\newcommand{\sThirtyTwo}{\ensuremath{^{32}}S}
\newcommand{\sbOneHundredTwentyThree}{\ensuremath{^{123}}S\lowercase{b}}
\newcommand{\scFortyFive}{\ensuremath{^{45}}S\lowercase{c}}
\newcommand{\seNat}{\ensuremath{^{\text{nat}}}S\lowercase{e}}
\newcommand{\seSeventySix}{\ensuremath{^{76}}S\lowercase{e}}
\newcommand{\seSeventyEight}{\ensuremath{^{78}}S\lowercase{e}}
\newcommand{\seEighty}{\ensuremath{^{80}}S\lowercase{e}}
\newcommand{\seEightyTwo}{\ensuremath{^{82}}S\lowercase{e}}
\newcommand{\siNat}{\ensuremath{^{\text{nat}}}S\lowercase{i}}
\newcommand{\siTwentyEight}{\ensuremath{^{28}}S\lowercase{i}}
\newcommand{\smOneHundredFortyFour}{\ensuremath{^{144}}S\lowercase{m}}
\newcommand{\smOneHundredFortyEight}{\ensuremath{^{148}}S\lowercase{m}}
\newcommand{\smOneHundredFifty}{\ensuremath{^{150}}S\lowercase{m}}
\newcommand{\snNat}{\ensuremath{^{\text{\lowercase{nat}}}}S\lowercase{n}}
\newcommand{\snOneHundredTwelve}{\ensuremath{^{112}}S\lowercase{n}}
\newcommand{\snOneHundredSixteen}{\ensuremath{^{116}}S\lowercase{n}}
\newcommand{\snOneHundredEighteen}{\ensuremath{^{118}}S\lowercase{n}}
\newcommand{\snOneHundredTwenty}{\ensuremath{^{120}}S\lowercase{n}}
\newcommand{\snOneHundredTwentyTwo}{\ensuremath{^{122}}S\lowercase{n}}
\newcommand{\snOneHundredTwentyFour}{\ensuremath{^{124}}S\lowercase{n}}
\newcommand{\srNat}{\ensuremath{^{\text{\lowercase{nat}}}}S\lowercase{r}}
\newcommand{\srEightySix}{\ensuremath{^{86}}S\lowercase{r}}
\newcommand{\srEightySeven}{\ensuremath{^{87}}S\lowercase{r}}
\newcommand{\srEightyEight}{\ensuremath{^{88}}S\lowercase{r}}

\newcommand{\taNat}{\ensuremath{^{\text{\lowercase{nat}}}}T\lowercase{a}}
\newcommand{\taOneHundredEightyOne}{\ensuremath{^{181}}T\lowercase{a}}

\newcommand{\teNat}{\ensuremath{^{\text{\lowercase{nat}}}}T\lowercase{e}}
\newcommand{\tiNat}{\ensuremath{^{\text{\lowercase{nat}}}}T\lowercase{i}}
\newcommand{\tiFortySix}{\ensuremath{^{46}}T\lowercase{i}}
\newcommand{\tiFortyEight}{\ensuremath{^{48}}T\lowercase{i}}
\newcommand{\tiFifty}{\ensuremath{^{50}}T\lowercase{i}}

\newcommand{\vNat}{\ensuremath{^{\text{\lowercase{nat}}}}V}
\newcommand{\vFiftyOne}{\ensuremath{^{51}}V}

\newcommand{\wNat}{\ensuremath{^{\text{\lowercase{nat}}}}W}
\newcommand{\wOneHundredEightyTwo}{\ensuremath{^{182}}W}
\newcommand{\wOneHundredEightyFour}{\ensuremath{^{184}}W}
\newcommand{\wOneHundredEightySix}{\ensuremath{^{186}}W}

\newcommand{\yEightyEight}{\ensuremath{^{88}}Y}
\newcommand{\yEightyNine}{\ensuremath{^{89}}Y}

\newcommand{\znSixtyFour}{\ensuremath{^{64}}Zn}
\newcommand{\znSixtySix}{\ensuremath{^{66}}Zn}
\newcommand{\znSixtyEight}{\ensuremath{^{68}}Zn}
\newcommand{\znSeventy}{\ensuremath{^{70}}Zn}
\newcommand{\zrNat}{\ensuremath{^{\text{\lowercase{nat}}}}Z\lowercase{r}}
\newcommand{\zrNinety}{\ensuremath{^{90}}Z\lowercase{r}}
\newcommand{\zrNinetyOne}{\ensuremath{^{91}}Z\lowercase{r}}
\newcommand{\zrNinetyTwo}{\ensuremath{^{92}}Z\lowercase{r}}
\newcommand{\zrNinetyFour}{\ensuremath{^{94}}Z\lowercase{r}}

\begin{document}

\begin{abstract}
    Optical-model potentials (OMPs) continue to play a key role in nuclear
    reaction calculations. However, the uncertainty of phenomenological OMPs in
    widespread use --- inherent to any parametric model trained on data --- has
    not been fully characterized, and its impact on downstream users of OMPs
    remains unclear. Here we assign well-calibrated uncertainties for two
    representative global OMPs, those of Koning-Delaroche and Chapel Hill '89,
    using Markov-Chain Monte Carlo for parameter inference. By comparing
    the canonical versions of these OMPs against the experimental data
    originally used to constrain them, we show how a lack of outlier rejection
    and a systematic underestimation of experimental uncertainties contributes
    to bias of, and overconfidence in, best-fit parameter values. Our updated,
    uncertainty-quantified versions of these OMPs address these issues and
    yield complete covariance information for potential parameters. Scattering
    predictions generated from our ensembles show improved performance both
    against the original training corpora of experimental data and against a
    new ``test'' corpus comprising many of the experimental single-nucleon
    scattering data collected over the last twenty years. Finally,
    we apply our uncertainty-quantified OMPs to two case studies of
    application-relevant cross sections. We conclude that, for many common
    applications of OMPs, including OMP uncertainty should become standard
    practice. To facilitate their immediate use, digital versions of our
    updated OMPs and related tools for forward uncertainty propagation are
    included as Supplemental Material.
\end{abstract}

\title{Uncertainty-quantified phenomenological optical potentials for
single-nucleon scattering}

\author{C.~D.~Pruitt}  \email[Corresponding author: ]{pruitt9@llnl.gov}
\author{J.~E.~Escher}
\affiliation{Lawrence Livermore National Laboratory, Livermore, CA 94550}
\author{R.~Rahman}
\altaffiliation{Present address: University of Tennessee, Knoxville, TN 37996}
\affiliation{Lawrence Livermore National Laboratory, Livermore, CA 94550}

\maketitle

\section{Introduction}
\label{sec_introduction}
For more than fifty years, optical-model potentials (OMPs) have played an
important role in nuclear scattering calculations by providing effective
projectile-target interactions. Early successes in fitting basic
phenomenological OMPs to elastic scattering data \cite{BecchettiGreenlees1965}
motivated continuing theoretical improvements on several fronts, including
construction of (semi-)microscopic OMPs via the local density approximation
\cite{Jeukenne1977, Jeukenne1976, Bauge1998, Bauge2001}, extension to deformed
and actinide systems \cite{Nobre2015, Soukhovitskii2016}, and formal connection
with the single-particle Green's function via application of relevant
dispersion relations \cite{MahauxSartor, Quesada2003, Morillon2004,
Morillon2007, Mueller2011, Mahzoon2014, Atkinson2019, Zhao2020}. The recent
development of a global microscopic OMP \cite{Whitehead2021} based on several
$\chi$EFT nucleon-nucleon (NN) potentials opens a promising new avenue for
making predictions of scattering on unstable nuclides with a minimum of
phenomenology. For recent reviews of OMP topics, see \cite{Dickhoff2017,
Dickhoff2018}.

Despite these advances, a number of basic questions remain about the
uncertainty and generality of OMPs. First are questions of interpolation and
extrapolation: how far can OMPs be trusted to generate reliable scattering
predictions where experimental data are not available, especially away from
$\beta$-stability? As new rare isotope beam facilities come online, reliable
estimates of scattering on unstable targets will be needed to make sense of the
wealth of new data that are anticipated. For meaningful comparison with these
new data, OMP predictions should come equipped with well-calibrated uncertainty
estimates, estimates that are typically absent from widely used
phenomenological OMPs, such as the Chapel-Hill 89' OMP \cite{CH89} (intended for $40 \leq
A \leq 209$ from 10 to 65 \MeV) and Koning-Delaroche OMP \cite{KoningDelaroche} (intended for
$24 \leq A \leq 209$ from 0.001 to 200 \MeV). In principle, a global microscopic OMP
based on a $\chi$EFT-derived NN potential, such as \cite{Whitehead2021}, come
``naturally'' equipped with uncertainties from truncation in the chiral
expansion and should be less prone to under- or overfitting problems that
affect phenomenological potentials. To date, however, microscopic models do not
achieve the accuracy of phenomenological OMPs in regions where experimental
data do exist, especially for inelastic scattering observables, which may
diminish their utility for nuclear data applications. Were it available,
knowledge of OMP uncertainties would help evaluators rank the relative
importance of OMPs among other sources of uncertainty that enter reaction
models, such as nuclear level densities and $\gamma$-ray strength functions
\cite{Koning2015}.

The second type of questions concern the functional form of potentials and
their capacity to realistically describe the underlying physics. As a simple
example, the Koning-Delaroche OMP includes an imaginary spin-orbit term, but
the Chapel Hill '89 OMP does not. Does inclusion of this term result in
meaningful differences in scattering predictions, and if so, which experimental
data actually constrain its parameters? The form of nonlocal terms
\cite{Arellano2018, Arellano2019, Arellano2021}, shape of the hole potential
and relation to dispersive correctness \cite{Mahzoon2014}, and the correct 
dependence of parameters on nuclear asymmetry \cite{Goriely2007} are important
open topics that would benefit from a firmer understanding of uncertainty in
extant OMPs.

To clarify these issues, several recent studies have investigated
uncertainty-quantification (UQ) techniques for phenomenological OMPs, including
direct comparisons of frequentist and Bayesian methods for model calibration
\cite{King2019, Lovell2020}, introduction of Gaussian process priors for
energy-dependent parameters \cite{Schnabel2021}, and introduction of an ad-hoc
dedicated ``model uncertainty'' term in a dispersive OMP analysis
\cite{Pruitt2020PRC}. The ambitious study of \cite{Koning2015} confronts the
mature reaction code TALYS \cite{Koning2012} with virtually the entire EXFOR
experimental reaction database \cite{EXFORDatabase} with the specific intent of
generating uncertainties for evaluations. Such theoretical studies are being
complemented by the creation of templates for experimentalists for capturing
the many (often undercharacterized) sources of uncertainty in experimental
measurement \cite{Smith2012, Neudecker2018, Neudecker2020}, designed
specifically so that newly collected data will be maximally useful for theory
and evaluation efforts going forward. Most recently, the work of
\cite{Schnabel2021} proposes a statistically sound, reproducible pipeline for
nuclear data evaluations, including characterization of OMP uncertainties,
demonstrating the potential to accelerate and standardize the challenging
process of evaluation.

Despite these methodological improvements over the last decade, many OMP users
do not yet consider the OMP contribution to the uncertainty budget of their
applications, either because it is assumed to be negligible or because tools to
do so are difficult to use. Those that do (e.g. \cite{Harissopulos2021,
Gyurky2001}) typically estimate uncertainty qualitatively by manually varying a
handful of parameters thought to be important and by comparing predictions from
a handful non-UQ OMPs against each other by eye. Even when OMP parameter
uncertainty estimates are available (e.g., \cite{Koning2012}), they are more
often based on hard-earned evaluator intuition rather than on detailed tests of
empirical performance. In the ideal case, each OMP would ship with complete covariance
information for potential parameters, be tested against trusted, easily-accessed data
libraries, be based on reproducible statistical practices and stated
assumptions, and make it easy for any downstream user to forward-propagate OMP
uncertainty into their research application. Robust OMP UQ of this type
would be a building block for larger UQ efforts such as the
evaluation efforts mentioned earlier \cite{Koning2015, Schnabel2021} or improved
experimental analysis pipelines. Motivating and demonstrating such a
framework for phenomenological OMP UQ is the main goal of the present work.

To demonstrate our approach, we apply it to both the Koning-Delaroche global
OMP (KD) \cite{KoningDelaroche} and the Chapel-Hill '89 OMP (CH89) \cite{CH89},
yielding two new uncertainty-quantified OMP ensembles we designate KDUQ and
CHUQ, respectively. To train these OMPs, we recompiled the same training data
corpora as used in the original treatments (we refer to our recompilations as
the CHUQ corpus and KDUQ corpus). The resulting UQ OMPs can be directly
inserted into existing user codes to incorporate the parametric
uncertainty of these OMPs. By applying our approach to multiple OMPs and
comparing with microscopic and semi-microscopic alternatives, we can develop
insight into how the next generation of uncertainty-equipped potentials can be
gainfully constructed. In particular, we will emphasize the importance of two
key steps in fitting phenomenological OMPs --- managing \textit{outliers} and
\textit{experimental uncertainty underestimation} --- that are paramount for
empirical UQ, both in OMPs and otherwise.

Our findings are organized in the following sections. Section \ref{sec_fitting}
introduces the generic parameter inference problem and its application to OMP
fitting, including challenges faced in the original CH89 and KD analyses.
Section \ref{sec_method} proposes a new likelihood function and inference
strategy based on Markov-Chain Monte Carlo (MCMC) that we argue is better
suited for generating realistic OMP uncertainties. Section \ref{sec_results}
applies this strategy to retrain the KD and CH89 OMP forms against their
original training data, yielding updated, uncertainty-quantified OMPs: KDUQ and
CHUQ. Section \ref{sec_impact} illustrates the impact of KDUQ and CHUQ both on
Hauser-Feshbach calculations for two radiative capture test cases and on the
reliability of OMP extrapolation along neutron-proton asymmetry. Section
\ref{sec_conclusion} summarizes our findings. Following the main text, further
technical details appear in the Appendix and three sections of Supplemental
Material \cite{SupplementalMaterial}, including explicit definitions of the
OMPs and scattering formulae, all experimental data used for training and
testing, and our recommended KDUQ and CHUQ parameter values for future use. We
hope that by providing thorough documentation, readers will be able to
reproduce or extend our results without guesswork.


\section{Challenges in OMP parameter inference}
\label{sec_fitting}

In this section, we first present a generic parameter inference problem,
illustrating some common challenges with a pedagogical example. We then turn to
the original KD and CH89 analyses, showing that certain assumptions, while
necessary for making these canonical analyses tractable, can result in
overconfidence in the fitted parameters.

\subsection{Generic parameter inference}
\label{sec_fitting_generic}

The goal of a parameter inference problem is to determine optimal parameters
for a given functional form, where ``optimal'' usually means best matching a
corpus of training data. In the specific case of OMP optimization, the OMP
constitutes a model $M$ with unknown, possibly correlated potential parameters
$\bm{\theta}$, and the task is to determine an optimal set of parameters
$\bm{\theta_{opt}}$ that minimizes the residuals between experimental
scattering data and scattering-code predictions made using $M$. (In these and
all following definitions, we use bold typeface to denote a vector or matrix
quantity.) A natural starting point for the probability density function of
$\bm{\theta}$ is to use a $k$-dimensional normal distribution:

\begin{equation} \label{MultivariateNormalPDF}
    \begin{split}
        p(\bm{\theta}) & = \frac{1}{\sqrt{(2\pi)^{k}|\bm{\Sigma}|}}e^{-\frac{1}{2}\bm{r}^{\intercal}\bm{\Sigma}^{-1}\bm{r}}\\
        \bm{r} & = \bm{\theta} - \bm{\theta_{opt}}.
    \end{split}
\end{equation}

Here, $\bm{\Sigma}$ is the $k \times k$ covariance matrix associated with
$\bm{\theta}$. If $\bm{\theta_{opt}}$ and $\bm{\Sigma}$ were known, the
inference problem would be solved (at least up to the assumption of
multivariate normality), with $\bm{\Sigma}$ holding the variance information
enabling downstream uncertainty propagation. Because we do not have direct
measurements of $\bm{\theta}$, only experimental scattering measurements, we
cannot use Eq. \ref{MultivariateNormalPDF} directly to train the model.
Instead, we need to construct a likelihood function that connects the
probability of observing a given experimental value given a candidate parameter
vector. For OMP optimization, this involves mapping a candidate $\bm{\theta}$
to predicted scattering observables via a scattering code, evaluated at
the relevant experimental indices (e.g, scattering energy, angle, target). This
mapping is highly nonlinear in $\bm{\theta}$ as it involves, among other
things, solving for the scattering matrix. Because the covariance matrix
between experimental measurements is rarely known (discussed in detail in
\cite{Koning2015}), connecting OMP parameters with experimental data via
selection of a likelihood function requires making certain assumptions about
the scattering data. The overwhelming majority of past OMP analyses (including
CH89 and KD) use a maximum likelihood approach based on some version of
weighted-least-squares for their likelihood function:

\begin{equation} \label{WeightedLeastSquaresEstimator}
    \begin{split}
        L(\bm{y}|\bm{x,\delta_{y}, \theta}) & = e^{-\frac{1}{2}\sum\limits_{i}\frac{r_{i}^{2}}{{{\delta_{y}}_{i}}^{2}}} \\
        r_{i} & = y_{i} - M(\bm{\theta}, x_{i}).
    \end{split}
\end{equation}

In this expression, for the $i^{th}$ training data point, $x_{i}$ are the experimental
conditions (such as energy, angle, etc.), $y_{i}$ is the observed value, such
as the cross section, and ${\delta_{y}}_{i}$ is the reported uncertainty of the
observed value. Thus ($\bm{x, y, \delta_{y}}$) denotes the entire training
corpus. Experimental data also often include an estimate of uncertainty in the
experimental conditions $\bm{\delta_{x}}$ but these are usually omitted from
the OMP analysis as they are more difficult to incorporate using standard
optimization approaches. The predicted values, $M(\bm{\theta},\bm{x})$, are an
output of the scattering code evaluated at each $\bm{x}$ and using the OMP
realization $M(\bm{\theta})$ for the projectile-target interaction.

If several conditions apply, including model linearity in the parameters,
experimental uncertainties characterized by a known, positive-definite
covariance matrix, and measurement samples being drawn from the same underlying
distribution, the weighted least-squares estimator (Eq.
\ref{WeightedLeastSquaresEstimator}) guarantees an analytic solution that
minimizes bias in $\bm{\theta_{opt}}$ \cite{Aitken1936}. Unfortunately for OMP
analysts, each of these conditions is violated in traditional OMP optimization
analyses that are concerned primarily with $\bm{\theta}$, and these violations
are especially problematic for the present UQ task ($\bm{\Sigma}$ estimation).
Most impactful is the weighted-least-squares assumption that experimentally
reported uncertainties are independent and complete (that is, that the vector
of individual data point uncertainties $\bm{\delta_{y}}$ fully represents the
true, unknown data covariance matrix).  In effect, this assumption assigns more
independent information to residuals than they actually have, making the
inference problem erroneously overdetermined and causing bias in
$\bm{\theta_{opt}}$ and underestimation of $\bm{\theta_{opt}}$ uncertainty.
Even if the full experimental data covariance were known, the OMP, by
definition, is a projection of the true projectile-target interaction onto a
reduced space of simple potential forms. As such we should expect it to suffer
at least somewhat from ``model defects'' that, if unaccounted for during
inference, may lead to overconfidence in an incorrect $\bm{\theta_{opt}}$, as
demonstrated for a simple physical model in \cite{Brynjarsdottir2014}. Further,
model nonlinearity in $\bm{\theta}$ means that the likelihood function surface
is not guaranteed to be convex, which can stymie simple optimization approaches
such as gradient descent but which may be tractable with other optimization
algorithms such as, e.g., simulated annealing. 

\subsection{A toy model}
\label{sec_fitting_toy}

To illustrate how outliers and uncertainty underestimation impact parameter
inference, we present a toy problem using a simple linear model. Imagine we
wish to describe some generic phenomenon, $T(x)$, that occurs on a domain $x
\in [-1,1]$. The true $T(x)$ is:
\begin{equation} \label{ToyModelTrueFunction}
    T(x) = 2.5P_{0}(x) + 2.0P_{1}(x) + 1.5P_{2}(x) + 1.0P_{3}(x),
\end{equation}
where $P_{n}$ is the $n^{th}$ Legendre polynomial. Suppose we know the
functional form of $T(x)$ but not the values of the coefficients, which we
would like to learn through inference against data. So we collect $i$
observations $\bm{y}$ at experimental conditions $\bm{x}$, using a device
subject to measurement uncertainty. Aware of this uncertainty, we estimate
measurement imprecision for each data point as $\bm{\delta_{y}}$.  We then
define a model, $M$, and compare model predictions $\bm{M(x,\theta)}$ to the
measured data, where $\bm{\theta}$ are the $n$ unknown coefficients that we
want to learn. Because our model is linear in $\bm{\theta}$ and our data
measurements are independent and uncorrelated, Eq.
\ref{WeightedLeastSquaresEstimator} provides the best unbiased estimator of the
true coefficients, denoted $\bm{\theta_{true}}$. We can find an optimum set of
parameter values $\bm{\theta_{opt}}$ analytically using maximum likelihood
estimation or numerically using, e.g., gradient descent until we reach some
threshold for convergence. The covariance matrix at $\bm{\theta_{opt}}$ is the
inverse of the Hessian matrix $\mathcal{H}(\bm{\theta_{opt}})$, which can be
easily assessed numerically.

\begin{figure*}
    \includegraphics[width=\textwidth]{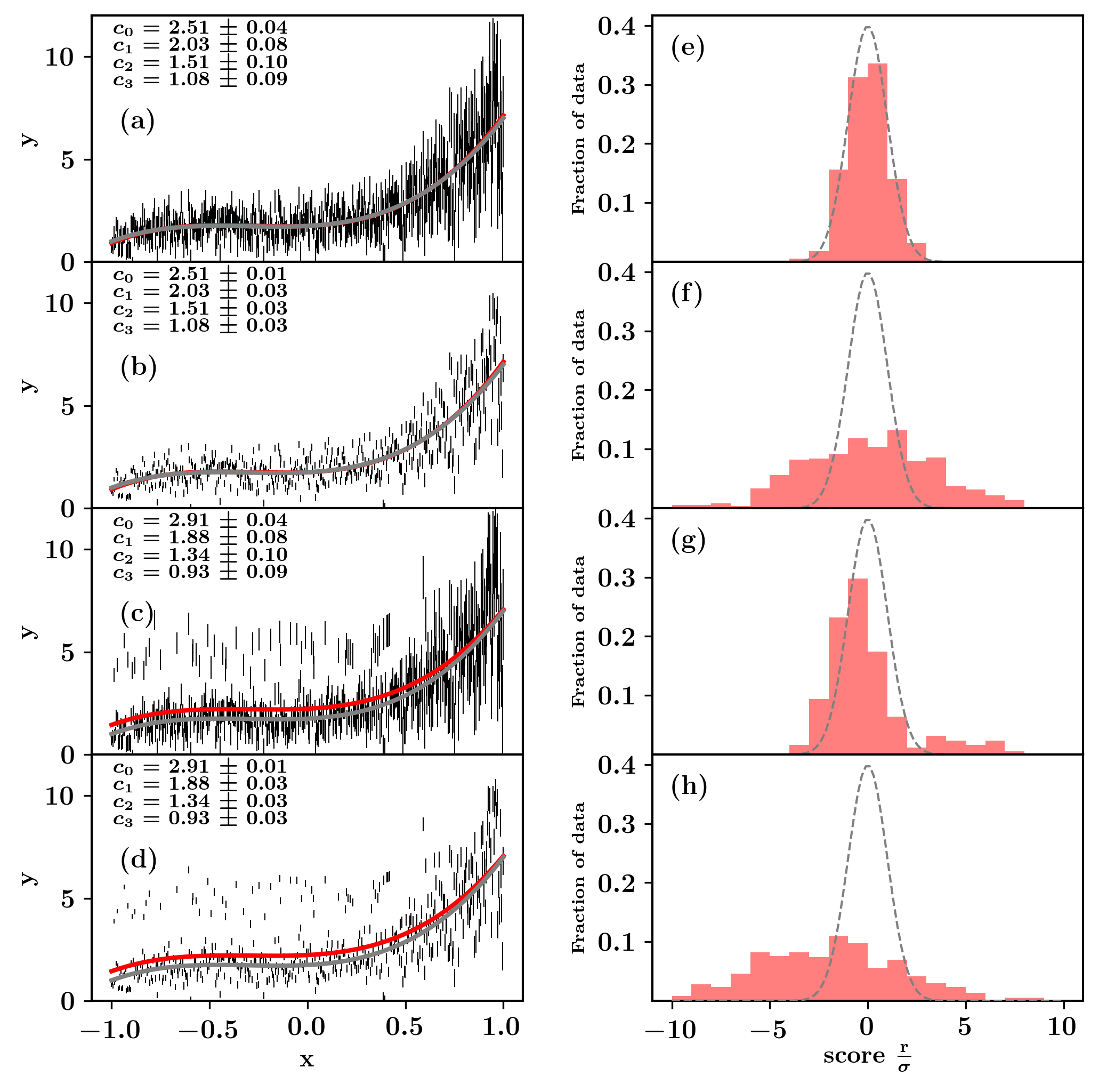}
    \caption{
        The four data-fitting scenarios for the toy model discussed in the text are compared.
        Panel (a) shows a fit to data with accurate uncertainty estimates and no
        outliers. Panel (b) shows a fit to data with underestimated
        uncertainties and no outliers. Panel (c) shows a fit to data with
        accurate uncertainty estimates but with outliers.  Panel (d) shows
        a fit using data with both underestimated uncertainties and with
        outliers. The simulated data used for
        fitting are shown as black bars, the ``true'' underlying function
        used to generate the data is shown as a gray line, and the
        fit to the ``experimental'' data is shown in red. In panels (e) to
        (h), the normalized residuals for data in the corresponding plots are
        histogrammed. A normal distribution with $\mu=0$ and $\sigma^{2}=1$
        (gray dashed line) is shown for reference.
    }
    \label{exampleFitFigure}
\end{figure*}

So far, we have described a simple, generic inference problem and its solution.
We now consider four possible scenarios for solving this problem, each
involving a different possible distribution for $\bm{y}$ and
$\bm{\delta_{y}}$. These differing distributions are plotted in panels (a) to
(d) of Fig. \ref{exampleFitFigure}, and defined according to:
\begin{equation} \label{Scenario1}
    \begin{split}
        y_{i} & \sim \mathcal{N}(T(x_{i}),0.32^{2}) \\
        {\delta_{y}}_{i} & = 0.32T(x_{i}) \\
    \end{split}
\end{equation}
\begin{equation} \label{Scenario2}
    \begin{split}
        y_{i} & \sim \mathcal{N}(T(x_{i}),0.32^{2}) \\
        {\delta_{y}}_{i} & = 0.10T(x_{i}) \\
    \end{split}
\end{equation}
\begin{equation} \label{Scenario3}
    \begin{split}
        y_{i} & \sim \mathcal{N}(T(x_{i}),0.32^{2}) + \alpha \\
        {\delta_{y}}_{i} & = 0.32T(x_{i}) \\
    \end{split}
\end{equation}
\begin{equation} \label{Scenario4}
    \begin{split}
        y_{i} & \sim \mathcal{N}(T(x_{i}),0.32^{2}) + \alpha \\
        {\delta_{y}}_{i} & = 0.10T(x_{i}),
    \end{split}
\end{equation}
for each $i$, where
\begin{equation*} \label{OutlierDefinition}
    \alpha 
    \begin{cases}
        \sim \mathcal{N}(3,0.6^{2}), & \text{ 10\% probability}, \\
        = 0, & \text{ 90\% probability}.
    \end{cases}
\end{equation*}
In this notation, $\sim \mathcal{N}(\mu,\sigma^{2})$ refers to sampling from 
a normal distribution of mean $\mu$ and variance $\sigma^{2}$.

We begin with the first scenario, shown in panel (a) of \ref{exampleFitFigure}.
This is the best-case scenario, given the assumptions appropriate for
weighted-least-squares: our measuring device suffers from zero bias and the
true mean measurement uncertainties $\bm{\delta_{y}}$ are known (Eq.
\ref{Scenario1}). For example, our measuring device exhibits independent
statistical and systematic uncertainties of $10\%$ and $30\%$, yielding a total
$32\%$ total uncertainty via addition in quadrature. Because both the measured
data and their uncertainties are faithful to the true underlying distribution,
our estimated $\bm{\theta_{opt}}$ match $\bm{\theta_{true}}$, up to the
estimated uncertainty of $\bm{\theta_{opt}}$. Panel (e) shows that the
distribution of standardized residuals between our model's predictions and the
corresponding experimental data are distributed according to a normal
distribution with unit variance.

Panel (b) of Fig. \ref{exampleFitFigure} shows the outcome of the second
scenario: our measuring device performs identically as in the first scenario,
but now our estimates of $\bm{\delta_{y}}$ are too small (Eq.
\ref{Scenario2}). This could arise if, for instance, both statistical and
systematic uncertainty contribute to the overall uncertainty of our measuring
device, but we have only recognized and reported the statistical uncertainty.
Because the minimum of our weighted-least-squares likelihood function is not
affected by overall rescalings of $\bm{\delta_{y}}$, we recover the same
$\bm{\theta_{opt}}$ as in the first scenario. However, our \textit{uncertainty}
estimates of $\bm{\theta_{opt}}$ have shrunk by a factor of three --- the same
factor by which we underestimated the measurement uncertainty -- because the
Hessian $\mathcal{H}(\bm{\theta_{opt}})$ scales proportionally with
$\bm{\delta_{y}}$.  Panel (f) shows that while the standardized residuals
remain normally distributed with a mean of zero, they are more dispersed than
the reference distribution. Thus, underestimation of experimental uncertainties
directly causes underestimation of parametric uncertainties. This is a generic
feature of parameter inference and, as we will show in the following section,
affects most previous OMP analyses.

Panel (c) of Fig. \ref{exampleFitFigure} presents a third scenario: as in the
first scenario, we have accurately estimated the experimental uncertainty
$\delta$, but now our experimental device occasionally returns anomalous
measurements (so-called ``outliers''). The simulated data $\bm{y}$ have been
drawn according to \ref{Scenario3}: each measurement has a 10\% chance of being
shifted upward by $\alpha$, which is an artificial ``outlier factor''.  This is
meant to represent a more realistic situation in which some fraction of
experimental data are inconsistent with the model, either because of model
defects or because of problems during experimental data collection. The
outliers ``pull'' on the likelihood function, causing our recovered
$\bm{\theta_{opt}}$ to differ from those of the previous scenarios, but,
because our $\bm{\delta_{y}}$ are the same as in the first scenario, our
\textit{uncertainty} estimates of $\bm{\theta_{opt}}$ do not change. The
parameter bias appears in panel (g) as asymmetry in the standardized residuals
with respect to the reference distribution, even as the variance of the
residuals is the same as in the first scenario. We note that even if our
measuring device returned no outliers, if our underlying model was incorrect
(i.e., model defect), certain data would appear to be outliers, and we would
obtain a similar result.

Finally, panel (d) of Fig. \ref{exampleFitFigure} combines the second and third
scenarios: $\bm{y}$ contains occasional outliers and $\bm{\delta_{y}}$ are
overconfident (Eq.  \ref{Scenario4}). Accordingly, our estimated
$\bm{\theta_{opt}}$ is biased and our uncertainty estimates of
$\bm{\theta_{opt}}$ are overconfident about the biased estimates. Both the bias
and the dispersion of the normalized residuals are visible in panel (h). This
scenario is the best analog to the OMP optimization task. For us to obtain
well-calibrated uncertainties that span the experimental data, our loss
function and optimization strategy must address both challenges: namely,
underestimation of experimental (co)variances, and fundamental discrepancies
between the model and data either due to model defects or problems with
experimental data collection (which we do not attempt to disentangle).

\subsection{Challenges for CH89 and KD}
\label{sec_fitting_OMP}

The difficulties of using weighted-least-squares estimators are well-known to
OMP designers, including those of CH89 and KD. A common symptom is that initial
fits to experimental data are often grossly unsatisfactory, clearly missing
``the physics'' present in the scattering data, leading to manual parameter
adjustment. The authors of CH89 comment that, early in their analysis, there were
often ``significant contributions from the data that the model is not able to
describe'' even when training to a single scattering data set. They tested
several alternative loss functions but found that in ``reduc[ing] the emphasis
of outlying points'' they ``lost sensitivity to even the good data''. After
testing various functions, their compromise was to introduce a weight factor to
their likelihood function for each data set $s$, equal to the minimum loss for
that data set obtained in a fit to \textit{only} that data set, i.e.,
\begin{equation} \label{CH89Likelihood}
    L(\bm{y}|\bm{x,\delta_{y},\theta}) = \sum_{s} \frac{L_{s}(\bm{\theta}|\bm{x_{s}},\bm{y_{s}})}{min(L_{s}^{loc}(\bm{\theta}|\bm{x_{s}},\bm{y_{s}}))},
\end{equation}
where $L_{s}$ is the contribution from data set $s$ to the overall
weighted-least-squares fit as in Eq.  \ref{WeightedLeastSquaresEstimator}. By
deemphasizing data sets that were poor matches to the form of their OMP, they
achieved a better visual fit to their training data. However, this solution
also introduces problems: the introduced weights are not easily interpreted nor
do they preserve the normalization of the likelihood function, which is
important for estimating $\bm{\Sigma}$. However, because finding
$\bm{\theta_{opt}}$ is insensitive to overall rescalings of $L$, most past
authors have been willing to sacrifice the possibility of accurately estimating
$\bm{\Sigma}$ in order to improve their single ``best-fit'' parameter vector.

Koning and Delaroche identified this issue in their global OMP
characterization as well and also provided extensive quantitative evidence that
traditional OMPs are incapable of reproducing the bulk of experimental data
within the range of reported experimental uncertainties. In Table 12 of
their OMP analysis \cite{KoningDelaroche}, they present sums of
uncertainty-weighted square residuals per degree of freedom (a
reduced-$\chi^{2}$ metric) for several prominent OMPs against a variety of
experimental data sets. In their analysis, a value near unity was taken as an
indication of good model-data agreement. For the widely used global OMPs they
considered, they found values of $\chi^{2}/N$ ranging from 6.3 to 11.2 for
differential elastic scattering cross sections and from 2.3 to 9.2 for neutron
total cross sections. Using their new potential (KD), they found values of
$\chi^{2}/N$ ranging from 4.5 to 7.4 for differential elastic scattering cross
sections and from 1.2 to 6.7 for neutron total cross sections, depending on the
experimental data corpus tested against.  They echoed the comments of the CH89
authors, noting that ``the optimization procedure is very sensitive'' to
underestimations in reported experimental uncertainties such that ``even a
slightly incorrect error estimation can easily vitiate an automated fitting
procedure''. In \cite{Koning2015}, Koning further analyzed model-experiment
discrepancies across the EXFOR database and combined several proposed remedies
into an ``evaluated'' $\chi^{2}$ expression meant to overcome the issues of
using na\"ive weighted least squares.

\begin{figure}[!htbp]
    \includegraphics[width=0.5\textwidth]{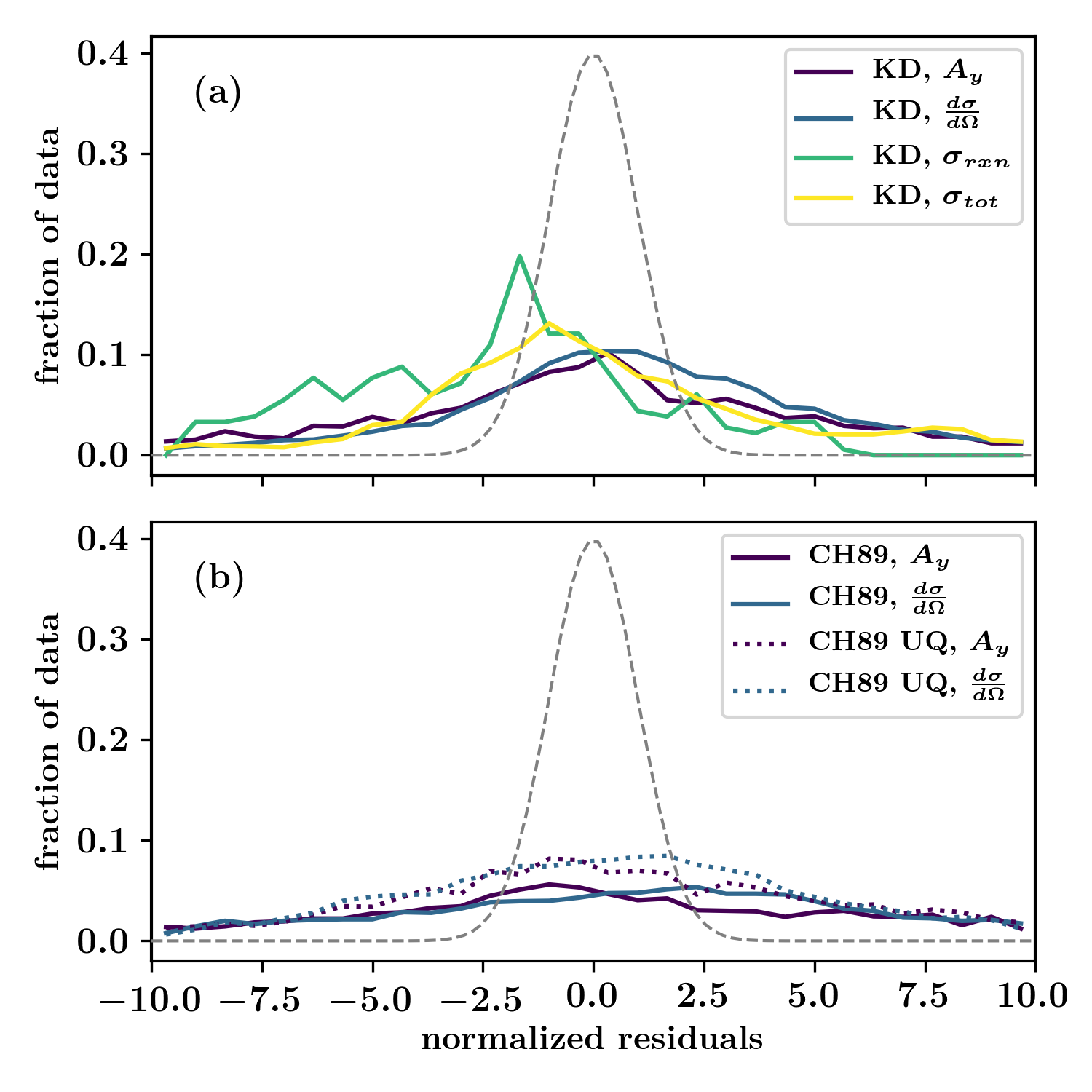}
    \caption{
        The distribution of normalized residuals ($r$/$\delta_{y}$) between
        the original CH89 and KD OMPs and their training data as reconstructed
        in this work are shown in panels (a) and (b), respectively. Residuals
        are histogrammed by data type, with all available proton and neutron data
        for that data type included (in contrast with Table \ref{OrigPotentialsResidualTable}
        which discriminates by projectile). The curves labeled CH89 UQ refer to
        the UQ assessment of the original CH89 work, which we have sampled here
        as $\bm{\theta} \sim \mathcal{N}(\bm{\theta}_{CH89},
        \bm{\Sigma}_{CH89})$, where $\bm{\theta}_{CH89}$ is the canonical CH89
        ``best fit'' parameter vector, and $\bm{\Sigma}_{CH89}$ is the
        canonical CH89 covariance matrix published in the original treatment.
    }
    \label{OrigPotentialNormalizedResiduals}
\end{figure}

To better understand these discrepancies between the trained model and training
data, we began by reproducing the original CH89 and KD analyses. Figure
\ref{OrigPotentialNormalizedResiduals} summarizes the performance of the
standard CH89 and KD potentials against the experimental data used to train
them, as reconstructed in the present work. For each experimental datum, the
normalized residual for that datum ($r/\delta_{y}$) was tabulated, then all
residuals histogrammed according to data type. In addition, in panel (b), two
dotted curves show the performance of CH89 when the CH89 parameters are
resampled according to the parameter covariance matrix presented in the
original publication. If the assumptions underpinning weighed-least-squares
were fulfilled, each line should follow the gray dashed line (a normal
distribution with unit variance), indicating that the CH89 and KD predictions
match the mean of the experimental data used to train them, and that the
training data are dispersed about the predictions in keeping with their
reported uncertainties. In reality, all types of scattering data show a
variance several times larger than unity, an indication either of
underestimation of experimentally reported uncertainties or of significant
model deficiencies, or both. The means of the distributions are offset to
varying degree, indicating that the canonical $\bm{\theta_{opt}}$ for these
OMPs retain some bias with respect to the underlying experimental data. Table
\ref{OrigPotentialsResidualTable} lists the mean, standard deviation, and
skewness of these observed distributions for each data type used to train the
CH89 and KD OMPs. This confirms the issues identified by past authors: clearly,
these OMPs do not span the variance of their training data, and for some data
types, predictions show systematic bias with respect to experiment.

\begin{table}[!htbp]
    \centering
    \caption{
        Mean ($\mu_{1}$), standard deviation ($\mu_{2}$), and skewness
        ($\mu_{3}$) for the distribution of standardized residuals between the
        original KD and CH89 OMPs and their training data, as reconstructed in
        this work. Results are tabulated separately for protons and neutrons.
        The columns labeled CH89 UQ refer to the UQ assessment of the
        original CH89 work, which we have sampled here as $\bm{\theta} \sim
        \mathcal{N}(\bm{\theta}_{CH89}, \bm{\Sigma}_{CH89})$, where
        $\bm{\theta}_{CH89}$ is the canonical CH89 ``best fit'' parameter
        vector, and $\bm{\Sigma}_{CH89}$ is the canonical CH89 covariance
        matrix estimate published in the original treatment.
    }
        \begin{tabular}{@{\extracolsep{4pt}}c c c c c c c c@{}}
         & \multicolumn{7}{c}{Proton data}
         \vspace{0.6em}\\
         & \multicolumn{2}{c}{CH89} & \multicolumn{2}{c}{CH89 UQ} & \multicolumn{3}{c}{KD} \\
         & $\frac{d\sigma}{d\Omega}$ & $A_{y}$ & $\frac{d\sigma}{d\Omega}$ & $A_{y}$ & $\frac{d\sigma}{d\Omega}$ & $A_{y}$ & $\sigma_{rxn}$ \\
         \cline{2-3} \cline{4-5} \cline{6-8}
         $\mu_{1}$ & 0.5 & -3.2 & 1.1 & 0.7 & 0.7 & -0.4 & -2.4 \\
         $\mu_{2}$ & 29.8 & 30.7 & 9.6 & 7.0 & 18.6 & 18.4 & 3.7 \\
         $\mu_{3}$ & -2.1 & -3.2 & -1.6 & 0.6 & -1.0 & -3.3 & -1.0 \\
    \end{tabular}
    \vspace{1.2em}\\
    \begin{tabular}{@{\extracolsep{4pt}}c c c c c c c c@{}}
         & \multicolumn{7}{c}{Neutron data}
         \vspace{0.6em}\\
         & \multicolumn{2}{c}{CH89} & \multicolumn{2}{c}{CH89 UQ} & \multicolumn{3}{c}{KD} \\
         & $\frac{d\sigma}{d\Omega}$ & $A_{y}$ & $\frac{d\sigma}{d\Omega}$ & $A_{y}$ & $\frac{d\sigma}{d\Omega}$ & $A_{y}$ & $\sigma_{tot}$ \\
         \cline{2-3} \cline{4-5} \cline{6-8}
         $\mu_{1}$ & -1.9 & 1.4 & -1.7 & 1.2 & -2.1 & 0.8 & -0.3 \\
         $\mu_{2}$ & 5.0 & 6.5 & 4.0 & 4.4 & 4.8 & 6.8 & 25.2 \\
         $\mu_{3}$ & -0.7 & 0.5 & -0.8 & 0.3 & -0.7 & -19.9 & -17.5 \\
    \end{tabular}

    \label{OrigPotentialsResidualTable}
\end{table}

\begin{figure}[tb]
    \includegraphics[width=0.5\textwidth]{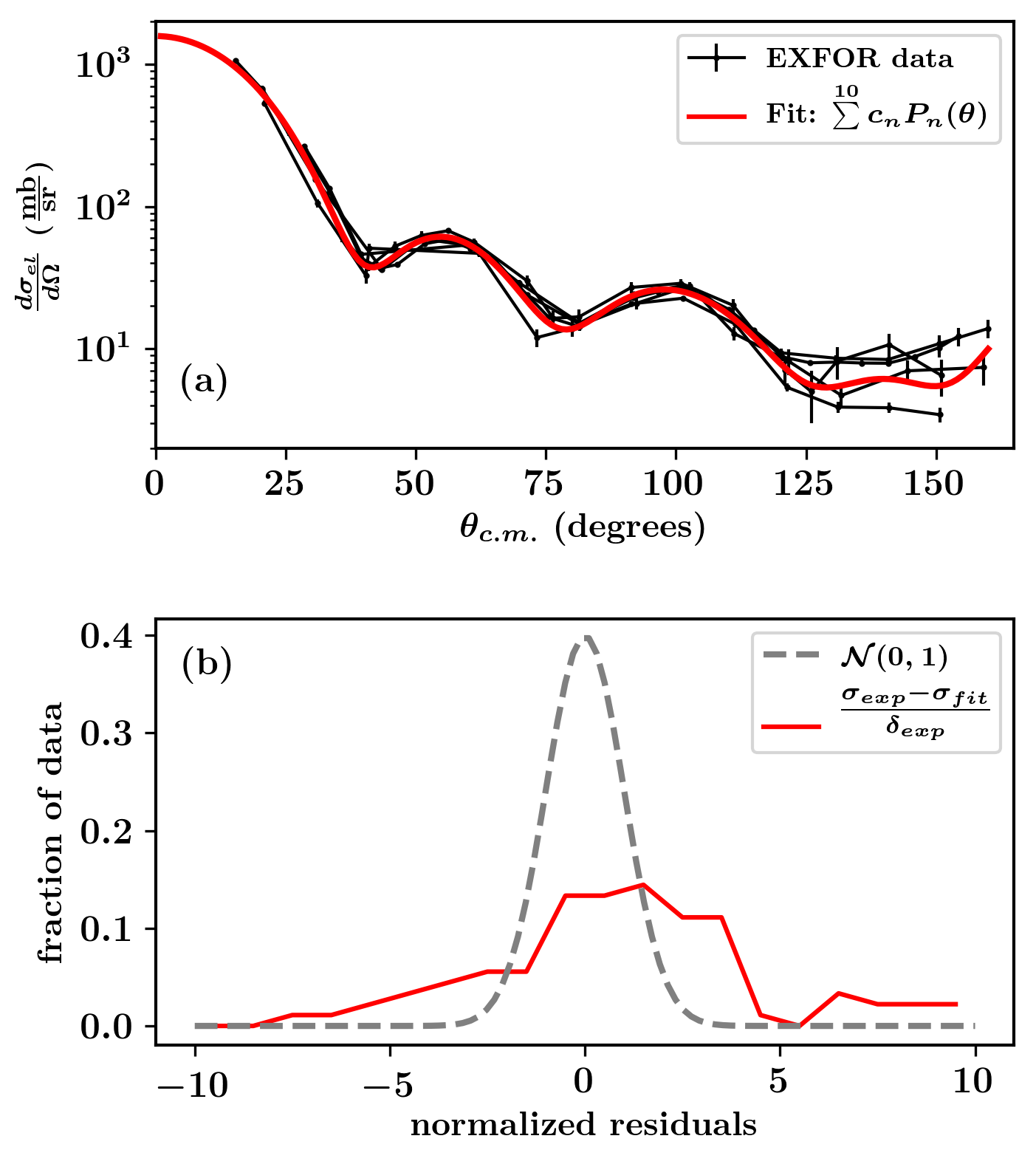}
    \caption{
        Five experimental data sets for \caForty(n,n)\caForty\ at 14 \MeV\ show
        significant variability, despite being collected under similar
        kinematic conditions. In panel (a), each data set is shown as a series of black
        points with the reported experimental uncertainties. A weighted-least-squares
        fit of all points, using the sum of the first ten Legendre polynomials
        as a model, is shown in gray. Panel (b) shows the normalized residuals for the
        experimental data points as a histogram (red line). A Gaussian
        distribution with unit variance is shown for reference (gray dashed
        line).
    }
    \label{ca40ExampleSpread}
\end{figure}

The comparison of these canonical OMPs with their training data led us to
investigate the self-consistency of the training data themselves. We discovered
that these training data sets were often inconsistent, in the sense that
\textit{no} plausible model could simultaneously fit them. This implies that for
data routinely used in OMP training, the reported experimental uncertainties
may be significantly underestimated. Figure \ref{ca40ExampleSpread} illustrates
the problem: in panel (a), five independent, representative elastic scattering
data sets for neutrons on \caForty\ at $14\pm0.1$ \MeV\ from the EXFOR database
\cite{EXFORDatabase} are shown. Each is comparable to the elastic scattering
data sets used to train the KD and CH89 OMPs. To facilitate comparison between
these data sets, which were measured at different angles, we describe their
mean behavior as:
\begin{equation} \label{LegendrePolynomialFit}
    f(\theta) = \sum\limits_{n=0}^{10} c_{n}p_{n}(\theta)
\end{equation}
where $p_{n}(x)$ are Legendre polynomials. A simple weighted-least-squares fit
was performed to optimize the polynomial coefficients $c_{n}$. When the fit and
training data are compared, the normalized residuals are inconsistent with one
another at the several-$\sigma$ level, as shown in panel (b) of the same
figure, due to underestimation of experimental uncertainties. Considering that
these data were all collected for the same projectile-target system and at the
same energy but are inconsistent at the several-$\sigma$ level, even larger
discrepancies may be expected when comparing many types of scattering
observables on different nuclei and energies during global OMP parameter
inference. (It is worth mentioning that, of the data types considered for training
OMPs, such experimental uncertainty underestimation appears to be most acute
for differential elastic scattering data.) To be reliable, any data-driven
assessment of OMP uncertainty must address this unaccounted-for dispersion of
the experimental data. Moreover, if we can determine how large such
unaccounted-for uncertainty must be to bring the optimized OMP and experimental
data into agreement, we gain insight into the degree of mutual consistency
between the OMP and the data libraries used to train the OMP.

\section{Improved inference for OMPs}
\label{sec_method}
In this section, we present our implementation for improved OMP parameter inference.
We propose a modified likelihood function that addresses the problems of
canonical OMP analysis as identified in the previous section. We then describe
our implementation of the CH89 and KD OMPs, our scattering code, and the MCMC
tools we used for performing parameter inference.

\subsection{Likelihood function}
\label{sec_method_likelihood}

For a training corpus consisting of $N$ experimental data, denoted ($\bm{x, y,
\delta_{y}}$), and an OMP with $k$ free parameters $\bm{\theta}$, we define our
likelihood function as follows:
\begin{equation} \label{LikelihoodFunction}
    \begin{split}
        L(\bm{y}|\bm{x,\delta_{y}, \theta, \delta_{u}}) & = \frac{1}{\sqrt{(2\pi)^{k}|\bm{\tilde{\Sigma}}|}} e^{-\frac{1}{2}\bm{r}^{\intercal}\bm{\tilde{\Sigma}}^{-1}\bm{r}}, \\
        \bm{r} & \equiv \bm{y} - M(\bm{\theta}, \bm{x}).
    \end{split}
\end{equation}
In place of the true (unknown) data covariance matrix $\bm{\Sigma}$, we have introduced
a diagonal covariance matrix ansatz $\bm{\tilde{\Sigma}}$:
\begin{equation} \label{DemocraticCovarianceAnsatz}
    \bm{\tilde{\Sigma}} \equiv \frac{k}{N} \begin{bmatrix} \Delta_{1} & & \\ & \ddots & \\ & & \Delta_{N} \\ \end{bmatrix}; \Delta_{1,\ldots,N} \in \bm{\Delta}.
\end{equation}
In this prescription, the augmented variances $\bm{\Delta}$ combine the experimentally
reported uncertainties $\bm{\delta_{y}}$ with an new unaccounted-for uncertainty for each datum, $\bm{\delta_{u}}$:
\begin{equation} \label{DeltaDefinition}
        \bm{\Delta} = \{(\delta_{y}^{2} + \delta_{u}^{2}) : \delta_{y} \in \bm{\delta_{y}}, \delta_{u} \in \bm{\delta_{u}}\},
\end{equation}
where each $\delta_{u}$ is calculated as follows:
\begin{equation} \label{DeltaUDefinition}
    \bm{\delta_{u}} = \{\frac{y + M(\bm{\theta}, x)}{2}\times\delta_{t} : x \in \bm{x}, y \in \bm{y}, \delta_{t} \in \bm{\delta_{T=\hat{t}(y)}}\}.
\end{equation}

To clarify these expressions, we start with the terms in Eq.
\ref{DeltaUDefinition}. As discussed in the previous section, the reported
uncertainties for experimental measurements are often too small to be
self-consistent, hindering robust OMP UQ assessment. As such, we need a way of
increasing our uncertainty in the experimental data that is consistent with the
expectation that different types of experimental data (differential elastic
cross sections, neutron total cross sections) will have different degrees of
uncertainty underestimation. At the same time, we want to respect the reported
experimental uncertainty, as it represents the measurer's informed judgement
about uncertainty affecting the measurement, even if in aggregate, they are
often underestimated. Our solution is to create a random variable $\delta_{T}$,
representing \textit{unaccounted-for uncertainty}, for each type of
experimental data appearing in the training corpus. For example, in the CHUQ
training corpus, there are four types of experimental data: differential
elastic scattering cross sections and analyzing powers, each for protons and
neutrons. As such, we create four random variables, each representing some
degree of unaccounted-for uncertainty for measured data of that type.  At
present, we do not know the value of these random variables $\bm{\delta_{T}}$,
so we treat them as parameters to be learned alongside the OMP parameters
$\bm{\theta}$.

Returning to Eq. \ref{DeltaUDefinition}, for each experimental datum in the
training set, we calculate a datum-specific unaccounted-for uncertainty term
$\delta_{u}$, which is the product of the average of the model prediction
$M(\bm{\theta},x)$ and the experimental datum value $y$ with the unaccounted-for
uncertainty $\delta_{T}$ of that datum's data type. (The term
$\bm{\delta_{T=\hat{t}(y)}}$ should be read as ``a $N$-long vector of
$\delta_{T}$ values, each corresponding to the data type of experimental
measurement $y$''.) Thus for each datum of the same type, the
individual unaccounted-for uncertainty $\delta_{u}$ is calculated using the 
same $\delta_{T}$.

With $\bm{\delta_{u}}$ defined, we proceed to Eq.
\ref{DeltaDefinition}. For each experimental training datum, the reported
uncertainty $\delta_{y}$ is added in quadrature with that datum's $\delta_{u}$
yielding the overall uncertainty $\Delta$ for that training datum. The vector
of these augmented uncertainties, $\bm{\Delta}$, enters Eq.
\ref{DemocraticCovarianceAnsatz}, which defines the covariance matrix ansatz.
The entries of $\bm{\tilde{\Sigma}}$ are scaled by $k/N$ in recognition that,
by replacing the $k \times k$-matrix $\bm{\Sigma}$ with an $N \times N$
covariance matrix ansatz $\bm{\tilde{\Sigma}}$, a scaling factor is required to
approximately preserve the matrix determinant that features in the overall
normalization. This is equivalent to saying that the $N$ training data cannot
all be independent random variables, as the information they contain can span,
most, the $k$ dimensions of $\bm{\theta}$.

In sum, our likelihood function (Eq. \ref{LikelihoodFunction}) replaces the unknown covariance matrix
$\bm{\Sigma}$ with a diagonal matrix of variance terms $\bm{\Delta}$, each of
which has been augmented based on the unaccounted-for-uncertainty $\delta_{T}$
for each data type. If reasonable values can be learned for unaccounted-for
uncertainties $\bm{\delta_{T}}$ in tandem with $\bm{\theta}$, this approach
will yield both a fitted OMP with good coverage of the training data and also
a sense of the missing uncertainty required to bring the experimental data
into agreement with the model. We remain agnostic about about the source of the
unaccounted-for uncertainty, be it underestimation of experimental uncertainty,
model deficiencies, errors in the tabulation of experimental results,
insufficient numerical precision during model calculations, or an ``unknown
unknown''. The practical effect of each $\Delta$ is the same as in
traditional weighted-least-squares, namely, to reduce the contribution of
residuals to the overall likelihood.

If we place the likelihood function in the
the log-likelihood form relevant for
optimization,
\begin{equation} \label{LogLikelihoodFunction}
    \log{L(\bm{y}|\bm{x,\delta_{y},\theta,\delta_{T}})} =
    -\frac{1}{2}\left[\frac{\bm{r^{T}r}}{|\bm{\tilde{\Sigma}}|}
        + \log{|\bm{\tilde{\Sigma}}|}
        + k \log{(2\pi)}
    \right].
\end{equation}
it becomes clear that minimizing the log-likelihood involves a competition between the first
and second terms inside the brackets. Larger $\bm{\delta_{T}}$ values make for larger
$\bm{\Delta}$ and a larger covariance determinant $|\bm{\tilde{\Sigma}}|$,
which reduces the first term but increases the second term. At the optimum,
where $\bm{\theta}$ minimizes the contribution from residuals, both terms should
be equal, 
\begin{equation}
    \bm{r^{T}r} = |\bm{\tilde{\Sigma}}|\log{|\bm{\tilde{\Sigma}}}|.
\end{equation}
This implies that, at the start of training our OMP, our unaccounted-for
uncertainty random variables $\bm{\delta_{T}}$ will grow rapidly, to
counterbalance the large residuals between model and data, but, as the fit
improves and the residuals shrink, $\bm{\delta_{T}}$ will grow smaller.

We note that the factor $k/N$ in the covariance ansatz is the simplest but not the only
choice to account for the unknown degree of correlation between individual
data. For example, one might expect \textit{a priori} that experimental data of
each type (such as proton reaction cross section, neutron analyzing powers,
etc.) will correlate strongly with each other, due to common features of the
experimental design or ease of certain types of measurement, but correlate more
weakly with data of other data types. Accordingly, one might want to ensure
that each data type contributes equally to the overall likelihood, independent
of how many data points it contains, so that data types with fewer data points
are not outvoted by data types with better experimental coverage. In that
case, $\bm{\tilde{\Sigma}}$ could be modified to be:
\begin{equation} \label{FederalCovarianceAnsatz}
    \bm{\tilde{\Sigma}} \equiv \frac{k}{n_{t}} \begin{bmatrix}
        \frac{1}{N_{1}} \Delta_{1} \\
        \vdots \\
        \frac{1}{N_{1}} \Delta_{N_{1}} \\
        \frac{1}{N_{2}} \Delta_{N_{1}+1} \\
        \vdots \\
        \frac{1}{N_{2}} \Delta_{N_{1}+N_{2}} \\
        \frac{1}{N_{3}} \Delta_{N_{1}+N_{2}+N_{3}} \\
        \vdots \\
        \frac{1}{N_{T}} \Delta_{N} \\
    \end{bmatrix} \times I_{N}
\end{equation}
where $n_{t}$ is the number of unique data types $t$, and $N_{t}$ is the number
of data points of type $t$, and $I_{N}$ is the identity matrix of dimension
$N$. With this choice for $\bm{\tilde{\Sigma}}$, all data points of a given
data type would be given equal influence for that type, and each data type
would be given equal influence on the overall likelihood. Any additional
information about the covariance structure of the experimental data, such as
knowledge of the systematic error for one or more specific data sets, can be
directly inserted to turn $\bm{\tilde{\Sigma}}$ into a more-realistic
block-diagonal matrix. We experimented with a handful of alternatives,
including Eq. \ref{FederalCovarianceAnsatz}, and found that their impact on
the final uncertainty-quantified OMPs was small except in situations where one
training data type had far fewer data points than the other types (see Fig.
\ref{DemocraticFederalComparison} in Section \ref{sec_results_discussion}).
Unless noted otherwise, all results in the following sections were generated
using Eq. \ref{LikelihoodFunction} as the likelihood function.

Finally, as discussed in toy-model scenarios two and four of Section
\ref{sec_fitting_toy}, we still need a way of identifying outliers in the
training data corpora. By outlier, we mean a datum that should not be used to
train the model, either because the model is missing physics that the data
capture (e.g., effects of deformation if the model assumes sphericity), or
because the data are erroneous. In either of these cases, training the model to
the datum would bias model parameters. To identify outliers, we implemented a
procedure similar to that by P\'erez, Amaro, and Arriola in their analysis of
the NN interaction via partial wave analysis of NN scattering data
\cite{Perez2013}, and first suggested by Gross and Stadler \cite{Gross2008}.
Briefly, in a standard NN scattering database they examined, they found that
certain data collected in similar kinematic conditions were mutually
inconsistent up to the experimentally reported uncertainties. Rather than
reject all inconsistent data as outliers, they used an iterative procedure to
simultaneously train a model to these data while updating the outlier status of each
datum used for training. In the initial step, their model was fit to the full
corpus of NN-scattering data. Any data lying $>3\sigma$ away from the model,
where $\sigma$ was taken to be the reported experimental uncertainty, were
flagged as outliers and not included in the following round of fitting. In the
second round, the model was fitted to the smaller set of ``inlier'' data, then
the outlier status of each datum was assessed again, based on the second fit.
The process was repeated until the model fit and the outlier status of each
data point became stable, yielding a mutually-consistent database, up to the
fitted model. Certain data that were initially incompatible with the others
were thus recovered as the model fit improved over multiple iterations. 

Our procedure was the same except in two respects. In our case, for
$\sigma$ we included both the variance of the model prediction from MCMC and
the experimental uncertainty, summed in quadrature:
\begin{equation}
    \bm{\sigma}^{2} = \{\delta_{y}^{2} + var[M(\bm{\theta}, x)]: \delta_{y} \in \bm{\delta_{y}}, x \in \bm{x}\}
\end{equation}
Second, because MCMC involves sampling noise, many walker steps are often 
required before walkers have time to react to changes in the outlier status of
the experimental data. Thus, we updated the outlier status of the training data
only at 100-step intervals during MCMC, rather than at every step.

\subsection{CH89 and KD implementation}
\label{OMPDescription}

We turn now to the implementation of the OMPs we retrained according to
our proposed approach. Both the CH89 and KD OMPs assume a spherical optical
potential, smooth in scattering energy $E_{lab}$ and target $A$, for modeling
the projectile-target interaction. CH89 \cite{CH89} was restricted to proton and
neutron elastic scattering cross sections and analyzing powers on nuclei ``in
the valley of stability'' with $40 \leq A \leq 209$ and for scattering energies
of $10 \leq E \leq 65$ \MeV\ (assumed to be the lab frame). The potential
consists of five terms: a real central potential, an imaginary central
potential, an imaginary surface potential, a real spin-orbit potential, and for
protons, a Coulomb potential (see the Appendix for detailed functional forms). In
all, these components employ 22 free parameters. To perform comparisons with
experimental data, the authors of CH89 used a joint scattering-optimization
code called MINOPT, a hybrid of the scattering code OPTICS \cite{Eastgate1973}
and the CERN optimization code MINUIT \cite{James1975}. For the wave equation,
the original treatment used the non-relativistic Schr\"odinger equation.
Because the lowest considered scattering data energy was $10$ \MeV, the
original treatment took the compound-nucleus contribution to be zero.

The KD global OMP \cite{KoningDelaroche} was fitted not only to proton and
neutron elastic scattering cross sections and analyzing powers, but also to
proton reaction (or ``non-elastic'') and neutron total cross sections. The
authors define its domain as ``(near)-spherical'' nuclei with $24 \leq A \leq
209$ for incident scattering energies of $0.001 \leq E \leq 200$ \MeV\ in the
lab frame. In addition to the potential component types used in CH89, KD adds
an imaginary spin-orbit component. Each component was made substantially more
flexible in energy- and asymmetry-dependence, bringing the total number of free
parameters to 46. To perform comparisons with experimental data, the developers
used the scattering code ECIS-97, as accessed through a visual interface called
ECISVIEW. For the wave equation, the authors ``[employed] the relativistic
Schr\"odinger equation throughout'', using ``the true masses of the projectile
and target expressed in atomic mass units''. To manage optimization in this
higher-dimensional space, they developed a new approach they called
``computational steering'': a user manually adjusted parameters in real time to
achieve a good visual fit, which was followed by an automated simulated
annealing procedure using the program SIMANN to achieve a quantitative optimum.

For our recharacterization of these OMPs, we adhered to the original potential
forms and scattering assumptions as described above but with a few minor
differences. First, scattering calculations for CH89 were performed according
to the same relativistic-equivalent Schr\"odinger equation used for KD
calculations rather than the non-relativistic treatment of the original. The
effect was to slightly improve the fidelity of calculated cross sections at the
highest scattering energies included in the CHUQ corpus (65 MeV). Second,
for differential elastic scattering cross sections at scattering energies below
roughly 10-15 \MeV\, the elastic contribution from the compound nucleus becomes
significant compared to the direct contribution from the OMP and must be
included for comparison to experimental data. The authors of CH89 restricted
their data corpus to scattering energies $\geq 10$\MeV\ for this reason. For
the KDUQ corpus, however, roughly 10\% of the elastic scattering data were
collected below $10$ \MeV. To enable comparison with these data, Koning and
Delaroche used the compound cross section values generated by ECIS-97, the same
code they used for direct scattering calculations. In our case, we generated
compound elastic cross sections using the LLNL Hauser-Feshbach code YAHFC
\cite{YAHFC}, using the canonical parameters of KD to generate the transmission
coefficients needed for the calculation.

\subsection{Scattering code and MCMC}

For scattering calculations and parameter inference, we combined the MCMC
utility \textsc{emcee} \cite{emcee} with a new, lightweight C++ and Python
library, \textsc{tomfool}, that we developed to perform single-nucleon
scattering calculations. Cross sections were generated via a
calculable-$R$-matrix Lagrange-mesh method after \cite{Descouvement2010,
Baye2015} detailed in the Appendix. The use of a Lagrange-Legendre basis
instead of a radial basis accelerates calculations severalfold but at the cost
of a small loss of precision, depending on the number of basis elements and
chosen $R$-matrix channel radius. To ensure fair comparison with the original
CH89 and KD analyses, we applied several measures to validate our calculation
pipeline. First, wherever possible, we drew mathematical functions from the Gnu
Scientific Library (GSL) \cite{GSL}. Any necessary functions unavailable in GSL
(such as optical potential functional forms and relativistic kinematics
equations) were subjected to a suite of unit and integration tests, including
comparison against results from the well-tested scattering code
\textsc{frescox} \cite{Thompson1988, FRESCOX} and \textsc{lise++}
\cite{Tarasov2016, LISE++}. For relativistic calculations, in addition to
treating scattering energies and angles relativistically, we use the wavenumber
and optical-potential rescaling approximations given by Eqs. 17 and 20/21 of
\cite{Ingemarsson1974}, the same formulae used for this purpose in
\textsc{frescox} and \textsc{talys}. Using \textsc{frescox} we prepared a set
of cross section benchmarks covering a range of scattering energies, angles and
targets representative of the KDUQ corpus. Using an $N=30$ Lagrange-Legendre
basis, an $R$-matrix channel radius of 15 \femto\meter, a maximum partial wave
angular momentum $l_{max}=80$, and a convergence threshold of $10^{-6}$ for the
magnitude of $S$-matrix elements, we achieved agreement with the
\textsc{frescox} benchmarks to 1\% or better, both for our relativistic and
non-relativistic implementations for CH89 and KD. This configuration was used
for all scattering calculation results in our analysis. Finally, we performed
numerous spot checks against the figures in the original CH89 and KD papers to
confirm that our implementation of their OMPs generates the same cross sections
to within the graphical resolution of the original publications.

For each OMP parameter, we assigned a weakly informative truncated Gaussian
prior centered on the canonical parameter value (that is, centered on the
parameter values from the original KD and CH89 publications). For each prior we
set the variance based on our estimates about the sensitivity of scattering
observables to changes in that type of parameter. For example, a change of 20\%
in a Woods-Saxon radius or diffuseness would result in large changes to the
location of elastic scattering diffraction minima and would thus be relatively
unlikely, but not impossible, given the level of consistency among the
experimental data. In contrast, the energy-dependence of the depth of the
imaginary spin-orbit potential is likely only very weakly sensitive to
available experimental data, so a deviation by a factor of 2 or more from the
canonical value in KD would not be surprising. Absolute upper and lower limits
of the truncated Gaussian priors were set to prevent any single parameter from
becoming non-physical, resulting in, for example, a negative radius. For the
unaccounted-for uncertainty random variables $\bm{\delta_{T}}$, we assigned
truncated Gaussian priors as
\begin{equation}
    \delta_{t} \sim \mathcal{N}(\mu=0.2, \sigma=0.2;\delta_{t}>0),
\end{equation}
for differential elastic observables and
\begin{equation}
    \delta_{t} \sim \mathcal{N}(\mu=0.02,\sigma=0.02;\delta_{t}>0),
\end{equation}
for integral observables $\sigma_{rxn}$ and $\sigma_{tot}$. This corresponds to
an expectation of 20\% unaccounted-for uncertainty in differential data types and 2\%
unaccounted-for uncertainty in integral data types. We based these priors on
the observed degree of agreement of the canonical KD and CH89 potentials
against their training corpora and on the typical range experimentally reported
uncertainties for these types of data. To begin MCMC, $8\times k$ walkers were
initialized according to
\begin{equation}
    \bm{\theta},\bm{\delta_{t}} \sim \mathcal{N}(\mu=\mu_{prior},\sigma=0.1\sigma_{prior}) 
\end{equation}
for CHUQ and
\begin{equation}
    \bm{\theta},\bm{\delta_{t}} \sim \mathcal{N}(\mu=\mu_{prior},\sigma=0.01\sigma_{prior}) 
\end{equation}
for KDUQ, with $k$ the number of parameters subject to inference. For the MCMC
proposal distribution, we used the \textsc{emcee} \cite{emcee} default proposal
distribution, the affine-invariant Goodman-Weare sampling prescription, but
with a scaling parameter $a=1.4$, reduced from the default value of $a=2$ to
improve the acceptance fraction given the high dimensionality of the parameter
space. Sampling continued for roughly 10,000 samples until ensemble means no
longer exhibited movement in any parameter dimension and the percentage of each
data type that were flagged as outliers ceased to change (excepting $\approx
0.1\%$ fluctuations due to the Monte Carlo nature of sampling). Due to our
expectation of very long autocorrelation times among walkers, we used only the
terminal sample from each walker for all results shown below.

Our reassembly of the training data corpora used to train the canonical OMPs
is detailed in the Supplemental Material \cite{SupplementalMaterial}. While we
were able to recompile and verify almost all of the training data as originally
used, there were a handful of discrepancies between data as reported in the
referenced literature, the data as listed in the canonical CH89 and KD
treatments, and the data as listed in the EXFOR experimental reaction database.
Details of these differences and references to the EXFOR accession number for
the data set in question (or, if the data were not available through EXFOR, to
the original literature) are provided in the Supplemental Material
\cite{SupplementalMaterial}. Because our approach involves outlier-rejection
and unaccounted-for uncertainties that were as large or larger than
experimentally reported uncertainties, the few discrepancies were unlikely to
have any appreciable effect on our analysis.

\section{Results} \label{sec_results}

Our results are organized in three parts. First, we compare the performance of
CHUQ against that of CH89 with respect to their training data. To assess
predictive power, CHUQ and CH89 are compared against a Test corpus of new
scattering data collected from 2003-2020 (after the publication of the original treatment).
Next, we present a similar comparison for KDUQ and KD. Last, we discuss the
comparative uncertainty of the potentials, including comparison
of volume integrals and how alternative likelihood functions could affect our results.

\subsection{CH89 vs CHUQ performance}
\begin{figure*}[!tbp]
    \includegraphics[width=\textwidth]{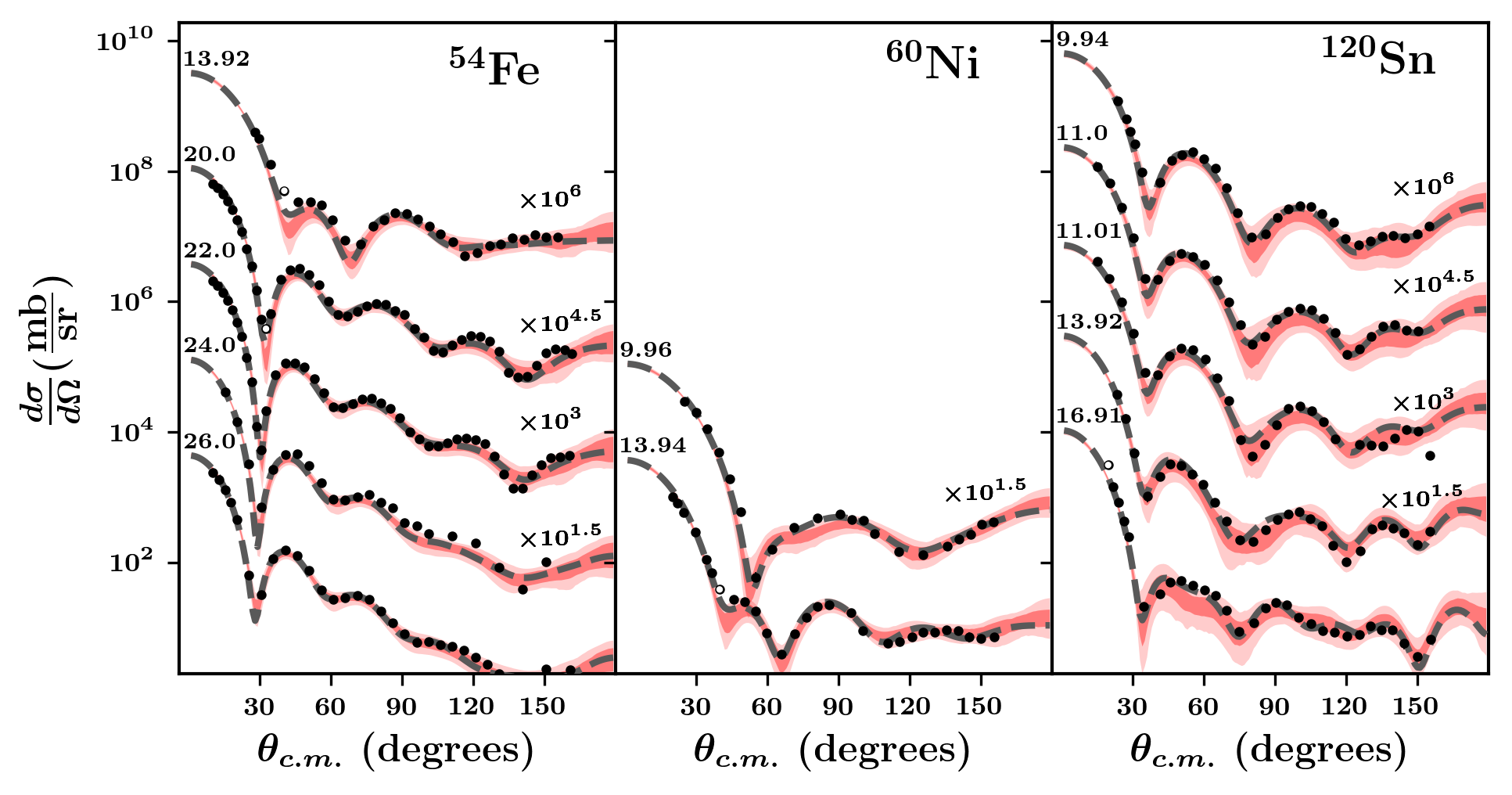}
    \caption{
        Representative experimental and calculated neutron differential elastic
        cross sections are plotted for \feFiftyFour, \niSixty\ and
        \snOneHundredTwenty\ at selected energies. Experimental data are shown
        as points with reported experimental uncertainties.  The outlier status
        of each point (as defined previously) is
        indicated by color: black points are inliers, and white points are
        outliers. Calculations from the canonical CH89
        parameters are shown as a gray dashed line. The CHUQ 68\% and 95\%
        uncertainty intervals are shown as dark and light red bands,
        respectively. The data sets are labeled by scattering energy (MeV, in
        the lab frame) and offset vertically for legibility.
    }
    \label{CHUQ_ECSDisplay}
\end{figure*}
\begin{figure*}[!tbp]
    \includegraphics[width=\textwidth]{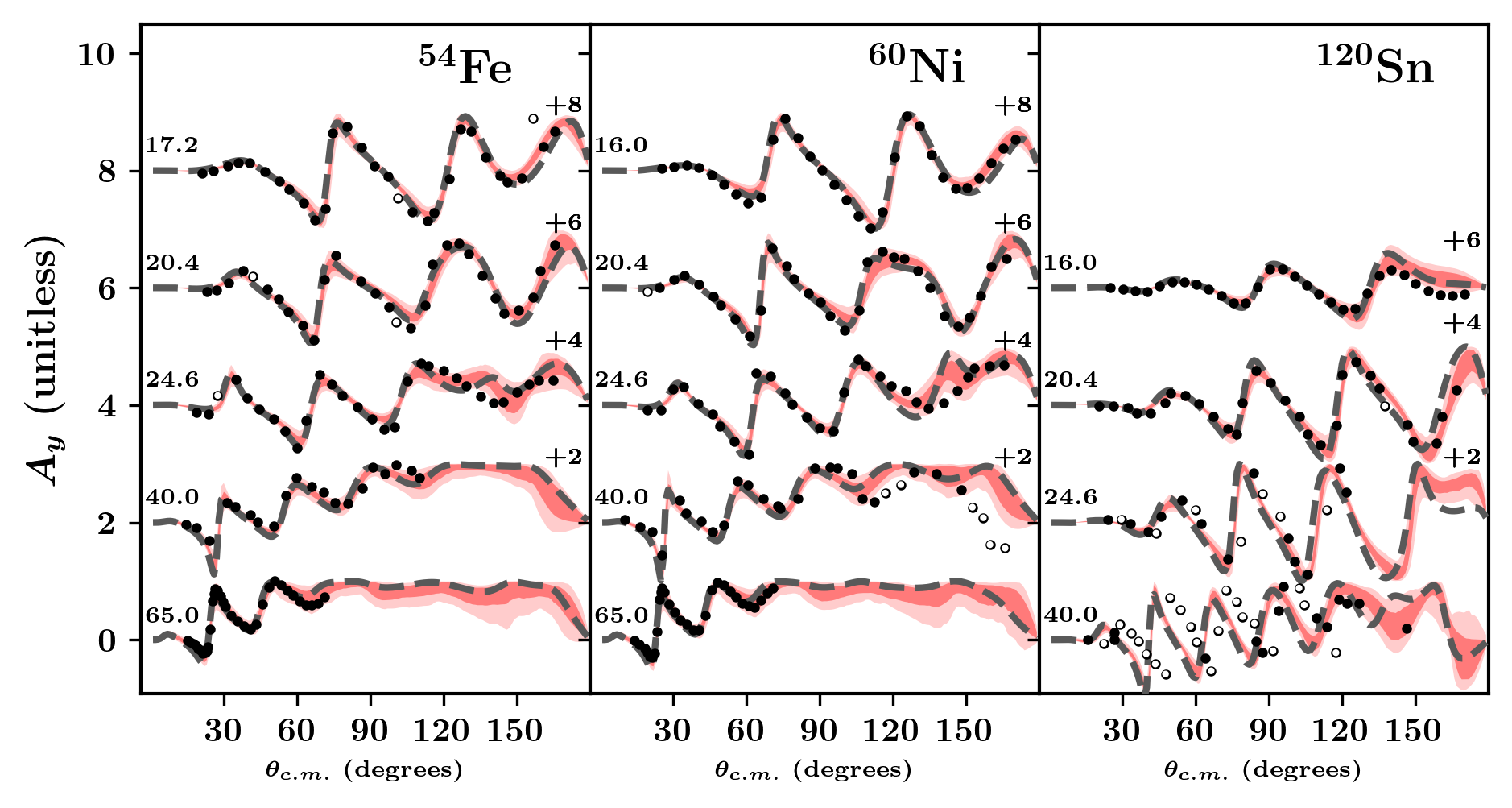}
    \caption{
        Representative experimental and calculated proton analyzing powers are
        plotted for \feFiftyFour, \niSixty\ and \snOneHundredTwenty\ at
        selected energies. See caption of Fig. \ref{CHUQ_ECSDisplay} for key.
    }
    \label{CHUQ_APowerDisplay}
\end{figure*}
\begin{figure}[!tbp]
    \includegraphics[width=0.5\textwidth]{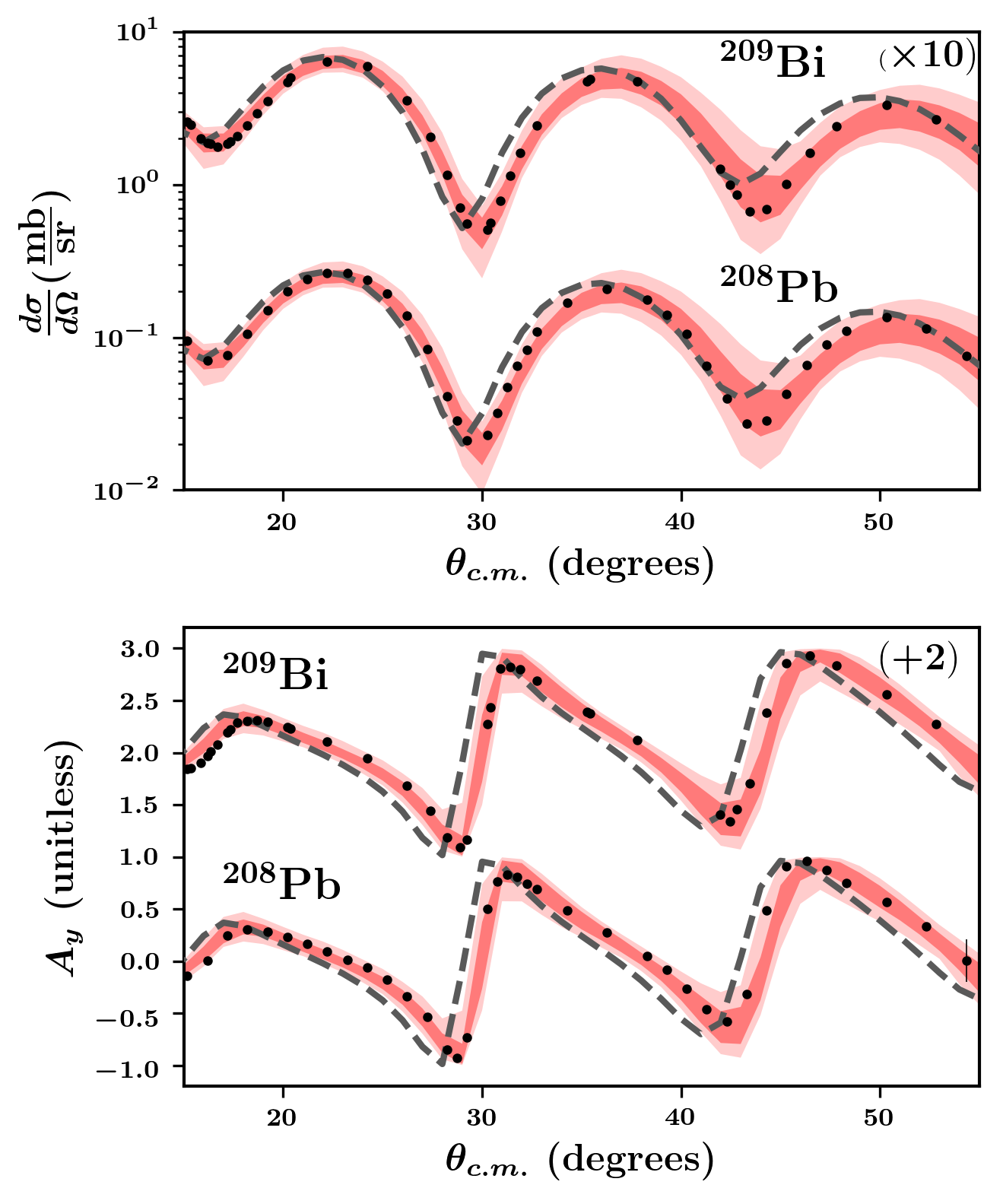}
    \caption{
        CH89 and CHUQ predictions are compared against experimental proton
        elastic scattering observables on \pbTwoHundredEight\ and
        \biTwoHundredNine\ at 65 \MeV. See caption of Fig.
        \ref{CHUQ_ECSDisplay} for key.
    }
    \label{CHUQAngularPhasing}
\end{figure}

\begin{figure}[!tbp]
    \includegraphics[width=0.5\textwidth]{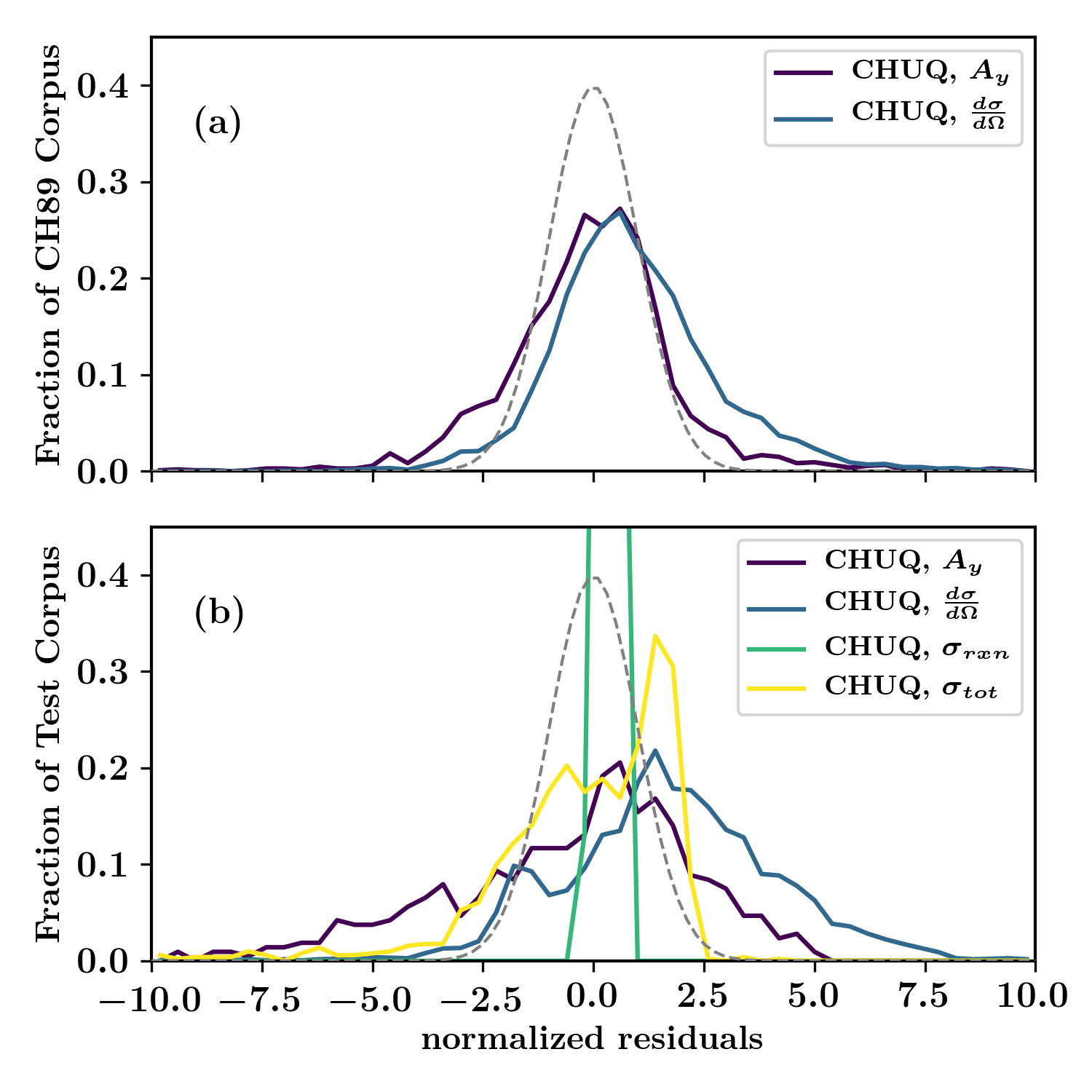}
    \caption{
           Normalized residuals ($r_{i}$/$\delta_{i}$) between CHUQ's
           predictions and the CHUQ corpus and Test corpus are histogrammed by data
           type. Panel (a) shows performance against the CHUQ corpus.
           Panel (b) shows performance against the Test corpus.
    }
    \label{CHUQNormalizedResiduals}
\end{figure}

Figures \ref{CHUQ_ECSDisplay} and \ref{CHUQ_APowerDisplay} show the performance
of CHUQ and the canonical CH89 OMP with respect to several representative
experimental data sets in the CHUQ training corpus. Figures comparing
CHUQ and CH89 over the entire CHUQ corpus are provided in the Supplemental
Material \cite{SupplementalMaterial}. Overall, the median predictions of CHUQ
are very similar to the canonical CH89 predictions, with the largest
differences being slightly lower predicted differential elastic cross sections
from CHUQ compared to those from CH89 around 10-11 \MeV, the lowest scattering
energies considered in the CHUQ corpus. Compared to the canonical CH89
analysis, our use of a fully relativistic-equivalent Schr\"odinger equation in
the present work and our relaxation of the fixed Coulomb radius parameters $r_{c}$
and $r_{c}^{(0)}$ for CHUQ improves the angular dependence of proton
differential elastic scattering predictions at higher energies on high-$A$
targets, as shown in Fig. \ref{CHUQAngularPhasing}.

Figure \ref{CHUQNormalizedResiduals} summarizes the overall performance of CHUQ
against the full CHUQ corpus and against the Test corpus. The means, standard
deviations, and skewnesses of the residual distributions shown in Fig.
\ref{CHUQNormalizedResiduals} are listed in Table \ref{CHUQResidualTable}.
Using the CHUQ corpus, we can directly compare the original treatment's
uncertainty estimation (CH89 UQ in Fig. \ref{OrigPotentialNormalizedResiduals}
and Table \ref{OrigPotentialsResidualTable}) and that of the present work (CHUQ
in Fig.  \ref{CHUQNormalizedResiduals}, panel (a), and Table
\ref{CHUQNormalizedResiduals}). Across the data types in the CHUQ corpus, CHUQ
yields similar mean residuals: between -1.0 and 1.0, versus -1.7 to 1.1 for
CH89 UQ. This suggests that both the canonical CH89 parameters and
CHUQ's central parameter values do well at reproducing average trends of
training data. In CHUQ, there is apparent tension between neutron differential
elastic scattering cross sections, which are slightly underpredicted
($\mu_{1}=-1.0$) and proton and neutron differential elastic scattering
analyzing powers, which are slightly overpredicted ($\mu_{1}=1.0$ and
$\mu{1}=0.9$).

The main difference is that compared to CH89 UQ, CHUQ yields
much smaller residual standard deviations: between 1.7 to 2.1 across data
types, versus 4.0 to 9.6 for CH89 UQ. That the variance of the residuals is
much closer to unity indicates that the larger parametric uncertainty of CHUQ
more faithfully represents the spread of the experimental data in the CHUQ
corpus. Further, the fact that the variance of CHUQ-corpus residuals remains
larger than unity shows that the priors we assigned to the unaccounted-for
uncertainties $\delta_{T}$ are preventing $\delta_{T}$ from becoming even larger,
which would further reduce the constraining power of the training data.

Panel (b) of Fig. \ref{CHUQNormalizedResiduals} illustrates performance of CHUQ
against the Test corpus. The Test corpus includes many scattering data far
beyond the prescribed range of validity given by the authors of CH89, including
data collected at scattering energies from 1-10 \MeV\ and from 65-295 \MeV,
proton $\sigma_{rxn}$ and neutron $\sigma_{tot}$ data, and data from targets
with $A<40$. The performance of CHUQ is moderately degraded on the Test corpus
compared to the CHUQ corpus, with mean residuals ranging from -1.8 to 2.0
across data types, and residual standard deviations ranging from 1.3 to 3.1 for
elastic observables. Though the CHUQ corpus used for training
did not include either proton $\sigma_{rxn}$ or neutron $\sigma_{tot}$ data,
CHUQ's average performance against the Test corpus in these data sectors is
surprisingly good, with mean residuals of 0.3 for proton $\sigma_{rxn}$ and
-0.3 for neutron $\sigma_{tot}$. This indicates that despite substantial
unaccounted-for uncertainty in the training data, fits that employ only elastic
scattering data can still provide meaningful constraints on the imaginary terms
in the potential.

\begin{table}[!tbp]
    \centering
    \caption{
        Mean ($\mu_{1}$), standard deviation ($\mu_{2}$), and skewness
        ($\mu_{3}$) for the distributions of standardized residuals between
        CHUQ and experimental data, as shown in Fig.
        \ref{CHUQNormalizedResiduals}. Here the distributions are tabulated
        separately for protons and neutrons (cf. with Tables
        \ref{KDUQResidualTable} and \ref{OrigPotentialsResidualTable}).
    }
        \begin{tabular}{@{\extracolsep{4pt}}c c c c c c@{}}
         & \multicolumn{5}{c}{Proton data}
         \vspace{0.6em}\\
         & \multicolumn{2}{c}{CHUQ Corpus} & \multicolumn{3}{c}{Test Corpus} \\
         & $\frac{d\sigma}{d\Omega}$ & $A_{y}$ & $\frac{d\sigma}{d\Omega}$ & $A_{y}$ & $\sigma_{rxn}$ \\
         \cline{2-3} \cline{4-6} 
         $\mu_{1}$ & 0.0 & 1.0 & -0.3 & -1.8 & 0.3 \\
         $\mu_{2}$ & 2.1 & 2.1 & 3.1 & 1.3 & 0.2 \\
         $\mu_{3}$ & -0.1 & 1.6 & -0.9 & -2.8 & 0.1 \\
    \end{tabular}
    \vspace{1.2em}\\
    \begin{tabular}{@{\extracolsep{4pt}}c c c c c c@{}}
         & \multicolumn{5}{c}{Neutron data}
         \vspace{0.6em}\\
         & \multicolumn{2}{c}{CHUQ Corpus} & \multicolumn{3}{c}{Test Corpus} \\
         & $\frac{d\sigma}{d\Omega}$ & $A_{y}$ & $\frac{d\sigma}{d\Omega}$ & $A_{y}$ & $\sigma_{tot}$ \\
         \cline{2-3} \cline{4-6} 
         $\mu_{1}$ & -1.0 & 0.9 & -1.2 & 2.0 & -0.3 \\
         $\mu_{2}$ & 1.7 & 1.9 & 1.8 & 2.3 & 2.2 \\
         $\mu_{3}$ & -0.7 & 0.5 & -0.2 & 0.1 & -2.1 \\
    \end{tabular}

    \label{CHUQResidualTable}
\end{table}


\subsection{KD vs KDUQ performance}

\begin{figure*}[!tbp]
    \includegraphics[width=\textwidth]{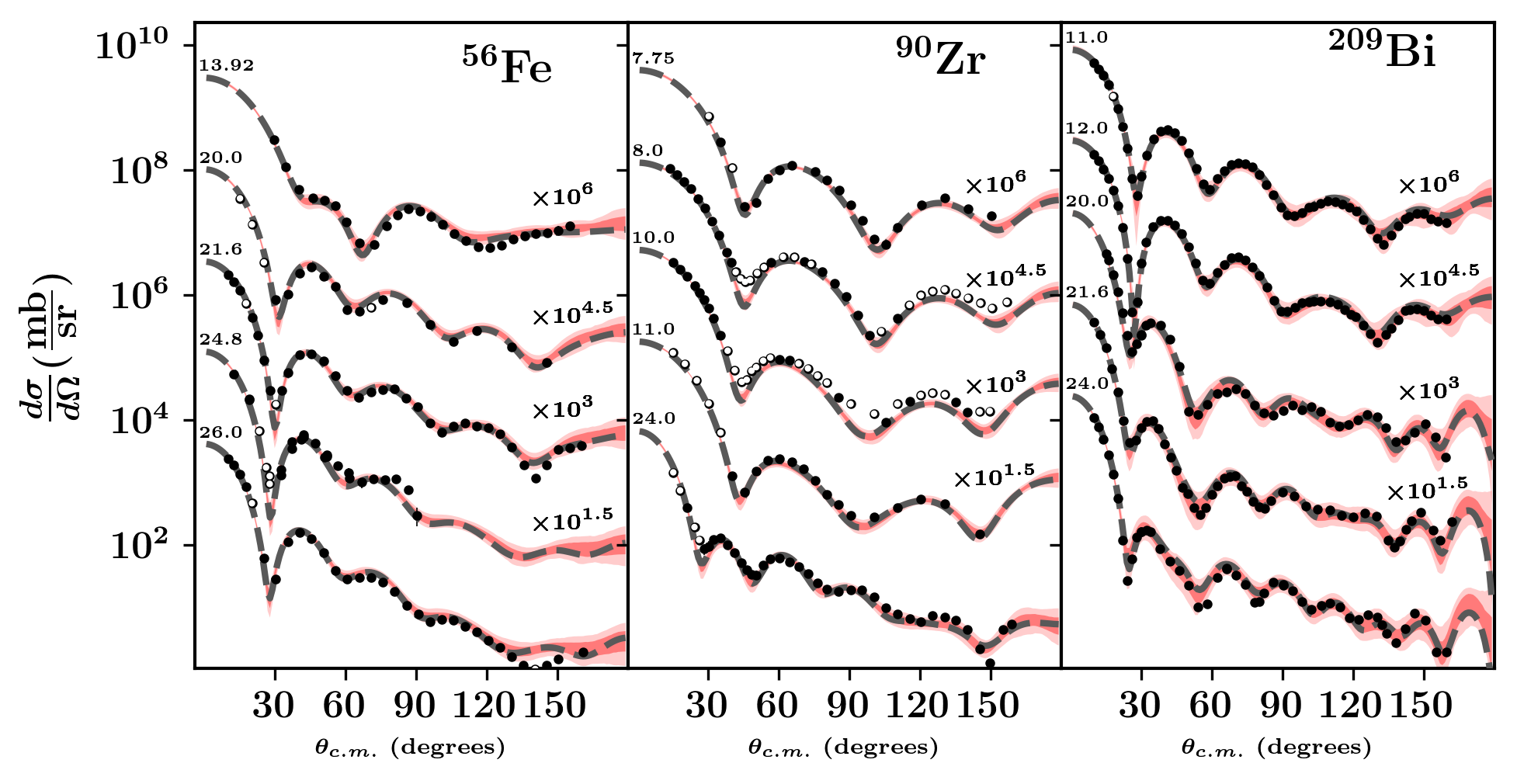}
    \caption{
        Representative experimental and calculated neutron differential elastic
        cross sections are plotted for \feFiftySix, \zrNinety\ and
        \biTwoHundredNine\ at selected energies. Experimental data are shown as
        points with associated uncertainties. The outlier status of each point
        (as defined previously) is indicated by
        color: black points are inliers, and white points are outliers.  Cross
        sections calculated using the original KD formulation are shown via
        gray dashed line. The KDUQ 68\% and 95\% uncertainty intervals are
        shown as dark and light red bands, respectively. The data sets are
        labeled by scattering energy (MeV, in the lab frame) and offset
        vertically for legibility.
    }
    \label{KDUQ_ECSDisplay}
\end{figure*}

\begin{figure*}[!tbp]
    \includegraphics[width=0.9\textwidth]{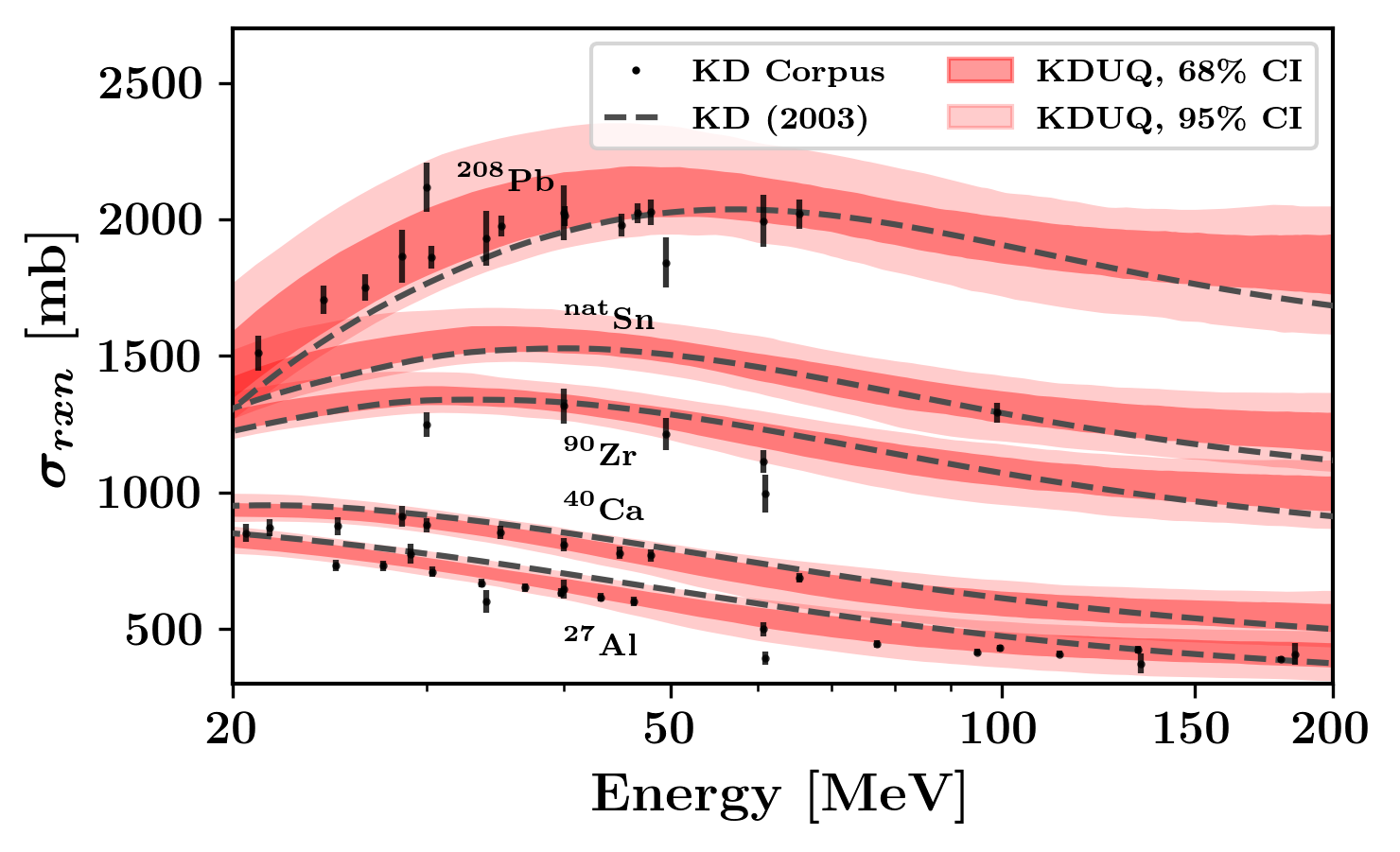}
    \caption{
        Representative experimental proton reaction cross section data and KD
        and KDUQ calculations are plotted for selected nuclei in the KDUQ corpus.
        See caption of Fig. \ref{KDUQ_ECSDisplay} for additional information on
        the legend.
    }
    \label{KDUQ_RCSDisplay}
\end{figure*}

\begin{figure*}[!tbp]
    \includegraphics[width=0.9\textwidth]{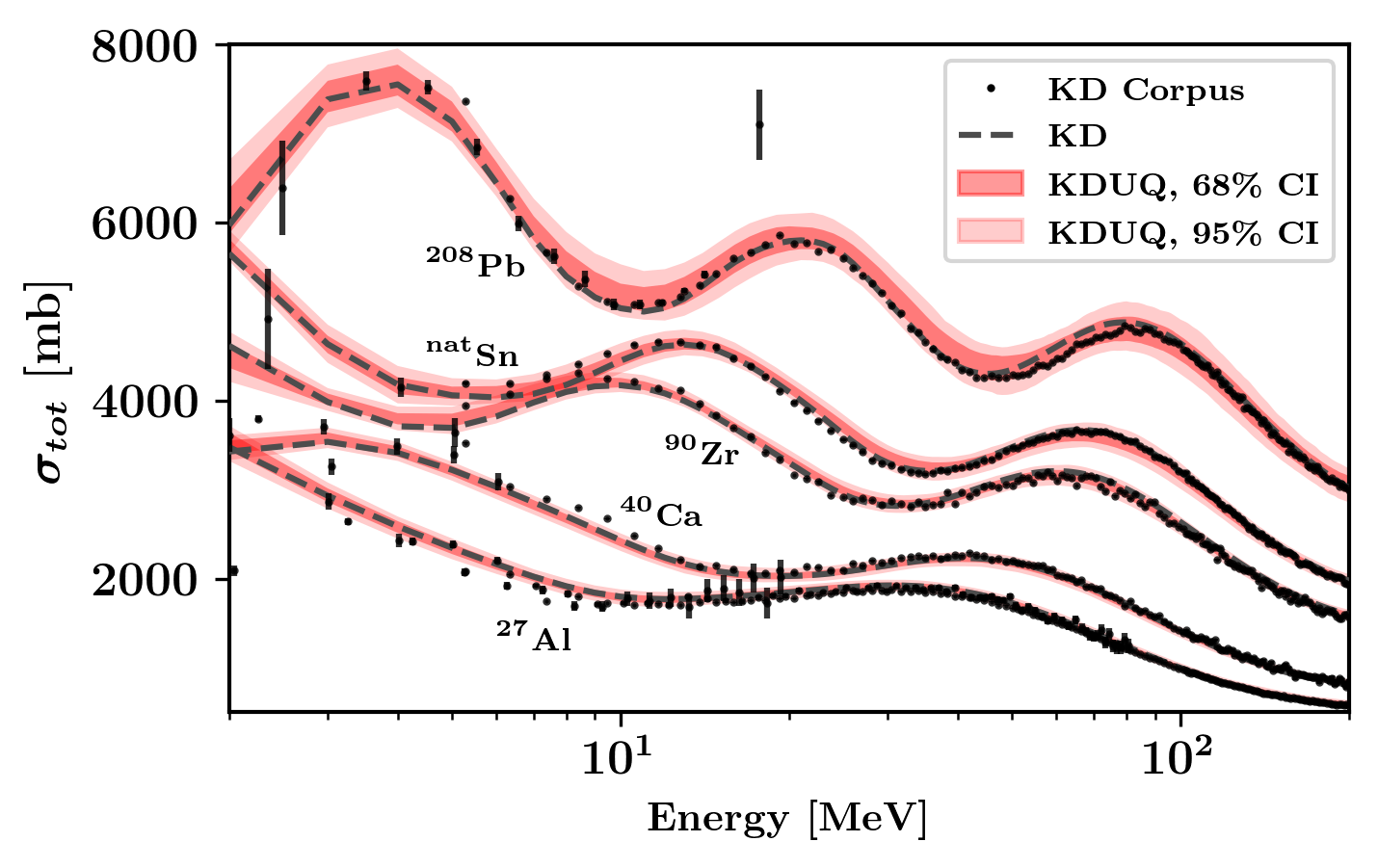}
    \caption{
        Representative experimental neutron total cross section data and KD and
        KDUQ calculations are plotted for selected nuclei in the KDUQ corpus. See
        caption of Fig. \ref{KDUQ_ECSDisplay} for additional information on
        the legend.
    }
    \label{KDUQ_TCSDisplay}
\end{figure*}

\begin{figure}[!tbp]
    \includegraphics[width=0.5\textwidth]{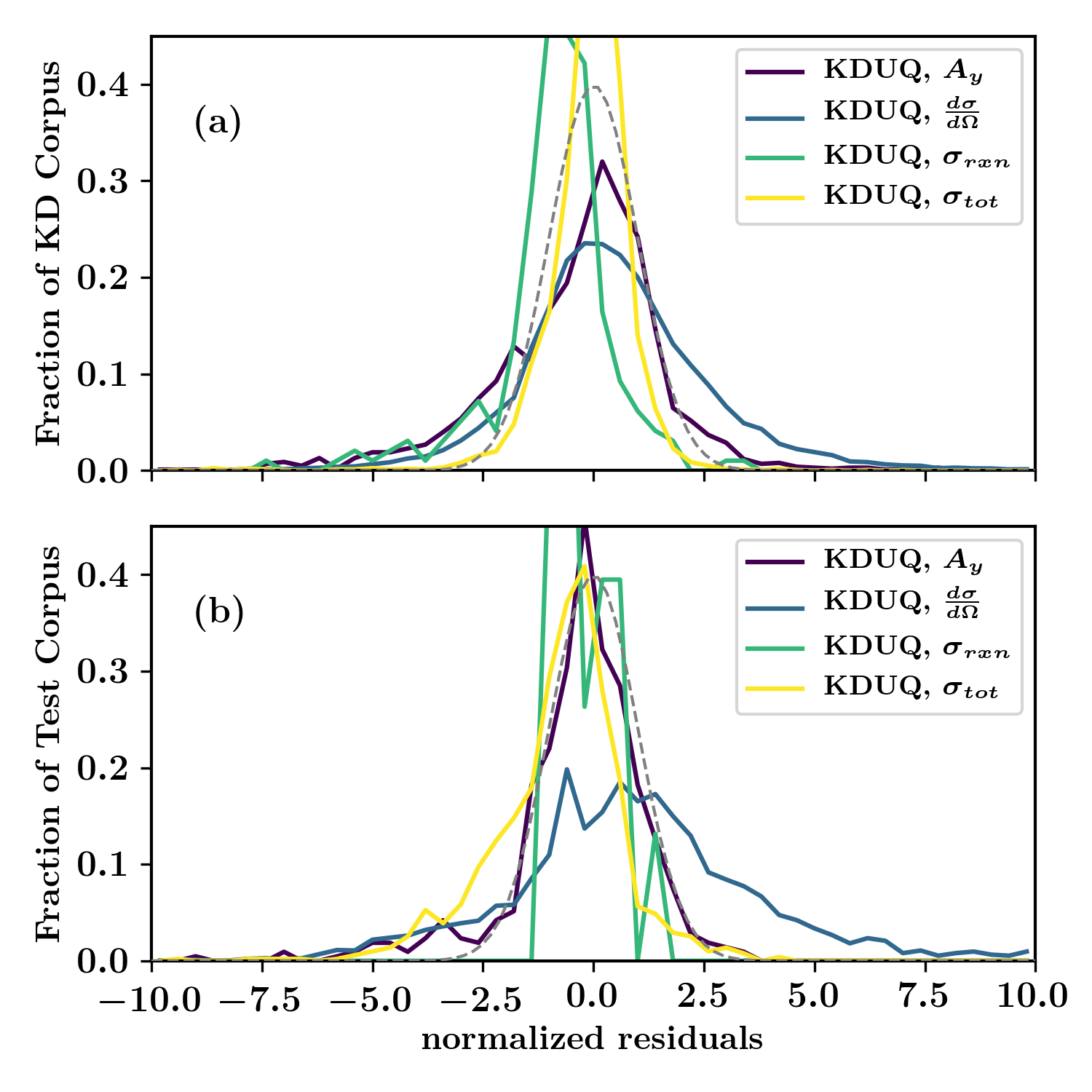}
    \caption{
           Normalized residuals ($r_{i}$/$\delta_{i}$) between KDUQ's predictions
           and the KDUQ corpus and Test corpus are histogrammed by data type.
           Panel (a) shows performance against the KDUQ corpus. Panel (b) shows 
           performance against the Test corpus.
    }
    \label{KDUQNormalizedResiduals}
\end{figure}

\begin{table}[!tbp]
    \centering
    \caption{
        Mean ($\mu_{1}$), standard deviation ($\mu_{2}$), and skewness
        ($\mu_{3}$) for the distributions of standardized residuals shown in
        Fig. \ref{KDUQNormalizedResiduals}, shown separately for protons and
        neutrons (cf. with Tables \ref{CHUQResidualTable} and
        \ref{OrigPotentialsResidualTable}).
    }
        \begin{tabular}{@{\extracolsep{4pt}}c c c c c c c@{}}
         & \multicolumn{6}{c}{Proton data}
         \vspace{0.6em}\\
         & \multicolumn{3}{c}{KDUQ Corpus} & \multicolumn{3}{c}{Test Corpus} \\
         & $\frac{d\sigma}{d\Omega}$ & $A_{y}$ & $\sigma_{rxn}$ & $\frac{d\sigma}{d\Omega}$ & $A_{y}$ & $\sigma_{rxn}$ \\
         \cline{2-4} \cline{5-7} 
         $\mu_{1}$ & -0.1 & 0.5 & -0.9 & 0.1 & -0.8 & -0.2 \\
         $\mu_{2}$ & 2.2 & 2.1 & 1.5 & 0.9 & 1.8 & 0.7 \\
         $\mu_{3}$ & -0.6 & 1.5 & -2.5 & 0.6 & 9.3 & 0.6 \\
    \end{tabular}
    \vspace{1.2em}\\
    \begin{tabular}{@{\extracolsep{4pt}}c c c c c c c@{}}
         & \multicolumn{6}{c}{Neutron data}
         \vspace{0.6em}\\
         & \multicolumn{3}{c}{KDUQ Corpus} & \multicolumn{3}{c}{Test Corpus} \\
         & $\frac{d\sigma}{d\Omega}$ & $A_{y}$ & $\sigma_{tot}$ & $\frac{d\sigma}{d\Omega}$ & $A_{y}$ & $\sigma_{tot}$ \\
         \cline{2-4} \cline{5-7} 
         $\mu_{1}$ & -1.1 & 0.5 & -0.1 & -1.5 & 1.2 & -0.8 \\
         $\mu_{2}$ & 2.1 & 2.3 & 1.2 & 2.0 & 3.3 & 1.5 \\
         $\mu_{3}$ & -0.6 & 0.3 & -5.3 & -0.8 & 1.1 & -0.7 \\
    \end{tabular}

    \label{KDUQResidualTable}
\end{table}

Figures \ref{KDUQ_ECSDisplay}, \ref{KDUQ_RCSDisplay}, and \ref{KDUQ_TCSDisplay}
show the performance of KDUQ and the canonical KD Global OMP with respect to
several representative experimental data sets in the KDUQ corpus used for
training. Figures comparing KDUQ and KD over the entire KDUQ corpus are
provided in the Supplemental Material \cite{SupplementalMaterial}. For elastic
scattering observables, the median predictions of KDUQ are very similar to the
canonical KD predictions at low angles, with moderate deviations appearing at
higher angles and scattering energies. Predicted neutron $\sigma_{tot}$ of KDUQ
and KD are nearly identical, and both achieve excellent agreement with the
training data above the resolved-resonance region. (At lower energies where
resonance structure is resolved, the OMP assumption of smooth,
resonance-averaged behavior is no longer expected to hold). The most
significant difference between KDUQ and KD is the improved reproduction of
proton $\sigma_{rxn}$ cross sections in KDUQ, where predictions are roughly 10\% smaller
for low-$A$ targets such as \alTwentySeven\ and \caForty\ compared to the
predictions of KD. In addition, at scattering energies $>100$ \MeV\
across all masses, the slope of predicted proton $\sigma_{rxn}$ cross sections
differs between KDUQ and KD, with KD predictions exhibiting a steeper decrease
with respect to energy, whereas KDUQ predictions remain roughly flat with
respect to energy. Past analyses with dispersive optical potentials have
connected the energy dependence of $\sigma_{rxn}$ cross sections in this region
with the behavior of deeply bound, highly correlated nucleons, as probed in
(e,e'p) reactions \cite{Atkinson2019}, and potentially correlated with neutron
skins in neutron-rich nuclei \cite{Atkinson2020, Pruitt2020PRL}. Such a
relationship could be quantitatively assessed with a global dispersive OMP
(\`{a} la \cite{Morillon2006}), but treated fully non-locally to maintain good
particle number and equipped with UQ as shown here.

Figure \ref{KDUQNormalizedResiduals} summarizes the performance of KDUQ against
both the KDUQ corpus training data and the Test corpus. The mean, standard
deviation, and skewness of the distribution of residuals shown in Fig.
\ref{KDUQNormalizedResiduals} are listed in Table \ref{KDUQResidualTable} for
both protons and neutrons. Overall, KDUQ performance differs little between the
KDUQ corpus and Test corpus, an indication that our MCMC-based approach has
avoided overfitting the training data. Compared to KD, KDUQ has a lower bias
with respect to proton reaction cross section data (mean normalized residual of
-0.9; cf. with -2.4 for KD in Table \ref{OrigPotentialsResidualTable}). Both KD
and KDUQ exhibit minimal bias for neutron total cross sections (mean normalized
residuals of -0.3 and -0.1, respectively). Apparently, our inclusion of
unaccounted-for uncertainty terms in KDUQ is sufficient to account for almost
all of the excess data variance seen for KD in Fig.
\ref{OrigPotentialNormalizedResiduals} (neutron $\sigma_{tot}$ normalized
residual standard deviations of 1.2 for KDUQ, compared to 25.2 for KD). For
differential elastic observables, the mean predictions from KDUQ perform
similarly to those of KD against both the KDUQ corpus and the Test corpus, with
the parametric uncertainty of KDUQ reducing the normalized residual standard
deviations to approximately 2 for both protons and neutrons. That the
normalized residual variances for differential elastic quantities are still
larger than one indicates additional variance among the experimental data that
the assumptions of our analysis are unable to account for.  One likely source
is assumption of sphericity leading to poorer agreement with differential data
on more-deformed targets in the KDUQ corpus. It is well-known that, especially
at low energies, only a deformation-cognizant, dispersive OMP such as those
introduced by Soukhovitskii et al. \cite{Soukhovitskii2016} and Capote et al.
\cite{Capote2008} will be capable of reproducing scattering behavior. Equipping
these deformed OMPs with UQ is a natural, if labor-intensive, extension. In the
meantime, by examining which data are flagged as outliers in our approach, one
could garner a quantitative idea of how where, and how badly, a spherical OMP
fails to capture the effects of deformation on scattering.

\subsection{Parameter comparison and discussion}
\label{sec_results_discussion}

\begin{table}[!htb]
    \centering
    \caption{
        The CH89 and CHUQ central parameter values and uncertainty intervals
        are listed. For CH89, the central values are the mean values reported
        in the original treatment, and the uncertainties are the estimated
        parameter standard deviations as calculated from a bootstrap analysis
        in the original treatment. For CHUQ, the central values are the
        posterior 50\textsuperscript{th} percentile value and the uncertainties
        are the difference between the central value and the posterior
        16\textsuperscript{th} and 84\textsuperscript{th} percentile values.
        The final row lists the determinant of the parameter covariance.
    }
    \begin{tabular}{c c c}
     & CH89 & CHUQ \\
     \hline   $V_{0}$ & $52.9_{-0.2}^{+0.2}$ & $56.19_{-1.82}^{+1.43}$ \\
   $V_{t}$ & $13.1_{-0.8}^{+0.8}$ & $13.82_{-5.25}^{+7.03}$ \\
   $V_{e}$ & $-0.299_{-0.004}^{+0.004}$ & $-0.36_{-0.02}^{+0.03}$ \\
   $r_{0}$ & $1.25_{-0.002}^{+0.002}$ & $1.20_{-0.03}^{+0.03}$ \\
   $r_{0}^{(0)}$ & $-0.225_{-0.009}^{+0.009}$ & $-0.20_{-0.13}^{+0.12}$ \\
   $a_{0}$ & $0.69_{-0.006}^{+0.006}$ & $0.73_{-0.02}^{+0.03}$ \\
   $r_{c}$ & $1.24_{-0}^{+0}$ & $1.25_{-0.12}^{+0.12}$ \\
   $r_{c}^{(0)}$ & $0.12_{-0}^{+0}$ & $0.13_{-0.12}^{+0.09}$ \\
   $V_{so}$ & $5.9_{-0.1}^{+0.1}$ & $5.58_{-0.58}^{+0.52}$ \\
   $r_{so}$ & $1.34_{-0.03}^{+0.03}$ & $1.29_{-0.11}^{+0.11}$ \\
   $r_{so}^{(0)}$ & $-1.2_{-0.1}^{+0.1}$ & $-1.12_{-0.51}^{+0.45}$ \\
   $a_{so}^{(0)}$ & $0.63_{-0.02}^{+0.02}$ & $0.61_{-0.04}^{+0.04}$ \\
   $W_{v0}$ & $7.8_{-0.3}^{+0.3}$ & $9.92_{-2.92}^{+4.63}$ \\
   $W_{ve0}$ & $35.0_{-1}^{+1}$ & $33.15_{-19.82}^{+25.03}$ \\
   $W_{vew}$ & $16.0_{-1}^{+1}$ & $24.00_{-9.52}^{+11.32}$ \\
   $W_{s0}$ & $10.0_{-0.2}^{+0.2}$ & $10.59_{-3.39}^{+3.99}$ \\
   $W_{st}$ & $18.0_{-1}^{+1}$ & $27.09_{-8.72}^{+12.28}$ \\
   $W_{se0}$ & $36.0_{-2}^{+2}$ & $20.00_{-20.82}^{+21.69}$ \\
   $W_{sew}$ & $37.0_{-2}^{+2}$ & $36.38_{-13.66}^{+23.75}$ \\
   $r_{ws}$ & $1.33_{-0.01}^{+0.01}$ & $1.32_{-0.08}^{+0.08}$ \\
   $r_{ws}^{(0)}$ & $-0.42_{-0.03}^{+0.03}$ & $-0.41_{-0.32}^{+0.36}$ \\
   $a_{ws}$ & $0.69_{-0.01}^{+0.01}$ & $0.69_{-0.05}^{+0.05}$ \\
    \hline    $|\bf{\Sigma}|$ & $5.76 \times 10^{-49}$ & $1.08 \times 10^{-12}$ \\
\end{tabular}

    \label{CH89vsCHUQParameters}
\end{table}

\begin{table}[!htb]
    \centering
    \caption{
        The KD and KDUQ parameter values are compared and the KDUQ uncertainties
        listed. For KDUQ, the listed values are the
        posterior 50\textsuperscript{th} percentile (median) value and the
        uncertainties are the difference between the median value and the
        posterior 16\textsuperscript{th} and 84\textsuperscript{th} percentile
        values.
    }
    \begin{tabular}{c c c}
     & KD & KDUQ \\
     \hline   $V_{1,0}$ & $5.93 \times 10^{1}$ & $5.86_{-0.18}^{+0.21} \times 10^{1}$ \\
   $V_{1,\alpha}$ & $2.10 \times 10^{1}$ & $1.34_{-0.47}^{+0.54} \times 10^{1}$ \\
   $V_{1,A}$ & $2.40 \times 10^{-2}$ & $2.61_{-0.99}^{+1.06} \times 10^{-2}$ \\
   $V_{2,0}^{n}$ & $7.23 \times 10^{-3}$ & $6.35_{-1.05}^{+0.71} \times 10^{-3}$ \\
   $V_{2,A}^{n}$ & $1.48 \times 10^{-6}$ & $1.82_{-4.74}^{+5.44} \times 10^{-6}$ \\
   $V_{3,0}^{n}$ & $1.99 \times 10^{-5}$ & $1.08_{-0.93}^{+0.88} \times 10^{-5}$ \\
   $V_{3,A}^{n}$ & $2.00 \times 10^{-8}$ & $1.45_{-2.77}^{+3.30} \times 10^{-8}$ \\
   $V_{2,0}^{p}$ & $7.07 \times 10^{-3}$ & $6.76_{-1.32}^{+1.12} \times 10^{-3}$ \\
   $V_{2,A}^{p}$ & $4.23 \times 10^{-6}$ & $2.91_{-8.20}^{+6.99} \times 10^{-6}$ \\
   $V_{3,0}^{p}$ & $1.73 \times 10^{-5}$ & $1.40_{-0.94}^{+1.00} \times 10^{-5}$ \\
   $V_{3,A}^{p}$ & $1.14 \times 10^{-8}$ & $1.43_{-4.47}^{+4.53} \times 10^{-8}$ \\
   $V_{4,0}$ & $7.00 \times 10^{-9}$ & $-4.30_{-20.30}^{+25.60} \times 10^{-9}$ \\
   $r_{0}$ & $1.30 \times 10^{0}$ & $1.27_{-0.04}^{+0.03} \times 10^{0}$ \\
   $r_{A}$ & $4.05 \times 10^{-1}$ & $3.61_{-1.34}^{+1.55} \times 10^{-1}$ \\
   $a_{0}$ & $6.78 \times 10^{-1}$ & $6.89_{-0.27}^{+0.24} \times 10^{-1}$ \\
   $a_{A}$ & $1.49 \times 10^{-4}$ & $-0.42_{-2.69}^{+2.56} \times 10^{-4}$ \\
   $r_{C,0}$ & $1.20 \times 10^{0}$ & $1.19_{-0.12}^{+0.11} \times 10^{0}$ \\
   $r_{C,A}$ & $6.97 \times 10^{-1}$ & $6.72_{-6.60}^{+7.36} \times 10^{-1}$ \\
   $r_{C,A2}$ & $1.30 \times 10^{1}$ & $1.30_{-1.26}^{+1.40} \times 10^{1}$ \\
   $V_{1,0}$ & $5.92 \times 10^{0}$ & $5.99_{-0.90}^{+0.96} \times 10^{0}$ \\
   $V_{1,A}$ & $3.00 \times 10^{-3}$ & $1.95_{-8.55}^{+9.63} \times 10^{-3}$ \\
   $V_{2,0}$ & $4.00 \times 10^{-3}$ & $4.75_{-2.17}^{+4.07} \times 10^{-3}$ \\
   $r_{0}$ & $1.19 \times 10^{0}$ & $1.21_{-0.06}^{+0.06} \times 10^{0}$ \\
   $r_{A}$ & $6.47 \times 10^{-1}$ & $7.35_{-2.58}^{+2.58} \times 10^{-1}$ \\
   $a_{0}$ & $5.90 \times 10^{-1}$ & $6.00_{-0.39}^{+0.39} \times 10^{-1}$ \\
   $W_{1,0}$ & $-3.10 \times 10^{0}$ & $-3.79_{-2.10}^{+2.08} \times 10^{0}$ \\
   $W_{2,0}$ & $1.60 \times 10^{2}$ & $2.19_{-0.89}^{+0.84} \times 10^{2}$ \\
   $W_{1,0}^{n}$ & $1.22 \times 10^{1}$ & $2.09_{-0.42}^{+0.39} \times 10^{1}$ \\
   $W_{1,A}^{n}$ & $1.67 \times 10^{-2}$ & $0.61_{-2.94}^{+3.35} \times 10^{-2}$ \\
   $W_{1,0}^{p}$ & $1.47 \times 10^{1}$ & $1.86_{-0.49}^{+0.56} \times 10^{1}$ \\
   $W_{1,A}^{p}$ & $9.63 \times 10^{-3}$ & $32.50_{-36.72}^{+45.92} \times 10^{-3}$ \\
   $W_{2,0}$ & $7.35 \times 10^{1}$ & $10.29_{-2.58}^{+3.45} \times 10^{1}$ \\
   $W_{2,A}$ & $7.95 \times 10^{-2}$ & $2.43_{-16.23}^{+19.45} \times 10^{-2}$ \\
   $D_{1,0}$ & $1.60 \times 10^{1}$ & $1.67_{-0.39}^{+0.72} \times 10^{1}$ \\
   $D_{1,\alpha}$ & $1.60 \times 10^{1}$ & $1.11_{-0.79}^{+1.01} \times 10^{1}$ \\
   $D_{2,0}$ & $1.80 \times 10^{-2}$ & $2.34_{-3.29}^{+2.56} \times 10^{-2}$ \\
   $D_{2,A}$ & $3.80 \times 10^{-3}$ & $3.73_{-26.67}^{+30.69} \times 10^{-3}$ \\
   $D_{2,A2}$ & $8.00 \times 10^{0}$ & $8.57_{-7.36}^{+7.31} \times 10^{0}$ \\
   $D_{2,A3}$ & $1.56 \times 10^{2}$ & $2.51_{-2.48}^{+1.21} \times 10^{2}$ \\
   $D_{3,0}$ & $1.15 \times 10^{1}$ & $1.38_{-0.31}^{+0.39} \times 10^{1}$ \\
   $r_{0}$ & $1.34 \times 10^{0}$ & $1.35_{-0.08}^{+0.07} \times 10^{0}$ \\
   $r_{A}$ & $1.58 \times 10^{-2}$ & $1.75_{-1.63}^{+1.72} \times 10^{-2}$ \\
   $a_{0}^{n}$ & $5.45 \times 10^{-1}$ & $5.43_{-0.38}^{+0.41} \times 10^{-1}$ \\
   $a_{A}^{n}$ & $1.66 \times 10^{-4}$ & $-2.14_{-4.51}^{+4.06} \times 10^{-4}$ \\
   $a_{0}^{p}$ & $5.19 \times 10^{-1}$ & $5.08_{-0.42}^{+0.42} \times 10^{-1}$ \\
   $a_{A}^{p}$ & $5.21 \times 10^{-4}$ & $14.10_{-6.57}^{+6.55} \times 10^{-4}$ \\
\end{tabular}

    \label{KDvsKDUQParameters}
\end{table}

In this section, we interpret the mean parameter values and uncertainties of
our new UQ OMPs. Besides providing a natural way to forward-propagate OMP
uncertainty via resampling, the parameter (co)variances provide information
about the extrapolability of CH89- and KD-like OMPs away from their training
data (e.g., away from $\beta$-stability). The optimized parameter estimates and
associated uncertainties are compared in Table \ref{CH89vsCHUQParameters} for
CH89 and CHUQ and in Table \ref{KDvsKDUQParameters} for KD and KDUQ. In
addition, for a metric for the overall degree of parametric uncertainty in CH89
and CHUQ, we list the determinants of the covariance matrices for CH89 UQ and
CHUQ (excluding the Coulomb radius parameters, which were fixed in the original
CH89 treatment) at the bottom of Table \ref{CH89vsCHUQParameters}.

\begin{figure}[!htb]
    \includegraphics[width=0.5\textwidth]{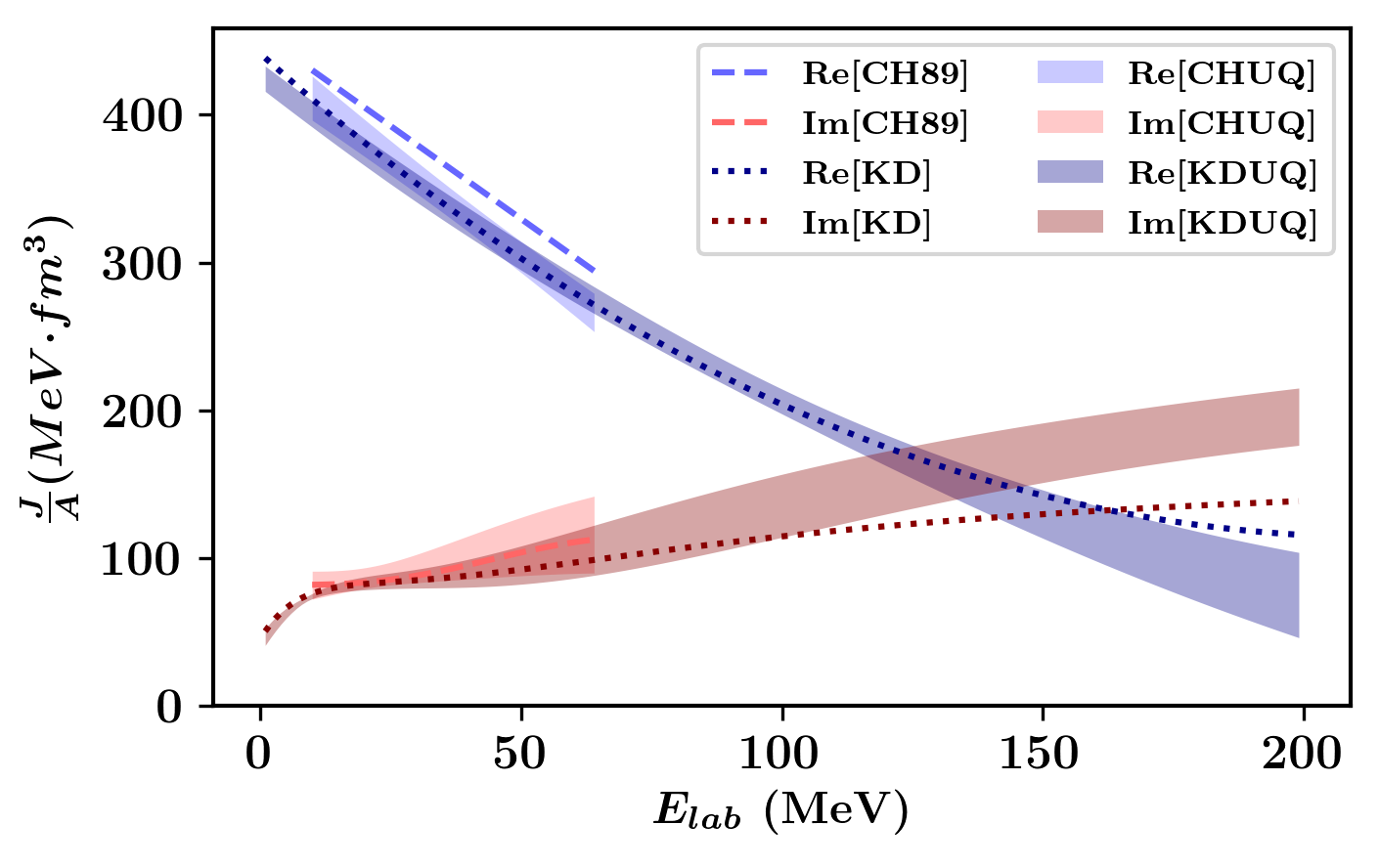}
    \caption{
        Volume integrals $J/A$ are plotted for CH89, KD, CHUQ, and KDUQ
        evaluated for neutron scattering on \zrNinety\ (all spin-orbit terms
        are excluded). The CHUQ and KDUQ bands show the 68\% uncertainty
        interval. The ranges of CH89 and CHUQ are restricted to the nominal
        validity range of CH89 of 10-65 MeV.
    }
    \label{VolumeIntegrals}
\end{figure}

Overall, the estimated central parameter values CHUQ are similar to the
original values of CH89, but in most cases, the median value from CHUQ lies
well outside the estimated uncertainty of CH89 UQ. In addition, CHUQ's
parametric uncertainty estimates are between two and twenty times larger than
the estimates from CH89 UQ. Most notable are changes in terms affecting the
potential magnitudes, including the asymmetry-dependent parameters $V_{t}$ and
$W_{st}$ and the imaginary central and surface terms' $A$-dependent parameters
$W_{ve0}$, $W_{vew}$, $W_{se0}$, and $W_{sew}$, all of which indicate far
greater uncertainty with respect to target asymmetry and $A$ than in the
canonical treatment. These increased uncertainties manifest as uncertainty in
the imaginary-part volume integrals as shown in Fig.  \ref{VolumeIntegrals}.

\begin{figure}[!htb]
    \includegraphics[width=0.5\textwidth]{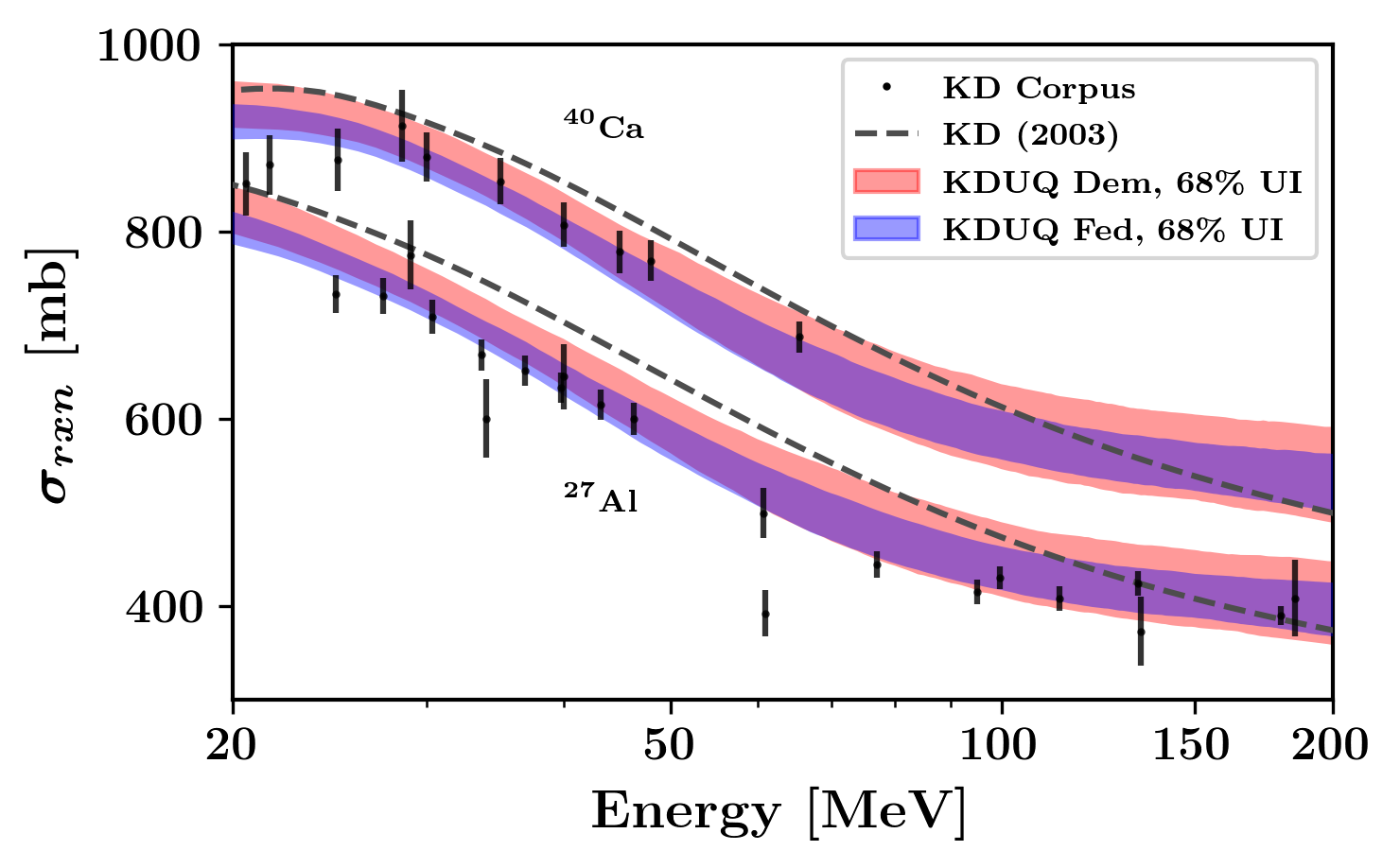}
    \caption{
        A comparison of KDUQ versions trained using Eq. \ref{DemocraticCovarianceAnsatz}
        and Eq. \ref{FederalCovarianceAnsatz} is shown for proton $\sigma_{rxn}$
        on \caForty. The 68\% and 95\% credible intervals of appear as
        the dark and light bands, respectively. The experimental data of
        \cite{Shane2010}, \cite{Pruitt2020PRC}, and \cite{Dietrich2003} appear
        as black points with associated errors.
    }
    \label{DemocraticFederalComparison}
\end{figure}

The much-larger uncertainty recovered in CHUQ vs CH89 UQ is indicative
of a better match of CHUQ to the breadth of the CHUQ corpus
compared to the canonical CH89. However, some important details of the CHUQ
corpus and Test corpus are still not captured by CHUQ, due to the relative
simplicity of the CH89 potential form. The choice of likelihood in the
canonical analysis (Eq. \ref{CH89Likelihood}) made for good performance of the
canonical CH89 parameters with respect to the experimental data, but the lack
of normalization in their likelihood function resulted in an underestimation of
parametric uncertainty. CHUQ performs moderately well against the Test corpus,
considering that the majority of Test corpus data lie outside the nominal
validity range of the CH89 potential form, but it is clear that other OMPs should
be preferred at energies below 10 \MeV.

We now turn to KD and KDUQ. For forty-two out of forty-six parameters, the the
canonical value of KD lies within one estimated standard deviation of the KDUQ
mean value; of the remaining four, three ($V_{1,\alpha}$, $V_{2,0}^{n}$, and
$V_{3,0}^{n}$) are within two estimated standard deviations, and the most
discrepant, $W_{1,0}^{n}$, lies just over two estimated standard deviations
away. Notably, many sub-term parameters which are coefficients in $E$- and
$A$-dependent polynomial expansions are strongly anti-correlated (see KDUQ
parameter correlogram in the Supplemental Material
\cite{SupplementalMaterial}), and their estimated uncertainties are many times
larger than their median values. Both these observations indicate
overparameterization of $E$- and $A$-dependence in those subterms, so some of
these higher-order expansion terms could likely be eliminated without impacting
observables. Taken as a whole, the parameter estimates we recover are highly
consistent with the canonical ones, which we take as evidence that our
replication attempt, though not identical to the canonical treatment, was
successful. Further, it confirms that even without the benefit of the
computational advances of the last twenty years, the canonical KD analysis was
remarkably close to global minimum we recover here.

In the KD/KDUQ functional form, Lane-like asymmetry-dependence appears only in
two terms: the first-order energy dependence of the depth of the real volume
potential as a function of asymmetry, $V_{1,\alpha}$, and the first-order
energy-dependence of the depth of the imaginary surface potential as a function
of asymmetry, $D_{1,\alpha}$. For each of these parameters, KDUQ recovers
significantly smaller median asymmetry-dependences than those from the
canonical treatment. This implies that KD's real and imaginary surface
asymmetry-dependences are weaker than previously assumed, and that the real and
imaginary-surface parts of the OMP may be more reliable than previously thought
when extrapolated to exotic (near-spherical) targets. At the same time, for
$D_{1,\alpha}$, the uncertainties we estimate are almost as large as the median
value we recover, indicating that the training data we used (coupled with our
analysis assumptions) provides only a weak constraint on the behavior of the
imaginary surface term away from the valley of $\beta$-stability. Considering
that many downstream applications, such as $r$-process nucleosythesis
calculations, fission neutron spectra modeling, and planned transfer and
knockout studies at NSCL and FRIB, rely on OMP-informed evaluations of
low-energy inelastic cross sections on neutron-rich targets, the fact that
$D_{1,\alpha}$ is poorly constrained is a pressing problem. A global,
UQ-equipped phenomenological OMP analysis that incorporates isovector-sensitive
observables, such as quasi-elastic charge exchange cross sections that have
already yielded insight into OMP isovector dependence (e.g.,
\cite{Danielewicz2017}), is a natural next step.

Besides these terms with explicit asymmetry-dependence, the imaginary volume
term, which is separately parameterized for protons and neutrons, contains
information about isovector dependence of imaginary strength. For both neutrons
and protons, our median-value estimates for first-order imaginary volume
strength ($W_{1,0}^{n}$ and $W_{1,0}^{p}$ terms) are moderately larger than the
canonical KD value, suggesting enhanced imaginary volume strength overall.
Coupled with the smaller overall imaginary surface depth $D_{1,0}$, these
result in a reduction in predicted neutron/proton reaction cross sections at
low energies (associated with the surface) and an increase at higher energies
associated with the volume --- in improved agreement with experimental trends
for protons shown in Fig. \ref{KDUQ_RCSDisplay}. This trend is also visible for
neutrons in Fig. \ref{VolumeIntegrals}, where above 100 \MeV, the imaginary
volume integral grows more rapidly for KDUQ than for KD. If verified, this
additional imaginary volume strength would further quench bound-state
spectroscopic factors available from dispersive optical models, as discussed in
\cite{Atkinson2019, Pruitt2020PRL}.

Lastly, to assess the effect of our data covariance matrix ansatz on these
interpretations, we compared two different versions of KDUQ:
one trained using the ``democratic'' covariance ansatz of Eq.
\ref{DemocraticCovarianceAnsatz} and one trained using the ``federal''
covariance ansatz of Eq. \ref{FederalCovarianceAnsatz}. Figure
\ref{DemocraticFederalComparison} shows results from both treatments on
predictions of proton $\sigma_{rxn}$ above 50 \MeV, where the differences are
largest. Overall, these different ansatzes have little effect on the mean
parameter values. However, in the case where one training data type has far
fewer data than others, the federal ansatz leads to a moderate reduction of
unaccounted-for uncertainty required to reproduce data of that type, which
leads to more precise predictions for data of that type. This agrees with our
expectation that the more realistic the experimental data covariance matrix
ansatz is, the less unaccounted-for uncertainty is required to achieve good
reproduction of the data.

\section{Impact} \label{sec_impact}

In this section, we apply our UQ-equipped OMPs to two case studies: predicting neutron $\sigma_{tot}$
evolution with respect to asymmetry, and propagation of OMP UQ into
Hauser-Feshbach calculations of of (n,$\gamma$) on \moNinetyFive\ and
(p,$\gamma$) on \srEightySeven.

\subsection{Case study 1: evolution of neutron total cross sections in isotopic pairs}

\begin{figure*}[hbt]
    \includegraphics[width=\textwidth]{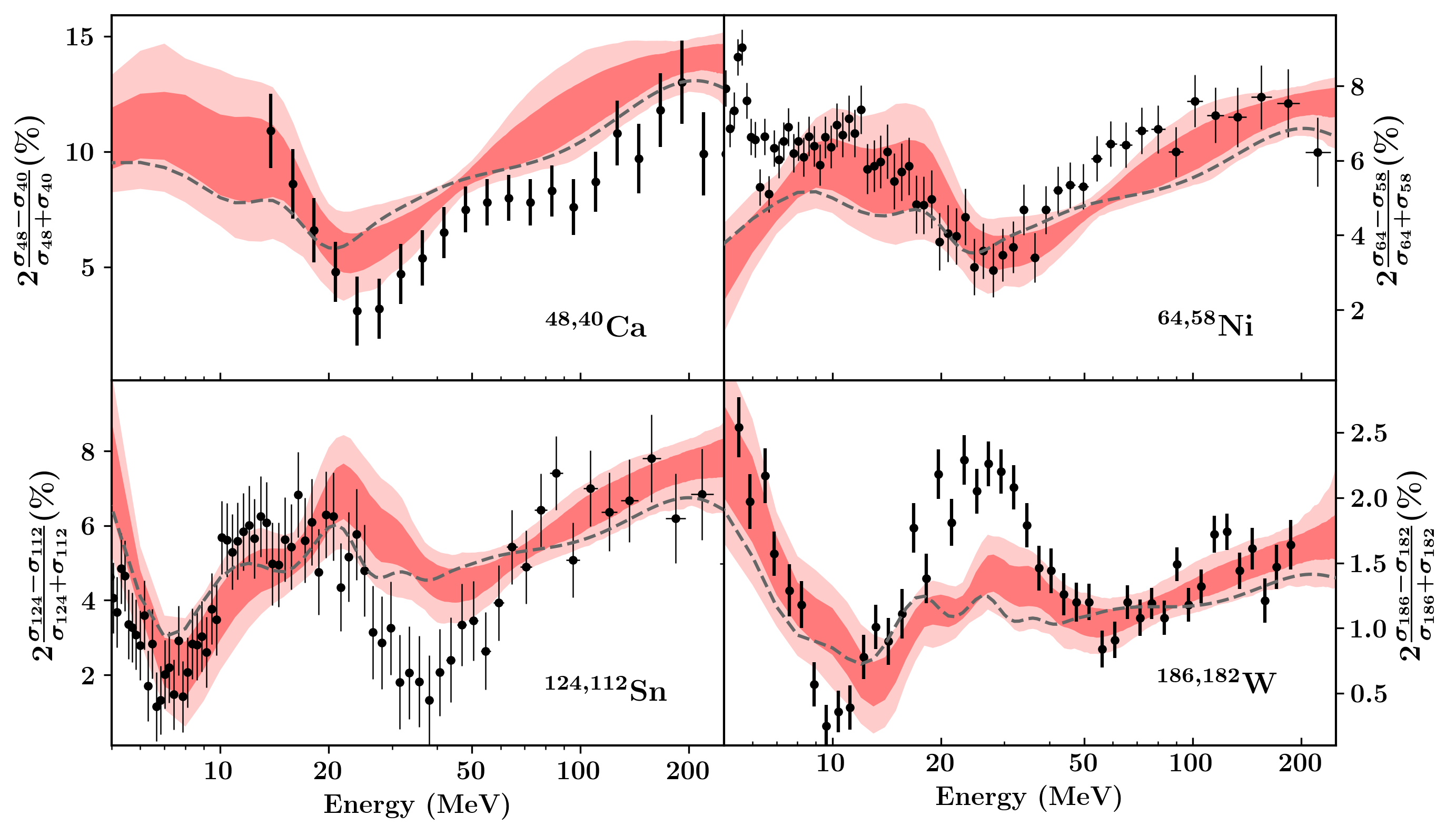}
    \caption{
        Relative differences of neutron total cross sections from 5 to 250
        \MeV\ for several isotopic pairs are plotted. Calculations using the
        standard Koning-Delaroche global potential are shown via the gray
        dashed line.  The 68\% and 95\% credible intervals of KDUQ appear as
        the dark and light red bands, respectively. The experimental data of
        \cite{Shane2010}, \cite{Pruitt2020PRC}, and \cite{Dietrich2003} appear
        as black points with associated errors.
    }
    \label{caseStudy1}
\end{figure*}

Cross sections for neutron-induced reactions on $\beta$-unstable targets are a
key input for several nuclear data applications, e.g., $r$-process
nucleosynthesis network calculations \cite{Surman2014}. Because of the
experimental difficulty in performing cross section measurements in this
regime, cross sections estimations rely on either (semi-)microscopic OMPs
\cite{Jeukenne1977, Bauge2001} or phenomenological potentials fitted solely to
stable-target data, such as the KD global OMP, that are then extrapolated
according to their assumed asymmetry-dependence. For incident neutrons at lower
energies ($<10$ \MeV), the asymmetry-dependence of the imaginary surface term
strongly affects capture cross sections \cite{Goriely2007}, but the magnitude
of this asymmetry-dependence remains poorly known. More broadly, such isovector
components of optical potentials are connected to other poorly constrained but
important nuclear quantities, such as neutron skins in finite nuclei
\cite{Dietrich2003, Atkinson2019, Pruitt2020PRL} and the density-dependence of
the symmetry energy, $L$, in nuclear matter, which influences both the
theoretical limit to neutron star radii and the dynamics of neutron star
mergers, among other properties \cite{Thiel2019}. As such, improving our
knowledge of the appropriate asymmetry-dependence of OMPs remains an important
task.

To constrain asymmetry-dependent terms of phenomenological OMPs, past analyses
have focused mainly on two types of experimental data: quasielastic charge
exchange cross sections to the isobaric analog state (as analyzed in
\cite{Danielewicz2017} using a KD-like potential), and ratios of neutron cross
sections measured on different isotopes along an isotopic or isotonic chain (as
studied by \cite{Camarda1984, Shamu1980, Dietrich2003, Shane2010,
Pruitt2020PRC}). Quasielastic charge exchange is an ideal probe in that
measured cross sections are sensitive specifically to isovector strength, but
analysis of these data may require more sophisticated theoretical machinery (as
compared to straightforward elastic and total cross section calculations) to
correct for contamination from $\Delta J^{\pi} \neq 0^{+}$ channels, as
demonstrated in \cite{Jon1997}. The second type, ratios of neutron cross
sections, has the advantage that by taking a cross section ratio, many
systematic uncertainties (such as detector efficiency) are divided out.
Further, if more than two isotopic targets are available, multiple ratios can
be constructed and additional quantities, such as degree of deformation, can be
extracted \cite{Camarda1984, Shamu1980}. In addition, because neutron total
cross sections can be simultaneously collected from a few to a few hundred \MeV
\cite{Finlay1993, Abfalterer2001} at precisions of $\approx1\%$, ratios of
neutron total cross sections can provide information about OMP isovector
features across broad regime of energies relevant for OMP construction and
application. The main drawback of this type of measurement is the
often-prohibitive expense of obtaining large, isotopically pure targets with
precisely known areal densities.  Even when isotopically pure targets are
available along an isobar or isotopic chain, because they must be stable or at
least long-lived to be suitable for target fabrication, they can span only a
small range of asymmetries, which diminishes the isovector signal in the cross
section ratio. 

The importance of constraining isovector terms warrants a future, global OMP
analysis including both of these data types as well as neutron strength
functions (as used by \cite{Bauge2001}) to characterize isovector dependence.
As a precursor to such an analysis, in Fig. \ref{caseStudy1} we consider
canonical KD OMP and KDUQ predictions against neutron total cross section
isotopic ratios on $^{40,48}$Ca \cite{Shane2010}, $^{58,64}$Ni, $^{112,124}$Sn,
\cite{Pruitt2020PRC}, $^{182,186}$W \cite{Dietrich2003}. In all instances, the
median predicted value from KDUQ closely follows the canonical KD predictions.
Due to the small reported uncertainties of the experimental data shown in Fig.
\ref{caseStudy1}, the canonical KD predictions are discrepant with the
experimental data at the several-$\sigma$ level in many places, e.g., the
\niSixtyFour-\niFiftyEight\ relative difference below 20 \MeV\ in panel (b). As
such, if one considers just the canonical KD curve, one might conclude that the
KD potential form is missing some important asymmetry-dependent physics that
are present in the data. However, when OMP parametric uncertainties are
considered (as shown in the KDUQ curve), it is clear that most of the
discrepancies between the canonical KD predictions and experimental data are
not statistically meaningful. That is, once parameter uncertainties are
included, the KD potential form is quite effective at predicting these cross
section ratios to which it was never trained. Moreover, any discrepancies that
remain after parametric uncertainty is considered (for example, the
overprediction of the Sn isotopic ratios, shown in panel (c), between 30 and 50
\MeV), become even more interesting: they \textit{do} indicate residual physics
that has been captured by the measurement, but not by the assumptions of our
OMP. In the specific case of Sn and W isotopic ratios, the likely cause for the
significant discrepancy between predictions and measurements is that the KD
form, by definition, neglects the differing density profiles for neutron and
protons in the Sn and W isotopes.  Indeed, Dietrich et al. \cite{Dietrich2003}
who collected the W data found that accurately reproducing the W isotopic ratio
data between 20 and 40 \MeV\ required a Jeukenne-Lejeune-Mahaux-inspired
coupled-channel OMP analysis that featured an increasing neutron skin from
\wOneHundredEightyTwo\ to \wOneHundredEightySix.  If more isotopically resolved
neutron total cross section ratios were available, a similar analysis across
many isotopic chains could provide neutron skin thicknesses and additional
information on $L$, though the potential would need to be deformation-aware and
not spherical, as assumed here. The apparent (but, in light of the parametric
uncertainty, insignificant) discrepancy between the canonical KD calculation
and the Ni isotopic ratio data is an example of how well-calibrated UQ helps
avoid mistaking noise in the experimental data for signal. At the very least,
the KDUQ predictions make clear that in order for neutron total cross section
ratios to constrain asymmetry-dependent OMP terms for a KD-like potential, the
relative difference measurement must achieve 1\% precision or better.

\subsection{Case study 2: \moNinetyFive(n,$\gamma$)\moNinetySix\ and \srEightySeven(p,$\gamma$)\yEightyEight\ cross sections}

\begin{figure*}[htb]
    \includegraphics[width=\textwidth]{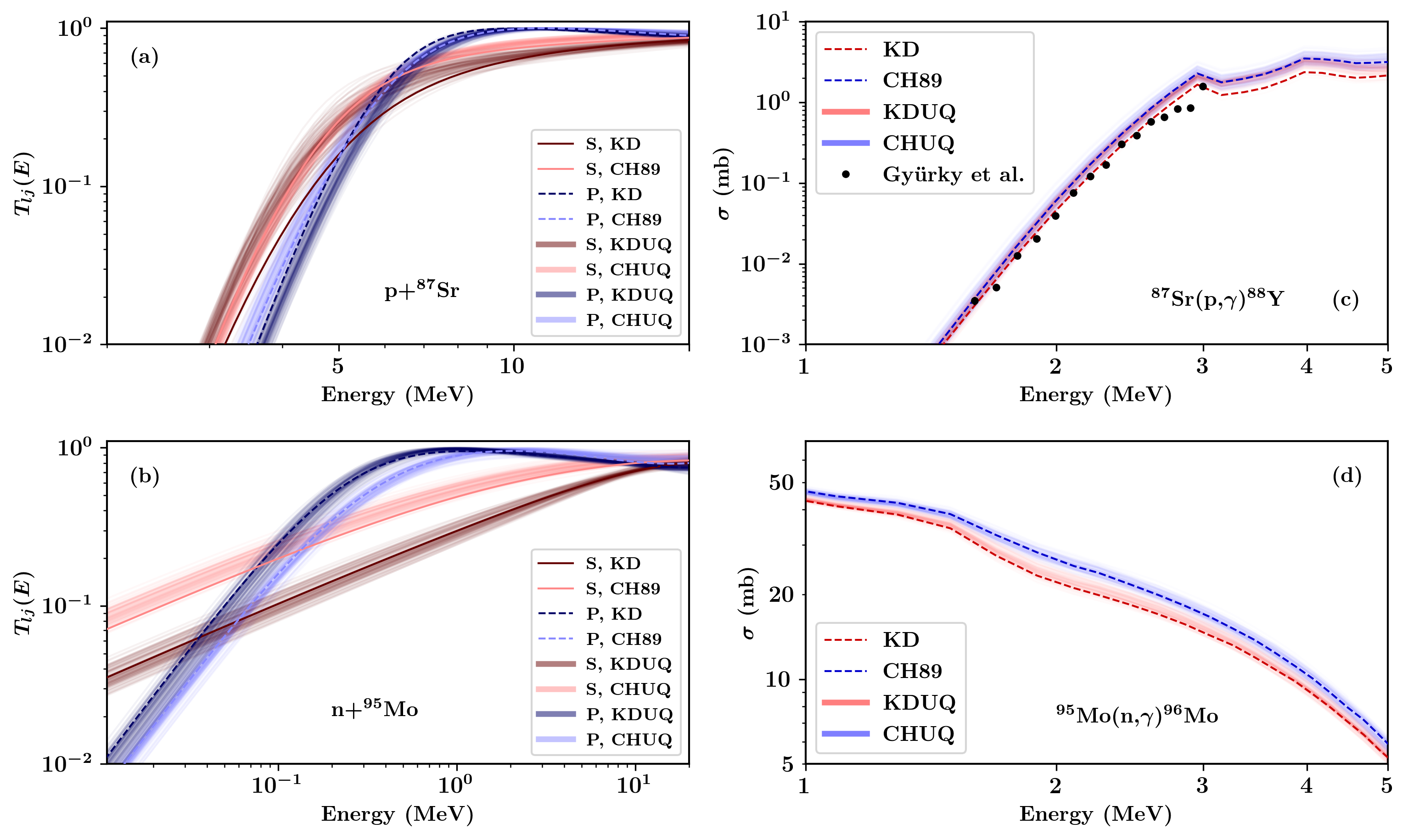}
    \caption{
        Transmission coefficients and cross sections for
        \srEightySeven(p,$\gamma$)\yEightyEight\ and
        \moNinetyFive(n,$\gamma$)\moNinetySix\ are plotted. All calculations
        were performed using the statistical reaction code YAHFC \cite{YAHFC}
        with default structure inputs. Calculations using the canonical
        Koning-Delaroche OMP are shown via blue lines; calculations using
        CH89 are shown as red lines. Calculations using 100 samples
        each from the KDUQ and CHUQ posterior distributions are shown as
        diffuse blue and red bands, respectively. In panels (a) and (b), both
        $S$-wave (L=0, J=$\frac{1}{2}$) and $P$-wave (L=1, J=$\frac{3}{2}$)
        transmission coefficient curves are shown. Panel (c) includes
        experimental data from Gy\"urky et al. \cite{Gyurky2001} (scaled
        upward by a factor of 2.5 to agree with the
        \srEightyEight(p,$\gamma$)\yEightyNine\ data of \cite{Galanopoulos2003},
        as indicated by Vagena et al. \cite{Vagena2021}).
    }
    \label{caseStudy2Tiled}
\end{figure*}

One of the most common applications for OMPs is as an input for radiative
capture calculations. While direct and pre-equilibrium capture
mechanisms play an important role for light and near-dripline nuclei
\cite{Xu2014}, the Hauser-Feshbach model, which assumes equilibration of the
excited composite nucleus before de-excitation, is appropriate for
most nucleon capture reactions. In this picture, both the probability of
creating a compound nucleus in the entrance channel and the evaporation of
nucleons from an excited nucleus depend on energy- and
angular-momentum-dependent transmission coefficients $T_{lj}(E)$ that are
determined by an OMP. To illustrate the relative impact of OMP
uncertainty on nucleon capture within the Hauser-Feshbach model, we propagated
CHUQ and KDUQ uncertainties through two representative reactions:
\moNinetyFive(n,$\gamma$)\moNinetySix\ and
\srEightySeven(p,$\gamma$)\yEightyEight\ at incident nucleon energies up to 5
\MeV. Calculations were carried out using the LLNL Hauser-Feshbach code YAHFC
\cite{YAHFC}, modified to accept KD-like and CH89-like potentials with
arbitrary parameters, and using YAHFC default configuration information,
discrete level data, nuclear level densities (LDs), and $\gamma$-ray strength functions
($\gamma$SFs). For each reaction, we ran YAHFC once using the canonical KD and
once using the canonical CH89 potential and then performed one hundred YAHFC
runs each for CHUQ and KDUQ, with each run using a unique sample of the OMP
parameter posterior. Results of these calculations are shown in Fig.
\ref{caseStudy2Tiled}. Panels (a) and (b) show transmission coefficients
$T_{lj}(E)$ generated by YAHFC's invocation of \textsc{frescox} \cite{FRESCOX}
for protons incident on \srEightySeven\ and for neutrons incident on
\moNinetyFive. Panels (c) and (d) display the corresponding capture cross
sections, where the uncertainty shown is due to the transmission coefficients
of panels (a) and (b). As YAHFC uses a Monte Carlo approach for de-exciting
compound nuclei, we drew $10^{6}$ samples at each scattering energy to ensure
that YAHFC's statistical uncertainty due to Monte Carlo sampling was less than
1\% for the calculated capture cross sections.

For p+\srEightySeven, the CH89, CHUQ, and KDUQ transmission coefficients show
overall consistency across all depicted energies, whereas the KD transmission
coefficients are slightly lower than the other OMPs between 3
and 10 \MeV. The principle difference for KD was reduced $s$-wave strength and
a more rapid rise in $p$-wave strength. Below 3 \MeV\, the Coulomb barrier
manifests as a steep reduction across the board. Above 10 \MeV\ (the minimum
energy included in the CHUQ corpus), the $T_{lj}(E)$ generated from all four
OMPs are consistent within approximately 10\%, an indication that the OMP
uncertainty is likely a minor source of uncertainty in cross section
predictions above this energy.

This reaction was one of those considered by Vagena et al. in their recent
study \cite{Vagena2021} of systematic effects of the proton OMP on $p$-process
nucleosynthesis. In their approach, using \textsc{talys} they sought to improve
the Bruy\'ere-le-Ch\^atel version of the Jeukenne-Lejeune-Mahaux
semi-microscopic proton OMP \cite{Bauge2001} by tuning its parameters to better
reproduce experimental cross sections for specific reactions. In the case of
\srEightySeven(p,$\gamma$)\yEightyEight, experimental data were available from
1.6 to 3 \MeV\ as collected by Gy\"urky et al., shown here in panel (c) of Fig.
\ref{caseStudy2Tiled}. Following Vagena et al., we have scaled the data up by a
factor of 2.5 from the original publication to comport with
\srEightyEight(p,$\gamma$)\yEightyNine\ data subsequently published by
\cite{Galanopoulos2003}. In their analysis, they argued that below roughly 3
\MeV, this reaction can be considered independent of the \srEightyEight\ LD and
$\gamma$SF, so any remaining discrepancy between predictions and measured data
serves as a basis for adjusting OMP parameters. In our case, while we did not
perform calculations using any microscopic OMPs, all four global
phenomenological OMPs we \textit{did} consider -- CH89, CHUQ, KD, and KDUQ --
generate predictions within a few tens of percent of the experimental cross
sections. This suggests that unless both the LD and $\gamma$SF are known within
a few tens of percent precision for a given reaction, constraining OMP
parameters by working backwards from measured capture cross sections may not be
feasible. A consistent joint treatment combining all of these sources of
uncertainty is a next step in which the yet-unknown correlations between OMPs,
LDs, and $\gamma$SFs will be critically important. We hope to engage in a
systematic study following the logic of \cite{Vagena2021} that compares
microscopic OMPs with UQ-equipped phenomenological ones for astrophysically
relevant reactions. At the very least, we argue that the intuition provided
here on standard phenomenological OMPs can guide analysts interested in
manually tuning microscopic OMP parameters to reproduce experimental scattering
observables. Given our finding that the CH89 and KD uncertainty effect on
capture cross sections between 1-5 \MeV\ that we examined is on the order of
tens of percent, a practitioner who encounters a larger discrepancy between
their prediction and experimental data should consider other sources of
uncertainty beyond the OMP parameters, such as deformation or level density
uncertainty.

Finally, we consider n+\moNinetyFive\ in panels (b) and (d). Throughout the depicted energy
range, CHUQ calculations are highly consistent with CH89 and KDUQ calculations
with KD, but both KD-type OMPs have a much slower rise in $s$-wave strength
with respect to energy than do the CH89-type OMPs. At energies above 100 \keV\,
the slower $s$-wave rise is offset by a correspondingly faster rise in $p$-wave
strength such that resulting neutron cross section predictions, which include
contributions over all incident partial waves, differ by only 20-30\%, highly
consistent with the degree of uncertainty seen for p+\srEightySeven. Importantly,
for any reactions at energies below 100 \keV\ involving primarily the
$s$-wave transmission coefficients, CH89 and CHUQ are expected to yield a cross section
two to three times higher than KD and KDUQ. In such case, the OMP
uncertainty should indeed dominate the cross section, as uncertainty in the LDs
and $\gamma$SFs have minimum impact at lower energies (again shown in Fig. 1
and 2 of Vagena et al. \cite{Vagena2021}). Such OMP-driven uncertainty could
impact both weak and strong $r$-process network calculations. Comparison of the
canonical KD OMP's $s$- and $p$-wave strength functions against experimental
data, as shown in Fig. 47 of Koning and Delaroche's original analysis, suggest
that at energies below 100 \keV, KD-type OMPs may have a more realistic energy
dependence than the CH89-type OMPs. A detailed study of OMPs uncertainty at
nucleosynthetic ``bottlenecks'' seems a worthy follow-up.

\section{Conclusions} \label{sec_conclusion}

Phenomenological OMPs continue to play an important role in nuclear reaction
calculations but lack well-calibrated UQ. Without reliable uncertainty
estimates, it is difficult to assess the relative importance of OMPs on the
overall uncertainty budget of applications dependent on reaction data. To
address this issue, we identified two main obstacles -- systematic
underestimation of experimental (co)variance and a lack of outlier rejection --
and developed a generic pipeline for performing UQ on phenomenological OMPs. We
then applied it to the widely-used CH89 and KD global OMPs, yielding two new
potential ensembles, CHUQ and KDUQ, with full covariance information between
potential parameters. CHUQ and KDUQ perform favorably against their training
corpora, with KDUQ showing superior performance on the Test corpus, especially
for proton $\sigma_{rxn}$ and neutron $\sigma_{tot}$. Accordingly, we recommend
using KDUQ over CHUQ for non-elastic calculations and for calculations below 10
\MeV\ (the stated threshold of validity for CH89). In the case of proton
$\sigma_{rxn}$ data, KDUQ shows improved performance compared to the canonical
KD global OMP.  Further, by training two versions of KDUQ with different
assumed forms of data covariance, we demonstrated how small changes in
underlying covariance assumptions can impact the uncertainty of predictions in
data-sparse regions, as shown for high-energy proton $\sigma_{rxn}$ in Fig.
\ref{DemocraticFederalComparison}. These results caution against na\"ive use of
a weighted-least-squares likelihood function when experimental data used for
training are known to have underestimated uncertainties and non-trivial
covariance structure. In the case we presented, an MCMC-based inference
strategy made sense so that we could include our unaccounted-for uncertainty
estimates as priors, but the need for a defensible likelihood function is just
as important in any approach, Bayesian or not, involving training a model to
data.

As a demonstration of their utility, we forward-propagated CHUQ and KDUQ's
parameter covariances in two case studies. In the first, we showed that KDUQ
accurately predicts neutron $\sigma_{tot}$ evolution with respect to asymmetry,
auguring well for neutron-scattering predictions beyond the valley of
$\beta$-stability, at least along closed shells in $Z$. Because our
uncertainty-quantified model was designed to incorporate the observed variance
of its training data, a discrepancy between our model and experimental data is
not easily explained away as arbitrariness in the model parameters. For
example, in our examination of isotopic relative differences of neutron
$\sigma_{tot}$, we saw KDUQ underpredicted the oscillations present in the
experimental relative differences for Sn and W isotopes between $20$ and $50$
\MeV\ (panels (c) and (d) of Fig. \ref{caseStudy1}). These oscillations can be
reproduced by an OMP analysis only if the different proton and neutron density
distributions of the target are taken into account, as shown in
\cite{Dietrich2003}. Although this physics is absent from the KD or CH89
pictures, it implies that, provided one uses an uncertainty-quantified OMP and
fits to relative $\sigma_{tot}$ differences rather than absolute cross
sections, neutron $\sigma_{tot}$ data are useful for extracting neutron skin
thickness information. As new reactions are pursued at modern radioactive beam
facilities, this kind of comparison between uncertainty-equipped data and
uncertainty-equipped models is important for calibrating our ``degree of
surprise'' to avoid chasing down spurious signals. Systematic comparison
against isovector data, including (p,n) cross sections and $\sigma_{tot}$
relative differences along isotopic and isotonic chains, is a promising
meeting-ground for phenomenological and microscopic OMPs.

Finally, we explored the impact of KDUQ and CHUQ on representative radiative
capture calculations for \srEightySeven\ and \moNinetyFive. The capture cross
sections between 1-5 \MeV\ computed using KDUQ are somewhat lower
($\approx20-30\%$) than those using CHUQ, though with substantial uncertainty
overlap. Given the systematic assessment of proton capture rate uncertainty of
\cite{Vagena2021}, we argue that in the few-\MeV\ range, the fraction of
overall cross section uncertainty due to the OMP is comparable to that in the
$\gamma$-ray strength function and level density, and at energies below 1 \MeV\
the OMP uncertainty may dominate. Moreover, while the partition of strength
between $s$- and $p$-wave below 10 \MeV\ are different, particularly for
neutrons, the contributions from each to the overall cross sections were
countervailing for \moNinetyFive. If angular momentum transfer is restricted to
a single partial wave, the differences between (and uncertainty in) OMPs can be
much larger, as shown for n+\moNinetyFive\ below 100 \keV, and the effect on
cross sections correspondingly larger. This is another region where comparison
between (semi)-microscopic and phenomenological OMPs is likely to be fruitful,
both for improving existing OMPs and for providing more stringent reaction
rates to astrophysical nucleosynthesis calculations. To support such efforts,
we enclose copies of CHUQ and KDUQ in the Supplemental Material
\cite{SupplementalMaterial}.

\section*{Acknowledgements}
We thank Robert Casperson and Ron Soltz for insightful discussion regarding
loss function selection and Kyle Beyer for identifying corrections to the manuscript. This
work was performed under the auspices of the U.S. Department of Energy by
Lawrence Livermore National Laboratory under Contract DE-AC52-07NA27344. We
gratefully acknowledge support from the High Energy Density Science program
(HEDP) and the Defense Science and Technology internship (DSTI) for R. Rahman.

\bibliography{references}

\clearpage

\appendix \section{Definition of optical potentials and scattering formulae}
Reproduced here are the definition of the Chapel Hill '89 \cite{CH89} and
Koning-Delaroche \cite{KoningDelaroche} optical potentials, starting from the
overall potential form and ending with the definitions for form subterms. Free
parameters (those subject to Bayesian inference via MCMC) are denoted in this section
using a \textbf{bold} typeface. For brevity we set $\hbar = c = 1$.

\subsection{CH89 definition}
The CH89 global optical potential for single-nucleon scattering consists of five terms:
\begin{equation} \label{CH89Potential}
    \begin{split}
        \mathcal{U}(r,E) = & \mathcal{V}_{r}(r,E) - i\mathcal{W}_{v}(r,E) -i\mathcal{W}_{s}(r,E) \\
        - & \mathcal{V}_{so}(r,E)(\ell\cdot\sigma) + \mathcal{V}_{C}(r),
\end{split}
\end{equation}
where
\begin{itemize}
    \item $\mathcal{V}_{r}$ is the real central potential,
    \item $\mathcal{W}_{v}$ is the imaginary central (or ``volume'') potential,
    \item $\mathcal{W}_{s}$ is the imaginary surface potential,
    \item $\mathcal{V}_{so}$ is the real spin-orbit potential, and
    \item $\mathcal{V}_{C}$ is the Coulomb potential (for protons only).
\end{itemize}

As with the Koning-Delaroche potential defined below, each component (except Coulomb)
consists of an energy-dependent \textit{depth} coupled with a radius-dependent
\textit{spatial form}:
\begin{equation} \label{CH89PotentialComponents}
    \begin{split}
        \mathcal{V}_{r}(r,E) & = V_{r}(E) \times f(r,R_{0},\mathbf{a_{0}}), \\
        \mathcal{W}_{v}(r,E) & = W_{v}(E) \times f(r,R_{w},\mathbf{a_{w}}), \\
        \mathcal{W}_{s}(r,E) & = W_{s}(E) \times
        -4\mathbf{a_{w}}\frac{d}{dr}f(r,R_{w},\mathbf{a_{w}}), \\
        \mathcal{V}_{so}(r,E) & = 2V_{so} \times
        \frac{-1}{r}\frac{d}{dr}f(r,R_{so},\mathbf{a_{so}}), \\
        \mathcal{V}_{C}(r) & = \begin{cases}
            \frac{Zze^{2}}{2R_{C}}\left(3-\frac{r^{2}}{R_{C}^{2}}\right), & \text{if
    } r < R_{C} \\
            \frac{Zze^{2}}{r}, & \text{if } r \geq R_{C}
        \end{cases}.
    \end{split}
\end{equation}
The spatial form $f(r, R, a)$ is the standard Woods-Saxon potential
\begin{equation} \label{WoodsSaxon}
    \begin{split}
        f(r, R, a) & = \frac{1}{1+e^{(r-R)/a}},\\
        \frac{d}{dr}f(r, R, a) & = \frac{1}{a}\left[\frac{-e^{(r-R)/a}}{(1+e^{(r-R)/a})^{2}}\right].
    \end{split}
\end{equation}
Here $R$ and $a$ are radius and diffuseness parameters, respectively.  The
usual $R=r_{0}A^{1/3}$ dependence is assumed (see Eq.
\ref{CH89RadialFormParameters} below for equations defining $r_{0}$ for each
component), with $A$ the nucleon number of the target.  We note that for a
natural-abundance target, the value that should be taken for $A$ is not
explicitly discussed in the original formulation of CH89 or KD. A
simple choice would be to use the $A$ of the most abundant isotope, which works
well for many elements but is unsatisfying in cases where the lightest or
heaviest isotope is most abundant.  For example, in \niNat\ the most abundant
isotope is \niFiftyEight, but the abundance-weighted nucleon number is 58.76 (a
difference of 1.3\% from 58). In this work, for natural targets we took for
$A$ the target's atomic weight, which for the targets we used agrees with the
abundance-weighted nucleon number to within $\approx0.1\%$.

The CH89 energy-dependent depths are given by:
\begin{equation} \label{CH89EnergyDependentDepths}
    \begin{split}
        V_{r}(E) & = \mathbf{V_{0}} + \mathbf{V_{e}} \Delta E \pm \alpha \mathbf{V_{t}} \\
        W_{v}(E) & = \mathbf{W_{v0}}\left[1+e^{\frac{\mathbf{W_{ve0}}- \Delta E}{\mathbf{W_{vew}}}}\right]^{-1} \\
        W_{s}(E) & = (\mathbf{W_{s0}}+\alpha \mathbf{W_{st}})\left[1+e^{\frac{\Delta E - \mathbf{W_{se0}}}{\mathbf{W_{sew}}}}\right]^{-1} \\
        V_{so}(E) & = \mathbf{V_{so0}}
    \end{split}
\end{equation}
The nuclear asymmetry $\alpha$ is defined $(N-Z)/A$. As with the definition of $A$, for 
natural targets a definition for $\alpha$ is not given in the original potential formulation.
For these targets, we took $\alpha = (A-2Z)/Z$, consistent with our definition
of $A$. The energy argument $\Delta E$ is the difference between the scattering energy and
the volume-averaged Coulomb energy:
\begin{equation} \label{CH89CoulombEnergy}
    \begin{split}
        \Delta E & = E_{lab} - E_{c} \\
        E_{c} & = \begin{cases}
            \frac{6Ze^{2}}{5R_{c}}, & \text{for protons} \\
            0, & \text{for neutrons}
        \end{cases}.
    \end{split}
\end{equation}

Lastly, the radial form parameters $R_{i}$ are defined as follows:
\begin{equation} \label{CH89RadialFormParameters}
    \begin{split}
        R_{0} = \mathbf{r_{0}}A^{1/3} + \mathbf{r_{0}^{0}}, \\
        R_{w} = \mathbf{r_{w}}A^{1/3} + \mathbf{r_{w}^{0}}, \\
        R_{so} = \mathbf{r_{so}}A^{1/3} + \mathbf{r_{so}^{0}}, \\
        R_{C} = \mathbf{r_{c}}A^{1/3} + \mathbf{r_{c}^{0}}.
    \end{split}
\end{equation}

In total there are 22 free potential parameters: 11 associated with the
potential depths and 16 associated with the radius-dependent spatial forms. We
comment that in the original CH89 treatment, only 20 parameters were free, as the
authors fixed the Coulomb parameters $r_{c}$ and $r_{c}^{(0)}$ based on a
separate assessment.

\subsection{Koning-Delaroche definition}
Similar to CH89, the Koning-Delaroche optical potential for single-nucleon
scattering is defined as a function of radius $r$ and energy $E$:
\begin{equation} \label{KoningDelarochePotential}
    \begin{split}
        \mathcal{U}(r,E) & = -\mathcal{V}_{V}(r,E) - i\mathcal{W}_{V}(r,E) - i\mathcal{W}_{D}(r,E) \\
                         & + \mathcal{V}_{SO}(r,E)(\ell\cdot\sigma) + i\mathcal{W}_{SO}(r,E)(\ell\cdot\sigma) + \mathcal{V}_{C}(r),
    \end{split}
\end{equation}
where
\begin{itemize}
    \item $\mathcal{V}_{V}$ is the real central potential,
    \item $\mathcal{W}_{V}$ is the imaginary central potential,
    \item $\mathcal{W}_{D}$ is the imaginary surface potential,
    \item $\mathcal{V}_{SO}$ is the real spin-orbit potential,
    \item $\mathcal{W}_{SO}$ is the imaginary spin-orbit potential, and
    \item $\mathcal{V}_{C}$ is the Coulomb potential (for protons only).
\end{itemize}

In the spin-orbit components, $\ell$ is the orbital angular momentum quantum number for each 
partial wave associated with the incident projectile and $\sigma$ is the spin of the incident 
projectile. Except Coulomb, each component consists of an energy-dependent 
\textit{depth} coupled with a radius-dependent \textit{spatial form}:
\begin{equation} \label{KoningDelarochePotentialComponents}
    \begin{split}
        \mathcal{V}_{V}(r,E) & = V_{V}(E) \times f(r,R_{V},a_{V}), \\
        \mathcal{W}_{V}(r,E) & = W_{V}(E) \times f(r,R_{V},a_{V}), \\
        \mathcal{W}_{D}(r,E) & = W_{D}(E) \times -4a_{D}\frac{d}{dr}f(r,R_{D},a_{D}), \\
        \mathcal{V}_{SO}(r,E) & = V_{SO}(E) \left(\frac{\hbar}{m_{\pi}c}\right)^{2} \times \frac{1}{r}\frac{d}{dr}f(r,R_{SO},a_{SO}), \\
        \mathcal{W}_{SO}(r,E) & = W_{SO}(E) \left(\frac{\hbar}{m_{\pi}c}\right)^{2} \times \frac{1}{r}\frac{d}{dr}f(r,R_{SO},a_{SO}), \\
        \mathcal{V}_{C}(r) & = \begin{cases}
            \frac{Zze^{2}}{2R_{C}}\left(3-\frac{r^{2}}{R_{C}^{2}}\right), & \text{if } r < R_{C} \\
            \frac{Zze^{2}}{r}, & \text{if } r \geq R_{C}
        \end{cases}.
    \end{split}
\end{equation}

The spatial form $f(r, R, a)$ is the same Woods-Saxon defined earlier in the
CH89 case (Eq. \ref{WoodsSaxon}), with $R=r_{0}A^{1/3}$. In the spin-orbit
subcomponent definitions, $m_{\pi}$ is the charged pion mass. In the Coulomb
component definition, $z$ is the projectile charge, $Z$ is the target charge,
and $e^{2}$ is the elementary charge squared ($\approx$ 1.44
\MeV$\cdot$\femto\meter).

Depending on whether the user is modeling neutron or proton scattering, the
energy-dependent depths appearing in Eq.
\ref{KoningDelarochePotentialComponents} are given by:
    \begin{equation} \label{EnergyDependentDepths}
        \begin{split}
            V_{V}(E) = v_{1}^{n,p} & [1-v_{2}^{n,p} \Delta E^{n,p} \\
                     & + v_{3}^{n,p} (\Delta E^{n,p})^{2} -v_{4}^{n,p} (\Delta E^{n,p})^{3}] \\
            + \overline{V}_{C} & \times v_{1}^{p} [v_{2}^{p} - 2v_{3}^{p}\Delta E^{p} + 3v_{4}^{p} (\Delta E)^{2}] \\
            W_{V}(E) & = w_{1}^{n,p}\frac{(\Delta E)^{2}}{(\Delta E)^{2} + (w_{2}^{n,p})^{2}}, \\
            W_{D}(E) & = d_{1}^{n,p}\frac{(\Delta E)^{2}}{(\Delta E)^{2} + (d_{3}^{n,p})^{2}} e^{-d_{2}^{n,p}\Delta E}, \\
            V_{SO}(E) & = v_{so1}^{n,p}e^{-v_{so2}^{n,p}\Delta E}, \\
            W_{SO}(E) & = w_{so1}^{n,p}\frac{(\Delta E)^{2}}{(\Delta E)^{2} + (w_{so2}^{n,p})^{2}},
        \end{split}
    \end{equation}
where the superscripts $n,p$ denote different parameters used for neutrons and
protons, respectively. The energy variable $\Delta E^{n,p}$ is the difference
between the incident scattering energy in \MeV\ in the lab frame and the Fermi
energy for neutrons or protons:
\begin{equation} \label{FermiEnergy}
    \begin{split}
        \Delta E^{n,p} & = E - E_{f}^{n,p} \\
        E_{f}^{n} & = -11.2814 + 0.02646 A \\
        E_{f}^{p} & = -8.4075 + 1.01378 A.
    \end{split}
\end{equation}

The potential depth parameters from Eq. \ref{EnergyDependentDepths} are defined as:
\begin{equation} \label{DepthParameters}
    \begin{split}
        v_{1}^{n,p} & = \mathbf{v_{1,0}} - \mathbf{v_{1,A}}A \pm \mathbf{v_{1,\bm{\alpha}}}\alpha \\
        v_{2}^{n,p} & = \mathbf{v_{2,0}^{n,p}} \pm \mathbf{v_{2,A}^{n,p}} \\
        v_{3}^{n,p} & = \mathbf{v_{3,0}^{n,p}} \pm \mathbf{v_{3,A}^{n,p}} \\
        v_{4}^{n,p} & = \mathbf{v_{4,0}} \\
        w_{1}^{n,p} & = \mathbf{w_{1,0}^{n,p}} + \mathbf{w_{1,A}^{n,p}}A \\
        w_{2}^{n,p} & = \mathbf{w_{2,0}} + \mathbf{w_{2,A}}A \\
        d_{1}^{n,p} & = \mathbf{d_{1,0}} \pm \mathbf{d_{1,\bm{\alpha}}}\alpha  \\
        d_{2}^{n,p} & = \mathbf{d_{2,0}} + \frac{\mathbf{d_{2,A}}}{1+e^{(A-\mathbf{d_{2,A3}})/\mathbf{d_{2,A2}}}} \\
        d_{3}^{n,p} & = \mathbf{d_{3,0}} \\
        v_{so1}^{n,p} & = \mathbf{v_{so1,0}} + \mathbf{v_{so1,A}}A \\
        v_{so2}^{n,p} & = \mathbf{v_{so2,0}} \\
        w_{so1}^{n,p} & = \mathbf{w_{so1,0}} \\
        w_{so2}^{n,p} & = \mathbf{w_{so2,0}} \\
        \overline{V}_{C} & = \frac{V_{C}Z}{r_{C}A^{1/3}} = \frac{6Ze^{2}}{5r_{C}A^{1/3}}.
    \end{split}
\end{equation}
In these expressions, $\pm$ should be taked as $-$ for neutrons and $+$ for
protons. Our definitions for $A$ and for the nuclear asymmetry $\alpha$ for
natural targets are the same as used above for CH89.

Finally, the radial form parameters entering Eqs.
\ref{WoodsSaxon} and \ref{KoningDelarochePotentialComponents} are defined
\begin{equation} \label{KoningDelarocheRadialFormParameters}
    \begin{split}
        r_{V} & = \mathbf{r_{V,0}} - \mathbf{r_{V,A}}A^{-1/3} \\
        a_{V} & = \mathbf{a_{V,0}} - \mathbf{a_{V,A}}A \\
        r_{D} & = \mathbf{r_{D,0}} - \mathbf{r_{D,A}}A^{-1/3} \\
        a_{D} & = \mathbf{a_{D,0}^{n,p}} \pm \mathbf{a_{D,A}^{n,p}}A \\
        r_{SO} & = \mathbf{r_{SO,0}} - \mathbf{r_{SO,A}}A^{-1/3} \\
        a_{SO} & = \mathbf{a_{SO,0}} \\
        r_{C} & = \mathbf{r_{C,0}} + \mathbf{r_{C,A}}A^{-2/3} + \mathbf{r_{C,A2}}A^{-5/3}.
    \end{split}
\end{equation}
As in Eq. \ref{DepthParameters}, $\pm$ should be taken as $-$ for neutrons and $+$ for protons.
In total there are 47 free potential parameters: 31 associated with the energy-dependent depths 
and 16 associated with the radius-dependent spatial forms.

\subsection{Scattering formulae}

In this section we list the expressions we used to calculate proton and
neutron scattering observables. Our procedure follows the calculable $R$-matrix
method outlined in Descouvement and Baye (DB) \cite{Descouvement2010}, but
modified (as discussed below) to be suitable for relativistic-equivalent
calculations. The scattering observables we considered can all be calculated
from the scattering matrix for incident partial waves. The $S$ matrix for the
incident projectile partial wave with angular momentum $l$ is (DB Eq. 3.24):
\begin{equation}
    S_{l} = e^{2i\delta_{l}} = e^{2i\phi_{l}}\frac{1-(L_{l}^{*}-B)R_{l}(E,B)}{1-(L_{l}-B)R_{l}(E,B)}.
\end{equation}
The $S$-matrix terms are equivalent to the partial wave phase shifts
$\delta_{l}$. Here $L_{l}$ is the logarithmic derivative of the outgoing
partial wave, evaluated at channel radius $a$. It can be expressed in terms of
Coulomb functions (DB Eqs. 3.28-3.30):
\begin{equation}
    \begin{split}
        L_{l} & = \frac{ka}{F_{l}(ka)^{2} + G_{l}(ka)^{2}} \\
              & \times \left[F_{l}(ka)F_{l}'(ka) + G_{l}(ka)G_{l}'(ka) + i\right].
    \end{split}
\end{equation}
The hard-sphere phase shift, $\phi_{l}$, is (DB Eq. 3.26)
\begin{equation}
    \phi_{l} = -\tan^{-1}(F_{l}(ka)/G_{l}(ka)).
\end{equation}
In these expressions, $F_{l}$ and $G_{l}$ are the regular and irregular Coulomb
functions, with $F'_{l}$ and $G'_{l}$ their derivatives. (In the Coulomb function
notation, we have omitted the implied Sommerfeld parameter $\eta$.) $R_{l}(E,B)$ are the
$R$-matrix elements, discussed below, and $B$ is a dimensionless
boundary parameter associated with the Bloch operator. As shown in Eq. 3.27 and
appendix B of DB, the scattering matrix is unaffected by the
choice of boundary parameter $B$, so $B$ can be set to 0 to simplify the $S$-
and $R$-matrix calculation algebra.

To calculate the $R$ matrix, we used the finite-basis approximation (Eq. 3.15
in DB):
\begin{equation}
    R_{l}(E,B) = \frac{1}{2\mu a} \sum^{N}_{i,j=1}\phi_{i}(a)(\bm{C}^{-1})_{ij}\phi_{j}(a).
\end{equation}
Here $E$ is the center-of-mass energy, $\mu$ is the reduced mass, $a$ is
the channel radius, $N$ is the number of basis states $\phi$, and $\bm{C}$ is the
symmetric matrix containing solutions to the inhomogenous Bloch-Schr\"odinger
equation (Eq. 3.7 in DB),
\begin{equation}
    \bm{C}_{i,j}(E,B) = \bra{\phi_{i}} T_{l} + \mathcal{L}(B) + V - E \ket{\phi_{j}}
\end{equation}
To solve this equation, we employed the Lagrange-mesh method of Baye
\cite{Baye2015} on an $N=30$ Legendre-polynomial mesh. The kinetic energy
$T_{l}$ and Bloch $\mathcal{L}(B)$ operators on this Lagrange-Legendre mesh
(which we do not reproduce here) are given by Eqs. 3.127 and 3.129 of
\cite{Baye2015}. In our case, $V$ is the optical potential, with $E$ the center-of-mass
energy. Note, however, that the energy argument of the optical potential,
e.g., $E$ in $U(r,E)$ of Eq. \ref{KoningDelarochePotential}, is the projectile
energy in the \textit{lab frame}, per the definition of CH89 and KD.

The above formulation is appropriate for the non-relativistic limit, but above
a few tens of \MeV, an approximate relativistic-equivalent version should be
used, requiring modification of several elements in the calculation. First, the
center-of-mass energies, angles, and the relative velocity appearing in the
Sommerfeld parameter should be calculated according to relativistic kinematics.
Second, in the relativistic picture the reduced mass and center-of-mass
wavenumber are no longer suitable to describe the relative motion between
projectile and target, so approximations are required. We used the relativistic
approximations of Eqs. 17 and 20 in Ingemarsson's topical study
\cite{Ingemarsson1974} that base the wavenumber on the relativistic momentum in
the center-of-momentum frame and treat the center-of-momentum motion of the
target as non-relativistic. These approximations modify the
wavenumber and reduced mass appearing throughout this section
as:
\begin{equation} \label{RelativisticEquivalent}
    \begin{split}
        k & \rightarrow \frac{m_{1}[E(E+2m_{2})]^{1/2}}{[(m_{1}+m_{2})^{2}+2m_{1}E)]^{1/2}} \\
        \mu & \rightarrow k^{2}\frac{E'}{{E'}^{2}-{m_{2}}^{2}}.
    \end{split}
\end{equation}
Here, $m_{1}$ is the target rest mass, $m_{2}$ is the projectile rest mass, $E$
is the incident projectile energy in the laboratory frame, and $E'$ is the sum
of center of mass energies of the target and projectile, plus the rest mass of
the projectile. These approximations for $k$ and $\mu$ can be inserted in the
preceding equations to yield the relativistic-approximate forms that we
actually used to perform calculations.

To generate scattering observables for spin-1/2 particles, two $S$-matrix terms,
corresponding to $j=l\pm1/2$, must be calculated for each partial
wave $l>0$. From these terms the non-spinflip amplitude $A(\theta)$ and
spinflip amplitude $B(\theta)$ can be calculated for scattering angle
$\theta$:
\begin{equation} \label{NonSpinflipAmplitude}
    \begin{split}
        A(\theta) & = \frac{i}{2k} \sum_{l=0}^{\infty} e^{2i\sigma_{l}} (2l+1 - (l+1)S_{l}^{+} - lS_{l}^{-})P_{l}(\cos{\theta}) \\
                  & - \frac{\eta}{2k\sin^{2}\frac{1}{2}\theta}e^{2i(\sigma_{0}-\eta \log \sin \frac{1}{2}\theta)}
    \end{split}
\end{equation}

\begin{equation} \label{SpinflipAmplitude}
    B(\theta) = \frac{i}{2k} \sum_{l=0}^{\infty} e^{2i\sigma_{l}} (S_{l}^{-} - S_{l}^{+})P_{l}^{1}(\cos{\theta}).
\end{equation}
Here, $S_{l}^{+}$ is the $S$-matrix element for $j=l+\frac{1}{2}$ and $S_{l}^{-}$
is the $S$-matrix element for $j=l-\frac{1}{2}$ (setting $S_{0}^{-}\equiv 0$).
$P_{l}$ is the Legendre polynomial of degree $l$, $P_{l}^{1}$ is the associated
Legendre polynomial of degree $l$ and order $m$, and $\sigma_{l}$ is the
Coulomb phase shift:
\begin{equation}
    \sigma_{l} = \arg \Gamma(l + 1 + i\eta),
\end{equation}
$\Gamma$ being the gamma function.
Equations \ref{NonSpinflipAmplitude} and \ref{SpinflipAmplitude} combine Eqs. 8
and 9 of Ingemarsson, which are for spin-1/2 neutral
particles, with the spinless, charged particle scattering amplitudes of DB Eq.
2.23. Specifically, the final term of Eq. \ref{NonSpinflipAmplitude} that
involves $\eta$ is the Coulomb scattering amplitude (DB Eq. 2.13).

Finally, from the scattering amplitudes, the differential elastic cross section
is simply
\begin{equation}
    \frac{d\sigma(\theta)}{d\Omega} = |A(\theta)|^{2} + |B(\theta)|^{2},
\end{equation}
and the analyzing power is
\begin{equation}
    A_{y} = \frac{A^{*}(\theta)B(\theta) + A(\theta)B^{*}(\theta)}{\frac{d\sigma(\theta)}{d\Omega}},
\end{equation}
per Eqs. 10 and 11 of Ingemarsson. The reaction (non-elastic)
and total cross sections can be computed directly from the $S$ matrix:
\begin{equation}
    \sigma_{rxn} = \frac{\pi}{k^{2}}\sum_{l=0}^{\infty}(l+1)(1-|S_{l}^{+}|^{2}) + l(1-|S_{l}^{-}|^{2})
\end{equation}
\begin{equation}
    \sigma_{tot} = \frac{2\pi}{k^{2}}\sum_{l=0}^{\infty}(l+1)(1-\text{Re}[S_{l}^{+}]) + l(1-\text{Re}[S_{l}^{-}]).
\end{equation}

\nocite{Carlson1996}
\nocite{Ingemarsson1999}

\newpage
\newpage

\section*{Supplemental Material A: Posterior samples for CHUQ and KDUQ}

This section of the supplemental material describes how to access and use 
CHUQ and KDUQ. For functional forms of the potentials and additional
information on their construction, see the main text.

CHUQ and KDUQ each consist of an ensemble of samples of potential parameters.
Each OMP sample is the terminal position of one MCMC walker at the end of
training against the appropriate training data corpus. For both CHUQ and KDUQ,
two versions of the uncertainty-quantified potential are provided: one trained
using the ``democratic" covariance ansatz, and one trained using the ``federal"
ansatz as described in the main text. The resulting potential ensembles are
denoted CHUQDemocratic/CHUQFederal and KDUQDemocratic/KDUQFederal. With the
exception of Fig. 13 in the main text, which compares the democratic and
federal ansatze, all figures were created with the democratic versions of the
potentials. Given that the federal ansatz is (arguably) a more realistic
representation of the unknown data covariance, we recommend that practitioners
use the federal versions of CHUQ and KDUQ for their applications, though
comparison of observables calculated using the two versions suggests that the
differences are small. (The largest differences were for proton reaction cross
sections at energies above 50 MeV, as shown in Fig. 13 in the main text).

In the attached file \texttt{supplement\_parameters.tar}, each ensemble is
placed in a separate directory. The samples comprising the ensembles are placed
in separate subdirectories, each labeled by the index of the corresponding
walker, starting with 0. Within each numbered subdirectory, there are two
files: \texttt{parameters.json} and \texttt{modelErrors.json}. The potential
parameters are stored in \texttt{parameters.json}. For example, the file
\texttt{KDUQFederal/10/parameters.json} includes the eleventh posterior sample
for the KDUQFederal ensemble.

In the parameter file, the potential subterms and their constituent parameters
represented as key-value pairs. For example, the canonical CH89 potential
parameter values would be represented as:

\begin{lstlisting}
{
    ``CH89RealCentral": {
        ``V_0": 52.90,
        ``V_t": 13.10,
        ``V_e": -0.299,
        ``r_o": 1.250,
        ``r_o_0": -0.225,
        ``a_0": 0.690
    },
    ``CH89Coulomb": {
        ``r_c": 1.24,
        ``r_c_0": 0.12
    },
    ``CH89SpinOrbit": {
        ``V_so": 5.9,
        ``r_so": 1.34,
        ``r_so_0": -1.2,
        ``a_so": 0.63
    },
    ``CH89ImagCentral": {
        ``W_v0": 7.8,
        ``W_ve0": 35.0,
        ``W_vew": 16.0,
        ``W_s0": 10.0,
        ``W_st": 18.0,
        ``W_se0": 36.0,
        ``W_sew": 37.0,
        ``r_w": 1.33,
        ``r_w0": -0.42,
        ``a_w": 0.69
    }
}
\end{lstlisting}
and the canonical KD potential parameter values would be represented as:
\begin{lstlisting}
{
    ``KDHartreeFock": {
        ``V1_0": 59.30,
        ``V1_asymm": 21.0,
        ``V1_A": 0.024,
        ``V2_0_n": 0.007228,
        ``V2_A_n": 1.48e-6,
        ``V3_0_n": 1.994e-5,
        ``V3_A_n": 2.0e-8,
        ``V2_0_p": 0.007067,
        ``V2_A_p": 4.23e-6,
        ``V3_0_p": 1.729e-5,
        ``V3_A_p": 1.136e-8,
        ``V4_0": 7e-9,
        ``r_0": 1.3039,
        ``r_A": 0.4054,
        ``a_0": 0.6778,
        ``a_A": 1.487e-4
    },
    ``KDCoulomb": {
        ``r_C_0": 1.198,
        ``r_C_A": 0.697,
        ``r_C_A2": 12.994
    },
    ``KDRealSpinOrbit": {
        ``V1_0": 5.922,
        ``V1_A": 0.0030,
        ``V2_0": 0.0040,
        ``r_0": 1.1854,
        ``r_A": 0.647,
        ``a_0": 0.59
    },
    ``KDImagSpinOrbit": {
        ``W1_0": -3.1,
        ``W2_0": 160
    },
    ``KDImagVolume": {
        ``W1_0_n": 12.195,
        ``W1_A_n": 0.0167,
        ``W1_0_p": 14.667,
        ``W1_A_p": 0.009629,
        ``W2_0": 73.55,
        ``W2_A": 0.0795
    },
    ``KDImagSurface": {
        ``D1_0": 16.0,
        ``D1_asymm": 16.0,
        ``D2_0": 0.0180,
        ``D2_A": 0.003802,
        ``D2_A2": 8.0,
        ``D2_A3": 156.0,
        ``D3_0": 11.5,
        ``r_0": 1.3424,
        ``r_A": 0.01585,
        ``a_0_n": 0.5446,
        ``a_A_n": 1.656e-4,
        ``a_0_p": 0.5187,
        ``a_A_p": 5.205e-4
    }
}
\end{lstlisting}

In addition to the potential parameters, the ``unaccounted-for uncertainties"
described in the main text are provided in the \texttt{modelErrors.json} file in a
similar format. For example, the following structure contains four such
uncertainties: differential elastic scattering unaccounted-for uncertainties of
50\% (both for protons and for neutrons) and analyzing power unaccounted-for
uncertainties of 60\% (both for protons and for neutrons):

\begin{lstlisting}
{
    ``ECS_p": 0.5,
    ``ECS_n": 0.5,
    ``APower_p": 0.6,
    ``APower_n": 0.6,
}
\end{lstlisting}

Two example Python scripts are included to serve as a reference for those
wishing to use CHUQ/KDUQ in their applications. The first,
\texttt{accessExample.py}, demonstrates access and inspection of KDUQDemocratic
samples. The second, \texttt{newSampleExample.py}, illustrates one approach for
calculating the KDUQDemocratic sample mean and covariance and generation of an
arbitrary number of new KDUQDemocratic samples using the mean and covariance.
To simplify this process, the script defines two flattening methods that map
the nested potential parameter structures (shown above) into one-dimensional
lists. For convenience, the second script also makes use of the \texttt{numpy}
and \texttt{pandas} Python modules, available at the
\href{https://pypi.org}{Python Package Index}.

Finally, in the \texttt{CHUQDemocratic} and \texttt{KDUQDemocratic}
directories, we include correlogram figures to illustrate the covariance
structure over ensemble samples. In each figure, the degree of correlation is
indicated by the cell size and color intensity, with blue indicating positive
correlation and red indicating negative correlation. The black dashed lines
denote groupings of related parameters (those belonging to the same potential
subterm).

\newpage

\section*{Supplemental Material B: KDUQ, CHUQ, and Test corpora}

In this section, we list the experimental data used to train and test the UQ
OMPs described in the main text. To enable direct comparison with the original
KD and CH89 treatments, we re-assembled (as faithfully as possible) the same
corpora of experimental training data listed in the original publications,
which we refer to the ``KD corpus'' and ``CH89 corpus'', respectively. To
distinguish these corpora as listed in the original publications from
our present reconstruction of these corpora, we use the names ``KDUQ corpus ''
and ``CHUQ corpus'' to refer to the corpora codified in this section. Also
tabulated are the data comprising the ``Test corpus'', all of which post-date
the publication of the original CH89 \cite{CH89} and KD \cite{KoningDelaroche}
treatments.

Before detailing the assembly of these corpora, a few methodological comments
are worth making. Wherever possible, for the data sets listed in the KD and
CH89 corpora, we drew on experimental data as tabulated in the EXFOR
experimental reaction database \cite{EXFORDatabase}. In the instances where we
could not locate the data in EXFOR, we turned to the original publication (as
listed in the canonical KD or CH89 papers) and retrieved the data from that
source. In the few remaining cases where the data either could not be found in
the original literature or were missing critical features (such as experimental
errors), we omitted the data set from our reconstructed corpora and documented
the discrepancy. For consistency, we prioritized the labels and metadata from
the EXFOR entries, as they existed at the time of our accession of those data.
If a data set was unavailable or incomplete in EXFOR, we then used the labels
and metadata as we interpreted from the original literature source. We justify
this as an attempt to make our analysis as reproducible as possible: because
EXFOR is regularly updated as errors are identified and corrected, our hope is
that future investigators wishing to use the same data corpus need only refer
to the EXFOR accession numbers we provide here and not start ``from scratch''.
A consequence of this approach is that our reconstructed corpora span slightly
fewer data compared to the the original corpora because we were unable to
locate, either in EXFOR or in the original literature, some data sets that were
nominally included in the original corpus. These few instances we attribute
either to typographic errors in the references as listed in the KD and CH89
publications or our own error in searching the available literature.
For construction of the Test corpus, all data were drawn from EXFOR.

Once the data sets were collected and metadata assigned, we applied a series of
munging steps to prepare the corpus for OMP analysis. First, units were
homogenized: scattering energies, scattering angles, integral cross sections,
and differential cross sections were transformed to \MeV\ (lab frame), degrees
(center-of-momentum frame), \milli\barn, and \milli\barn/\steradian,
respectively. All proton differential elastic cross section data sets that were
provided relative to Rutherford scattering were rescaled into absolute terms
(assuming relativistic kinematics). Any datum that was missing a ``necessary
feature'' was flagged and removed from the KDUQ/CHUQ corpora. For example, for
differential elastic cross sections, these necessary features were: scattering
energy, scattering angle, cross section, and cross section error. Many
EXFOR-based data list more than one type of error, for instance, digitization
error, statistical error, etc. For a given datum, we assigned the overall error
according to the following list of EXFOR error labels, in order of preference:
ERR-T (total error), (+DATA-ERR + -DATA-ERR)/2 (average of positive and
negative error), ERR-S (statistical error), ERR-DIG (digitization error),
ERR-SYS (systematic error). Absent a covariance matrix for each data set (none
were available for the data sets in the KDUQ/CHUQ corpus), our assignment of
overall error is not unique, but is at least reproducible. Additionally, in the
KD corpus, many of the neutron total cross section data sets had an
unnecessarily large number of energy bins for the requirements of optical
potential optimization. For instance, the data set from 250 \keV\ to 20 \MeV\
for \alTwentySeven\ (EXFOR Acc. No.
\href{https://www-nds.iaea.org/exfor/servlet/X4sGetSubent?subID=22331004}{22331004})
includes nearly 50,000 unique energy bins. Calculating cross sections for each
of these energies in our analysis would be computationally expensive for
negligible benefit, given the slowly-varying nature of the total cross section
in this energy regime. For these neutron total cross section data sets, rather
than include the full complement of neutron total cross section data, we
downsampled each data set to have one datum per \MeV\ --- an energy step still
sufficiently fine under the smoothness assumptions inherent to a global optical
potential.

Due to rounding and discrepancies in tabulation, the scattering energies listed
here do not always exactly match those reported from the KD corpus or the
references therein. The scattering energies listed here are as provided in
EXFOR, or, if not available there, transcribed directly from the literature
reference. To facilitate others' use of the KDUQ corpus for future optical model
work, we list the EXFOR accession number the data sets that we drew from
EXFOR. For the few data sets that were not available in EXFOR, this field is
filled with ``-" in the following tables, and the original reference listed in
the Comments following each table. The Comments also document special cases and
notable differences between the original corpora and our reconstructions. In
the tables, all listed energies are in units of \MeV.

\subsection*{Assembling the KDUQ corpus}
Tables 1, 2, and 7 of \cite{KoningDelaroche} list five
of the six sectors of experimental data used in the original Koning-Delaroche
analysis: neutron differential elastic scattering cross sections and analyzing
powers, neutron total cross sections, and proton differential elastic
scattering cross sections and analyzing powers. The sixth sector, proton
reaction cross sections, was drawn from \cite{Carlson1996} and
\cite{Ingemarsson1999}, following the original analysis.

During data assembly we noted two general types of discrepancy between the
labeling and citations given in the original Koning-Delaroche analysis, the
data currently available in EXFOR, and the labeling as reported in the original
experimental literature. The first type of discrepancy was a difference in
target nucleus listed for a given data set. In some cases, this involved a data
set being labeled as being for an elemental (natural abundance) target in one source
but for an isotopic target in another (e.g., \cuSixtyThree\ in EXFOR and the
original reference, but \cuNat\ in the KD corpus). In one case, the KD
corpus lists a neutron differential elastic scattering data set on \laNat\
at 7.8 MeV, but the cited reference lists the scattering target as \alNat,
likely a typographic error. Because both the target $A$ and asymmetry $(N-Z)/A$
enter the potential definition, errors in these metadata could bias the
potential parameters, especially those with exponential dependence on $A$.
The second type of discrepancy concerned several data sets listed in the
original KD corpus as having scattering energies we were unable to locate
either in EXFOR or in the original literature. In a few of these cases, the
issue was inaccessibility of the original literature source (e.g., unpublished
work, conference proceedings, or older PhD theses). In other cases, we located
experimental data in the original source, but could not find data points or
sets at the specified scattering energies. These data sets were omitted from
the KDUQ corpus. While the number of data sets affected by this issue was
small, because many KD potential depth parameters depend on the scattering
energy to the second and third power, any incorrect scattering energy labels
could have significant effect unless an outlier procedure, such as that
detailed in the main text, is applied.

\subsection*{Assembling the CHUQ corpus}
Table 5 of \cite{CH89} lists the four sectors of
experimental data used in the original CH89 analysis: differential elastic
scattering cross sections and analyzing powers for both neutrons and protons.
Proton data were collected at four facilities: the Triangle Universities
Nuclear Laboratory (TUNL), the Eindhoven University of Technology, Oak Ridge
National Laboratory (ORNL), and the Research Center for Nuclear Physics (RCNP)
in Osaka. Neutron data were collected at two facilities: TUNL and the Ohio
University Accelerator Laboratory (OUAL). As the authors of CH89 point out, the
fact that only a few laboratories were used reduces the likelihood of
systematic discrepancies between data sets, a potential advantage for
their OMP analysis. During assembly, the main discrepancy between the metadata
as listed the original analysis and the data in EXFOR were a handful of missing
(p,p) data sets on \niFiftyEight, \ndOneHundredFortyTwo,
\smOneHundredFortyEight, and \smOneHundredFifty\ from Eindhoven. Neither EXFOR
nor the primary literature cited for these data sets refer to these targets, or
if the target is included, these sources give a different scattering energy for
the target in question. We struck these data sets from the CHUQ corpus.
Given that these data sets comprise a small fraction of the overall
corpus, it is unlikely their omission impacted our analysis.

A more difficult obstacle was faithful assignment of experimental
uncertainties. For thirty-six of the proton data sets and twelve of the neutron
data sets used for CH89, the corresponding EXFOR entry included multiple
columns of partial experimental uncertainty. For many proton data sets, one or
more error columns consisted of almost all ``null'' values and no additional
information was available on whether the multiple columns of experimental
uncertainties were correlated or independent. For consistency across all data
sets, we dropped entirely any mostly-``null'' columns and accepted the
uncertainties from the other column as the overall uncertainty. Consequently,
for a few of the proton scattering data in the CHUQ corpus, the associated
experimental uncertainties may differ than those used in the original CH89
corpus. In the 12 neutron data sets with multiple experimental uncertainty
columns, neither column had ``null'' values.  We chose to combine the columns
in quadrature into a new ``overall uncertainty'' column. Such cases are
documented in the notes following the CHUQ corpus tables below. Regardless of
these potential differences, because we employed both outlier rejection and a
residual unaccounted-for uncertainty term in our OMP characterizations, we
expect that these differences had negligible impact on the present analysis.

Lastly, rather than fit the analyzing power data directly (as in
Koning-Delaroche), the authors of CH89 chose to fit proton analyzing power data
as ``analyzing cross sections" $A_{y}\times\sigma_{el}$ (analyzing powers times
elastic cross sections). They explained that ``this observable is preferable to
$A_{y}$, since theoretically $A_{y}$ is computed from it by dividing by
$\sigma$.'' They estimated the uncertainty of $A_{y}\times\sigma_{el}$ by
combining the individual uncertainties of $A_{y}$ and $\sigma_{el}$ according
to uncorrelated propagation of uncertainties, which they noted was justified by
a separate analysis. The practical effect was to give ``polarization
observables in the $A_{y}\times\sigma_{el}$ form\dots a more equitable
weighting in the fitting process.'' In our analysis, we chose to fit analyzing
powers directly to avoid treating neutron and proton analyzing data
differently. At low angles, the low-angle proton differential elastic cross
sections are dominated by Coulomb and are often orders of magnitude larger than
those cross sections at higher angles. Thus, low-angle proton cross sections
typically have far smaller reported statistical uncertainties than neutron
cross sections at the same angles. As such, it is possible that using
``analyzing cross sections'' could result in bias to asymmetry-dependent terms,
so we elected to fit analyzing powers directly.  However, due to our employment
of outlier-rejection and unaccounted-for uncertainties, our judgment was that
treating the analyzing powers directly versus rescaling them into ``analyzing
cross sections'' would not appreciably impact our results.

\subsection*{Assembling the Test corpus}
Our general criteria for populating the Test corpus were as follows:
\begin{itemize}
    \item Data should be directly measured, not a derived or calculated quantity
    \item Data collection should post-date the publication of the
        original KD and CH89 treatments
    \item Data should be of the same types (e.g., absolute neutron total cross sections,
        differential elastic cross sections) used in the training corpora
    \item Data should be plausibly describable by a low-to-intermediate-energy optical
        potential treatment (excluding, for example, scattering measurements at 1 \GeV,
        or proton elastic scattering measurements probing resonances below the Coulomb barrier)
    \item Data should have the minimum essential features and meta-data enabling consistent
        treatment beside the CH89 and KD corpora, including experimental uncertainties,
        and be free of plain typographic or methodological issues
\end{itemize}

We canvassed the EXFOR database for data sets subject to the above constraints
and recovered roughly 250 data sets. Among data fulfilling these criteria,
several special cases were encountered; our treatment of these data is detailed
in the Comments following each data table. Overall, we note that compared to
the four decades prior to the publication of the KD global potential
(1960-2003), the rate of scattering data measurements has declined in the last
two decades, at least as indicated by the number of new EXFOR entries. As the
reach and number of radioactive beam facilities expand, smaller cyclotron
facilities are closed, digital data acquisition becomes the norm, and funding
preferences change, the selection of scattering targets and energies has also
shifted. Due to these factors it seems likely that the data comprising the Test
Corpus represent a somewhat different underlying distribution than those
comprising the KD and CH89 corpora; analysts wishing to compare the performance
of their OMPs against these corpora should keep this in mind.

\clearpage 
\section*{KDUQ corpus}
\subsection*{Neutron differential elastic cross sections}
\twocolumngrid 
\begin{longtable}{c c c}
    Isotope & Energy & EXFOR Acc. \\ 
\mgTwentyFour & 3.4 & \href{https://www-nds.iaea.org/exfor/servlet/X4sGetSubent?subID=30463008}{30463008} \\ 
 & 6.1 & \href{https://www-nds.iaea.org/exfor/servlet/X4sGetSubent?subID=21490002}{21490002} \\ 
 & 9.76 & \href{https://www-nds.iaea.org/exfor/servlet/X4sGetSubent?subID=21773002}{21773002} \\ 
 & 14.83 & \href{https://www-nds.iaea.org/exfor/servlet/X4sGetSubent?subID=21773010}{21773010} \\ 
\hline 
\mgNat & 1.969 & \href{https://www-nds.iaea.org/exfor/servlet/X4sGetSubent?subID=11493004}{11493004} \\ 
 & 2.958 & \href{https://www-nds.iaea.org/exfor/servlet/X4sGetSubent?subID=11493004}{11493004} \\ 
 & 3.948 & \href{https://www-nds.iaea.org/exfor/servlet/X4sGetSubent?subID=11493004}{11493004} \\
\hline 

        \end{longtable}
        \label{ECS_n_bi209}

\subsection*{Comments}
\begin{itemize}
    \item[] \siTwentyEight: The data sets at 30.3 and 40 MeV from (Phys. Rev. C 28 p.2530 (1983)) listed as being for \siTwentyEight\ are actually from \siNat; we assigned them to \siNat\ for the KDUQ corpus.
    \item[] \siNat: There are 2 unique datasets at 6.4 MeV from ORNL Report 4517 (1970); which was used in the KD corpus is not specified. We kept both in the KDUQ corpus.
    \item[] \caForty: The data set at 11.0 MeV from NP/A 286 (1977) 232 was not available in EXFOR. In original literature, it was presented in a figure that was difficult to digitize; we omitted this dataset from the KDUQ corpus.
    \item[] \cuNat: We were unable to locate the data set at 155 MeV in EXFOR. For these data we used the same reference as listed in the KD corpus: R. S. Harding (Physical Review, Vol.111, p.1164 (1958)).
    \item[] \srNat: The data set at 0.9 MeV from Cox and Dowling (ANL Report No. ANL-7935 (1972)) includes no errors; we omitted this dataset from the KDUQ corpus.
    \item[] \yEightyNine: The data set at 3.8 MeV is apparently from Budtz-Jorgensen et al. (Zeitschrift f\"ur Physik A, Hadrons and Nuclei, Vol.319, p.47 (1984)), not PRC 34 (1986) 1599 as listed in the KD corpus.
    \item[] \nbNinetyThree: The data sets 2.6 and 2.9 MeV ascribed to Smith, Guenther, and Whalen (Nucl. Phys. A vol 415 issue 1, page 1 (1984) are apparently available only in an earlier, separate report by the same authors (ANL Report No. 70 (1982)). For the KDUQ corpus, we used the data as listed in the earlier report. Additionally, the data set at 14.6 MeV does not appear to exist in the reference cited by the KD corpus (Pedroni et al. (PRC 43 2336 (1991))); we omitted this dataset from the KDUQ corpus.
    \item[] \cdNat: The data set at 7.5 MeV from NP/A 568 221 (1994) is actually the sum of elastic scattering and inelastic scattering to the first excited state (this state is low-lying and could not be resolved from the ground state in the scattering experiment). This quantity was described as ``pseudo-elastic" and apparently analyzed as elastic in the original KD corpus. For consistency, we included it in the KDUQ corpus.
    \item[] \snOneHundredTwenty: The data set at 6.04 MeV from Wilenzick et al., Nucl. Phys. vol. 62, p. 511 (1965) listed as being for \snOneHundredTwenty\ is actually from \snNat; we assigned them to \snNat\ for the KDUQ corpus. The data set at 11 MeV listed in the KD Corpus as being for \snOneHundredTwenty\ apparently does not exist in Guss et al. (PRC 39 (1989) 405); we omitted this dataset from the KDUQ corpus.
    \item[] \laNat: The data set at 7.8 MeV is listed in the KD corpus as being from Dagge et al. (PRC 39 (1989) 1768), but this reference reports scattering measurements on \alTwentySeven, not \laNat. As we were unable to locate any data set in EXFOR
        at this scattering energy for \laNat\ that might correspond to the experimental data shown in Fig. 28 of the canonical KD treatment, we omitted this dataset from the KDUQ corpus.
    \item[] \ndOneHundredFortyTwo: The data set at 2.5 MeV from Bull. Am. Phys. Soc. 24 (1979) 854 includes no errors; we omitted this dataset from the KDUQ corpus.
    \item[] \ndOneHundredFortyFour: The data set at 2.5 MeV from Bull. Am. Phys. Soc. 24 (1979) 854 includes no errors; we omitted this dataset from the KDUQ corpus.
    \item[] \auOneHundredNinetySeven: The data set at 2.5 MeV from Day (Priv. comm. (1965)) includes no errors; we omitted this dataset from the KDUQ corpus.
    \item[] \hgNat: The data set listed in the KD corpus as having a scattering energy of 14.8 MeV from Nauta et al. (Nucl. Phys. 2 (1956) 124) is actually at 14.1 MeV, as reported in the original publication. We assigned a scattering energy of 14.1 \MeV\ for the KDUQ corpus.
    \item[] \pbNat: We were unable to locate the data set at 155 MeV in EXFOR. For these data we used the same reference as listed in the KD corpus: R. S. Harding (Physical Review, Vol.111, p.1164 (1958)).
    \item[] \pbTwoHundredEight: The data set at 21.6 MeV from Olsson et al. (NP/A 472 (1987) 237) was apparently from a target 88.2\% enriched in \pbTwoHundredSix, and already appears as a dataset in the KD corpus under \pbTwoHundredSix. As such, we omitted it from \pbTwoHundredEight\ in the KDUQ corpus. The data set at 155 MeV from Phys. Rev. 111 (1958) 1164) listed in the KD corpus as being for \pbTwoHundredEight\ is actually from \pbNat; we assigned it to \pbNat\ for the KDUQ corpus.
    \item[] \biTwoHundredNine: For the data sets from 2-3.53 MeV from Das and Finlay (PRC 42 (1990) 1013), we used data from Nucl. Sci. Engrg. 75 69 (1980) (reference 220 in the KD Corpus) for the KDUQ corpus. For the data sets from 4.0-7.0 MeV from Das and Finlay (PRC 42 (1990) 1013), we used data from NP/A 443 249 (1985) (reference 23 in the KD Corpus) for the KDUQ corpus.
\end{itemize}

\clearpage 
\subsection*{Neutron analyzing powers}
\twocolumngrid 
\begin{longtable}{c c c}
    Isotope & Energy & EXFOR Acc. \\ 
\alTwentySeven & 7.62 & \href{https://www-nds.iaea.org/exfor/servlet/X4sGetSubent?subID=22532004}{22532004} \\ 
 & 14 & \href{https://www-nds.iaea.org/exfor/servlet/X4sGetSubent?subID=13684002}{13684002} \\ 
 & 17 & \href{https://www-nds.iaea.org/exfor/servlet/X4sGetSubent?subID=13684002}{13684002} \\ 
\hline 
\caForty & 9.91 & \href{https://www-nds.iaea.org/exfor/servlet/X4sGetSubent?subID=12785003}{12785003} \\ 
 & 11.91 & \href{https://www-nds.iaea.org/exfor/servlet/X4sGetSubent?subID=12785003}{12785003} \\ 
 & 13.9 & \href{https://www-nds.iaea.org/exfor/servlet/X4sGetSubent?subID=12785003}{12785003} \\ 
 & 16.923 & \href{https://www-nds.iaea.org/exfor/servlet/X4sGetSubent?subID=12996004}{12996004} \\ 
\hline 
\feFiftyFour & 9.941 & \href{https://www-nds.iaea.org/exfor/servlet/X4sGetSubent?subID=12997004}{12997004} \\ 
 & 13.937 & \href{https://www-nds.iaea.org/exfor/servlet/X4sGetSubent?subID=12997004}{12997004} \\ 
 & 16.93 & \href{https://www-nds.iaea.org/exfor/servlet/X4sGetSubent?subID=12997004}{12997004} \\ 
\hline 
\niFiftyEight & 9.906 & \href{https://www-nds.iaea.org/exfor/servlet/X4sGetSubent?subID=12997008}{12997008} \\ 
 & 13.94 & \href{https://www-nds.iaea.org/exfor/servlet/X4sGetSubent?subID=12997008}{12997008} \\ 
 & 16.934 & \href{https://www-nds.iaea.org/exfor/servlet/X4sGetSubent?subID=12997008}{12997008} \\ 
\hline 
\cuSixtyFive & 9.96 & \href{https://www-nds.iaea.org/exfor/servlet/X4sGetSubent?subID=12844003}{12844003} \\ 
 & 13.9 & \href{https://www-nds.iaea.org/exfor/servlet/X4sGetSubent?subID=12844003}{12844003} \\ 
\hline 
\yEightyNine & 9.954 & \href{https://www-nds.iaea.org/exfor/servlet/X4sGetSubent?subID=12994003}{12994003} \\ 
 & 13.934 & \href{https://www-nds.iaea.org/exfor/servlet/X4sGetSubent?subID=12994003}{12994003} \\ 
 & 16.93 & \href{https://www-nds.iaea.org/exfor/servlet/X4sGetSubent?subID=12994003}{12994003} \\ 
\hline 
\nbNinetyThree & 9.941 & \href{https://www-nds.iaea.org/exfor/servlet/X4sGetSubent?subID=12995003}{12995003} \\ 
 & 13.915 & \href{https://www-nds.iaea.org/exfor/servlet/X4sGetSubent?subID=12995003}{12995003} \\ 
\hline 
\snOneHundredTwenty & 9.907 & \href{https://www-nds.iaea.org/exfor/servlet/X4sGetSubent?subID=13158002}{13158002} \\ 
 & 13.894 & \href{https://www-nds.iaea.org/exfor/servlet/X4sGetSubent?subID=13158002}{13158002} \\ 
\hline 
\pbTwoHundredEight & 1.8 & \href{https://www-nds.iaea.org/exfor/servlet/X4sGetSubent?subID=400750051}{400750051} \\ 
 & 5.969 & \href{https://www-nds.iaea.org/exfor/servlet/X4sGetSubent?subID=13531003}{13531003} \\ 
 & 6.967 & \href{https://www-nds.iaea.org/exfor/servlet/X4sGetSubent?subID=13531003}{13531003} \\ 
 & 7.962 & \href{https://www-nds.iaea.org/exfor/servlet/X4sGetSubent?subID=13531003}{13531003} \\ 
 & 8.958 & \href{https://www-nds.iaea.org/exfor/servlet/X4sGetSubent?subID=13531003}{13531003} \\ 
 & 9.95 & \href{https://www-nds.iaea.org/exfor/servlet/X4sGetSubent?subID=13531003}{13531003} \\ 
\hline 
\biTwoHundredNine & 4.5 & \href{https://www-nds.iaea.org/exfor/servlet/X4sGetSubent?subID=10855002}{10855002} \\ 
 & 6 & \href{https://www-nds.iaea.org/exfor/servlet/X4sGetSubent?subID=13680002}{13680002} \\ 
 & 9 & \href{https://www-nds.iaea.org/exfor/servlet/X4sGetSubent?subID=13680002}{13680002} \\ 
\hline 

        \end{longtable}
        \label{APower_n_bi209}

\subsection*{Comments}
No discrepancies were identified between the KD and KDUQ corpora for neutron
analyzing powers. For related information, see the Comments for neutron
differential elastic cross sections above (note that the formatting of the
original KD corpus did not explicitly distinguish between inclusion of
differential elastic cross section data sets and analyzing power data sets).

\clearpage 
\subsection*{Neutron total cross sections}
\twocolumngrid 
\begin{longtable}{c c c}
    Isotope & Energy & EXFOR Acc. \\ 
    \mgNat & 5.293-297.8 & \href{https://www-nds.iaea.org/exfor/servlet/X4sGetSubent?subID=13753010}{13753010} \\ 
           & 0.008-39.807 & \href{https://www-nds.iaea.org/exfor/servlet/X4sGetSubent?subID=10791002}{10791002} \\ 
    \hline 
    \alTwentySeven & 1.999-80.62 & \href{https://www-nds.iaea.org/exfor/servlet/X4sGetSubent?subID=12882005}{12882005} \\ 
                   & 5.293-297.772 & \href{https://www-nds.iaea.org/exfor/servlet/X4sGetSubent?subID=13569008}{13569008} \\ 
                   & 0.935 & \href{https://www-nds.iaea.org/exfor/servlet/X4sGetSubent?subID=21926003}{21926003} \\ 
                   & 0.25-19.286 & \href{https://www-nds.iaea.org/exfor/servlet/X4sGetSubent?subID=22331004}{22331004} \\ 
    \hline 
    \siNat & 0.187-47.68 & \href{https://www-nds.iaea.org/exfor/servlet/X4sGetSubent?subID=10377005}{10377005} \\ 
           & 1.996-79.828 & \href{https://www-nds.iaea.org/exfor/servlet/X4sGetSubent?subID=12882006}{12882006} \\ 
           & 5.293-297.772 & \href{https://www-nds.iaea.org/exfor/servlet/X4sGetSubent?subID=13569009}{13569009} \\ 
    \hline 
    \sNat & 2.259-14.888 & \href{https://www-nds.iaea.org/exfor/servlet/X4sGetSubent?subID=10047018}{10047018} \\ 
          & 0.102 & \href{https://www-nds.iaea.org/exfor/servlet/X4sGetSubent?subID=11540002}{11540002} \\ 
          & 5.293-297.8 & \href{https://www-nds.iaea.org/exfor/servlet/X4sGetSubent?subID=13753012}{13753012} \\ 
    \hline 
    \caForty & 0.04-6.058 & \href{https://www-nds.iaea.org/exfor/servlet/X4sGetSubent?subID=10721002}{10721002} \\ 
             & 5.293-297.772 & \href{https://www-nds.iaea.org/exfor/servlet/X4sGetSubent?subID=13569010}{13569010} \\ 
    \hline 
    \tiFortyEight & 0.974-4.025 & \href{https://www-nds.iaea.org/exfor/servlet/X4sGetSubent?subID=10669002}{10669002} \\ 
                  & 5.293-297.8 & \href{https://www-nds.iaea.org/exfor/servlet/X4sGetSubent?subID=13753015}{13753015} \\ 
    \hline 
    \crNat & 0.185-29.306 & \href{https://www-nds.iaea.org/exfor/servlet/X4sGetSubent?subID=10342004}{10342004} \\ 
    \hline 
    \crFiftyTwo & 1.0-4.15 & \href{https://www-nds.iaea.org/exfor/servlet/X4sGetSubent?subID=12750002}{12750002} \\ 
                & 5.293-297.8 & \href{https://www-nds.iaea.org/exfor/servlet/X4sGetSubent?subID=13753017}{13753017} \\ 
                & 0.021 & \href{https://www-nds.iaea.org/exfor/servlet/X4sGetSubent?subID=20435003}{20435003} \\ 
    \hline 
    \feNat & 2.268 & \href{https://www-nds.iaea.org/exfor/servlet/X4sGetSubent?subID=13727002}{13727002} \\ 
    \hline 
    \feFiftySix & 0.187-48.608 & \href{https://www-nds.iaea.org/exfor/servlet/X4sGetSubent?subID=10377007}{10377007} \\ 
                & 5.293-297.8 & \href{https://www-nds.iaea.org/exfor/servlet/X4sGetSubent?subID=13753019}{13753019} \\ 
                & 1.0-18.026 & \href{https://www-nds.iaea.org/exfor/servlet/X4sGetSubent?subID=22258002}{22258002} \\ 
    \hline 
    \niFiftyEight & 0.003-67.481 & \href{https://www-nds.iaea.org/exfor/servlet/X4sGetSubent?subID=12972008}{12972008} \\ 
                  & 5.293-297.8 & \href{https://www-nds.iaea.org/exfor/servlet/X4sGetSubent?subID=13753021}{13753021} \\ 
                  & 0.5-19.54 & \href{https://www-nds.iaea.org/exfor/servlet/X4sGetSubent?subID=22314006}{22314006} \\ 
    \hline 
    \cuNat & 1.2-4.5 & \href{https://www-nds.iaea.org/exfor/servlet/X4sGetSubent?subID=12869002}{12869002} \\ 
           & 5.293-297.772 & \href{https://www-nds.iaea.org/exfor/servlet/X4sGetSubent?subID=13569011}{13569011} \\ 
    \hline 
    \yEightyNine & 1.822-19.52 & \href{https://www-nds.iaea.org/exfor/servlet/X4sGetSubent?subID=12853010}{12853010} \\ 
                 & 5.293-297.8 & \href{https://www-nds.iaea.org/exfor/servlet/X4sGetSubent?subID=13753022}{13753022} \\ 
    \hline 
    \zrNinety & 0.433-1.54 & \href{https://www-nds.iaea.org/exfor/servlet/X4sGetSubent?subID=10338004}{10338004} \\ 
              & 0.933-5.054 & \href{https://www-nds.iaea.org/exfor/servlet/X4sGetSubent?subID=10468002}{10468002} \\ 
              & 5.293-297.772 & \href{https://www-nds.iaea.org/exfor/servlet/X4sGetSubent?subID=13569012}{13569012} \\ 
              & 2.349 & \href{https://www-nds.iaea.org/exfor/servlet/X4sGetSubent?subID=13736003}{13736003} \\ 
    \hline 
    \nbNinetyThree & 0.75-3.963 & \href{https://www-nds.iaea.org/exfor/servlet/X4sGetSubent?subID=12797002}{12797002} \\ 
                   & 0.215-1.32 & \href{https://www-nds.iaea.org/exfor/servlet/X4sGetSubent?subID=12853018}{12853018} \\ 
                   & 5.293-297.772 & \href{https://www-nds.iaea.org/exfor/servlet/X4sGetSubent?subID=13569013}{13569013} \\ 
    \hline 
    \moNat & 2.253-14.713 & \href{https://www-nds.iaea.org/exfor/servlet/X4sGetSubent?subID=10047047}{10047047} \\ 
           & 0.103-1.105 & \href{https://www-nds.iaea.org/exfor/servlet/X4sGetSubent?subID=10277034}{10277034} \\ 
           & 0.215-1.32 & \href{https://www-nds.iaea.org/exfor/servlet/X4sGetSubent?subID=12853021}{12853021} \\ 
           & 1.822-19.28 & \href{https://www-nds.iaea.org/exfor/servlet/X4sGetSubent?subID=12853022}{12853022} \\ 
           & 0.985-4.195 & \href{https://www-nds.iaea.org/exfor/servlet/X4sGetSubent?subID=12853023}{12853023} \\ 
           & 5.293-297.8 & \href{https://www-nds.iaea.org/exfor/servlet/X4sGetSubent?subID=13753023}{13753023} \\ 
    \hline 
    \snNat & 0.002 & \href{https://www-nds.iaea.org/exfor/servlet/X4sGetSubent?subID=10639005}{10639005} \\ 
           & 0.215-1.32 & \href{https://www-nds.iaea.org/exfor/servlet/X4sGetSubent?subID=12853042}{12853042} \\ 
           & 5.293-297.772 & \href{https://www-nds.iaea.org/exfor/servlet/X4sGetSubent?subID=13569014}{13569014} \\ 
    \hline 
    \ceNat & 2.26-14.866 & \href{https://www-nds.iaea.org/exfor/servlet/X4sGetSubent?subID=10047062}{10047062} \\ 
           & 17.5-27.8 & \href{https://www-nds.iaea.org/exfor/servlet/X4sGetSubent?subID=11108033}{11108033} \\ 
           & 2.253-56.92 & \href{https://www-nds.iaea.org/exfor/servlet/X4sGetSubent?subID=12891002}{12891002} \\ 
           & 0.182 & \href{https://www-nds.iaea.org/exfor/servlet/X4sGetSubent?subID=20819002}{20819002} \\ 
           & 160.0-280.0 & \href{https://www-nds.iaea.org/exfor/servlet/X4sGetSubent?subID=22117011}{22117011} \\ 
           & 1.013 & \href{https://www-nds.iaea.org/exfor/servlet/X4sGetSubent?subID=30134005}{30134005} \\ 
    \hline 
    \auOneHundredNinetySeven & 0.048-4.389 & \href{https://www-nds.iaea.org/exfor/servlet/X4sGetSubent?subID=10935003}{10935003} \\ 
                             & 5.293-297.8 & \href{https://www-nds.iaea.org/exfor/servlet/X4sGetSubent?subID=13753026}{13753026} \\ 
    \hline 
    \hgNat & 2.255-14.873 & \href{https://www-nds.iaea.org/exfor/servlet/X4sGetSubent?subID=10047087}{10047087} \\ 
           & 0.04 & \href{https://www-nds.iaea.org/exfor/servlet/X4sGetSubent?subID=11953009}{11953009} \\ 
           & 5.293-297.8 & \href{https://www-nds.iaea.org/exfor/servlet/X4sGetSubent?subID=13753027}{13753027} \\ 
           & 1.007-2.007 & \href{https://www-nds.iaea.org/exfor/servlet/X4sGetSubent?subID=30134010}{30134010} \\ 
    \hline 
    \pbTwoHundredEight & 2.491-14.137 & \href{https://www-nds.iaea.org/exfor/servlet/X4sGetSubent?subID=10047089}{10047089} \\ 
                       & 0.717-1.719 & \href{https://www-nds.iaea.org/exfor/servlet/X4sGetSubent?subID=12215003}{12215003} \\ 
                       & 17.665 & \href{https://www-nds.iaea.org/exfor/servlet/X4sGetSubent?subID=13735002}{13735002} \\ 
                       & 5.293-297.772 & \href{https://www-nds.iaea.org/exfor/servlet/X4sGetSubent?subID=13569018}{13569018} \\ 
    \hline 
    \biTwoHundredNine & 2.376-13.977 & \href{https://www-nds.iaea.org/exfor/servlet/X4sGetSubent?subID=10047093}{10047093} \\ 
                      & 0.011-1.013 & \href{https://www-nds.iaea.org/exfor/servlet/X4sGetSubent?subID=10449003}{10449003} \\ 
                      & 1.237-3.353 & \href{https://www-nds.iaea.org/exfor/servlet/X4sGetSubent?subID=108460021}{108460021} \\ 
                      & 69.547 & \href{https://www-nds.iaea.org/exfor/servlet/X4sGetSubent?subID=13199004}{13199004} \\ 
                      & 5.293-297.772 & \href{https://www-nds.iaea.org/exfor/servlet/X4sGetSubent?subID=13569019}{13569019} \\ 
    \hline 

\end{longtable}
\label{TCS_n_bi209}

\subsection*{Comments}
\begin{itemize}
    \item[] \mgNat: a data set from the \keV\ range to 39 MeV is referenced in the KD corpus as being from Lawson et al. (ORNL Report 6420), but these data were not available on EXFOR. For the KDUQ corpus, we used what are ostensibly the same data as published by Weigmann et al. (PRC 14 (1976) 1328).
    \item[] \alTwentySeven: The data for this nucleus we sourced from EXFOR, where they are assigned slightly different references compared to the references given in the KD corpus (the scattering energy ranges
        listed in EXFOR match those as shown in the KD corpus).
    \item[] \siNat: The data set from Larson et al. (ORNL Report No. ORNL-TM-5618 (1976)) only has data up to 0.7 MeV, making it of limited value to OMP analysis that assumes a degree of smoothness in the imaginary strength not fulfilled in the resolved-resonance region. As such, we omitted this from the KDUQ corpus. The remaining TCS data we sourced from EXFOR, where they are listed under slightly different references compared to the KD corpus.
    \item[] \sNat: The KD corpus refers to Finlay et al. (PRC 47 (1993) 237) for this data set, but this reference has no data for natural S. We assume that the KD corpus meant to reference Abfalterer et al. (PRC 63 (2001) 44608), the other large-scale experimental campaign of neutron total cross sections conducted at LANSCE. In the KDUQ corpus, we use the data from that source.
    \item[] \caForty: the data set from Johnson et al. (Bull. Am. Phys. Soc. 23 (1978) 636) is on EXFOR listed under Int. Conf. on Nucl. Phys., Munich 1 525 (1973); we used the latter reference and data set for the KDUQ corpus.
    \item[] \crFiftyTwo: the data set from Perey (Priv. Comm. (1973)) listed as being for \crFiftyTwo\ in the KD corpus is actually for \crNat; we assigned it to \crNat\ for the KDUQ corpus.
    \item[] \feFiftySix: the data set from U.S. D.O.E. Nuclear Data Committee Reports, No.33, p.142 (1984) listed as being for \feFiftySix\ is actually for \feNat; we assigned it to \feNat\ for the KDUQ corpus.
    \item[] \niFiftyEight: EXFOR doesn't appear to contain the data set with maximum scattering energy of 67.5 \MeV\ from Perey et al. (ORNL Report No. ORNL-TM-10841 (1988)); for the KDUQ corpus, we used the data set from Perey et al. (PRC 47 (1993) 1143).
    \item[] \cuNat: the data set from Pandey et al. (PRC 15 (1977) 600) only has data up to 1.12 MeV for \cuSixtyThree and \cuSixtyFive, making it of limited value to OMP analysis that assumes some degree of smoothness in the imaginary strength not fulfilled in the resonance region. As such, we omitted these from the KDUQ corpus.
    \item[] \ceNat: We sourced these data from EXFOR, where they are assigned slightly different references compared to the KD corpus.
\end{itemize}

\clearpage 
\subsection*{Proton differential elastic cross sections}
\twocolumngrid 
\begin{longtable}{c c c}
    Isotope & Energy & EXFOR Acc. \\ 
\alTwentySeven & 17.0 & \href{https://www-nds.iaea.org/exfor/servlet/X4sGetSubent?subID=O0262004}{O0262004} \\ 
 & 28.0 & \href{https://www-nds.iaea.org/exfor/servlet/X4sGetSubent?subID=C1208002}{C1208002} \\ 
 & 35.2 & \href{https://www-nds.iaea.org/exfor/servlet/X4sGetSubent?subID=O0083019}{O0083019} \\ 
 & 61.4 & \href{https://www-nds.iaea.org/exfor/servlet/X4sGetSubent?subID=O0211003}{O0211003} \\ 
 & 142.0 & \href{https://www-nds.iaea.org/exfor/servlet/X4sGetSubent?subID=O0247006}{O0247006} \\ 
 & 156.0 & \href{https://www-nds.iaea.org/exfor/servlet/X4sGetSubent?subID=O0049003}{O0049003} \\ 
 & 160.0 & \href{https://www-nds.iaea.org/exfor/servlet/X4sGetSubent?subID=D0283004}{D0283004} \\ 
 & 183.0 & \href{https://www-nds.iaea.org/exfor/servlet/X4sGetSubent?subID=O0365008}{O0365008} \\ 
\hline 
\siTwentyEight & 17.8 & \href{https://www-nds.iaea.org/exfor/servlet/X4sGetSubent?subID=O0254009}{O0254009} \\ 
 & 28.0 & \href{https://www-nds.iaea.org/exfor/servlet/X4sGetSubent?subID=C1208003}{C1208003} \\ 
 & 30.4 & \href{https://www-nds.iaea.org/exfor/servlet/X4sGetSubent?subID=O0124002}{O0124002} \\ 
 & 40.0 & \href{https://www-nds.iaea.org/exfor/servlet/X4sGetSubent?subID=O0162021}{O0162021} \\ 
 & 65.0 & \href{https://www-nds.iaea.org/exfor/servlet/X4sGetSubent?subID=E0166008}{E0166008} \\ 
 & 80 & \href{https://www-nds.iaea.org/exfor/servlet/X4sGetSubent?subID=C0084002}{C0084002} \\ 
 & 100 & \href{https://www-nds.iaea.org/exfor/servlet/X4sGetSubent?subID=C0084002}{C0084002} \\ 
 & 135 & \href{https://www-nds.iaea.org/exfor/servlet/X4sGetSubent?subID=C0084002}{C0084002} \\ 
 & 179 & \href{https://www-nds.iaea.org/exfor/servlet/X4sGetSubent?subID=C0084002}{C0084002} \\ 
 & 180.0 & \href{https://www-nds.iaea.org/exfor/servlet/X4sGetSubent?subID=C0152002}{C0152002} \\ 
\hline 
\caForty & 16.0 & \href{https://www-nds.iaea.org/exfor/servlet/X4sGetSubent?subID=C0893002}{C0893002} \\ 
 & 40.0 & \href{https://www-nds.iaea.org/exfor/servlet/X4sGetSubent?subID=O02080061}{O02080061} \\ 
 & 45.5 & \href{https://www-nds.iaea.org/exfor/servlet/X4sGetSubent?subID=C0076002}{C0076002} \\ 
 & 61.4 & \href{https://www-nds.iaea.org/exfor/servlet/X4sGetSubent?subID=O0211012}{O0211012} \\ 
 & 65.0 & \href{https://www-nds.iaea.org/exfor/servlet/X4sGetSubent?subID=O0032002S}{O0032002S} \\ 
 & 75.0 & \href{https://www-nds.iaea.org/exfor/servlet/X4sGetSubent?subID=O0553010}{O0553010} \\ 
 & 80.0 & \href{https://www-nds.iaea.org/exfor/servlet/X4sGetSubent?subID=T0101002}{T0101002} \\ 
 & 135.0 & \href{https://www-nds.iaea.org/exfor/servlet/X4sGetSubent?subID=T0101002}{T0101002} \\ 
 & 152.0 & \href{https://www-nds.iaea.org/exfor/servlet/X4sGetSubent?subID=O0553004}{O0553004} \\ 
 & 156.0 & \href{https://www-nds.iaea.org/exfor/servlet/X4sGetSubent?subID=O0049004}{O0049004} \\ 
 & 160.0 & \href{https://www-nds.iaea.org/exfor/servlet/X4sGetSubent?subID=T0101002}{T0101002} \\ 
 & 181.5 & \href{https://www-nds.iaea.org/exfor/servlet/X4sGetSubent?subID=T0108005}{T0108005} \\ 
 & 201.4 & \href{https://www-nds.iaea.org/exfor/servlet/X4sGetSubent?subID=C0148006}{C0148006} \\ 
\hline 
\feFiftyFour & 9.69 & \href{https://www-nds.iaea.org/exfor/servlet/X4sGetSubent?subID=O03930021}{O03930021} \\ 
 & 12.0 & \href{https://www-nds.iaea.org/exfor/servlet/X4sGetSubent?subID=C1024002}{C1024002} \\ 
 & 16.0 & \href{https://www-nds.iaea.org/exfor/servlet/X4sGetSubent?subID=C0893008}{C0893008} \\ 
 & 17.2 & \href{https://www-nds.iaea.org/exfor/servlet/X4sGetSubent?subID=O0091002}{O0091002} \\ 
 & 18.6 & \href{https://www-nds.iaea.org/exfor/servlet/X4sGetSubent?subID=D0286015}{D0286015} \\ 
 & 19.6 & \href{https://www-nds.iaea.org/exfor/servlet/X4sGetSubent?subID=O0079006}{O0079006} \\ 
 & 20.4 & \href{https://www-nds.iaea.org/exfor/servlet/X4sGetSubent?subID=O0091004}{O0091004} \\ 
 & 24.6 & \href{https://www-nds.iaea.org/exfor/servlet/X4sGetSubent?subID=O0091006}{O0091006} \\ 
 & 30.4 & \href{https://www-nds.iaea.org/exfor/servlet/X4sGetSubent?subID=O12430031}{O12430031} \\ 
 & 35.2 & \href{https://www-nds.iaea.org/exfor/servlet/X4sGetSubent?subID=O1198020}{O1198020} \\ 
 & 39.8 & \href{https://www-nds.iaea.org/exfor/servlet/X4sGetSubent?subID=O0456002}{O0456002} \\ 
 & 49.35 & \href{https://www-nds.iaea.org/exfor/servlet/X4sGetSubent?subID=O0788007}{O0788007} \\ 
 & 65.0 & \href{https://www-nds.iaea.org/exfor/servlet/X4sGetSubent?subID=O0032061S}{O0032061S} \\ 
\hline 
\feNat & 182.4 & \href{https://www-nds.iaea.org/exfor/servlet/X4sGetSubent?subID=O0365012}{O0365012} \\ 
\hline 
\feFiftySix & 10.93 & - \\ 
 & 11.7 & - \\ 
 & 16.0 & \href{https://www-nds.iaea.org/exfor/servlet/X4sGetSubent?subID=C0893010}{C0893010} \\ 
 & 18.6 & \href{https://www-nds.iaea.org/exfor/servlet/X4sGetSubent?subID=D0286016}{D0286016} \\ 
 & 19.1 & \href{https://www-nds.iaea.org/exfor/servlet/X4sGetSubent?subID=O0167006}{O0167006} \\ 
 & 30.3 & \href{https://www-nds.iaea.org/exfor/servlet/X4sGetSubent?subID=O0142005}{O0142005} \\ 
 & 35.2 & \href{https://www-nds.iaea.org/exfor/servlet/X4sGetSubent?subID=O0083016}{O0083016} \\ 
 & 39.8 & \href{https://www-nds.iaea.org/exfor/servlet/X4sGetSubent?subID=O0456003}{O0456003} \\ 
 & 49.35 & \href{https://www-nds.iaea.org/exfor/servlet/X4sGetSubent?subID=O0788006}{O0788006} \\ 
 & 65.0 & \href{https://www-nds.iaea.org/exfor/servlet/X4sGetSubent?subID=E1201002}{E1201002} \\ 
 & 156.0 & \href{https://www-nds.iaea.org/exfor/servlet/X4sGetSubent?subID=O0049006}{O0049006} \\ 
\hline 
\niFiftyEight & 10.7 & \href{https://www-nds.iaea.org/exfor/servlet/X4sGetSubent?subID=O0446002}{O0446002} \\ 
 & 11.7 & - \\ 
 & 14.4 & \href{https://www-nds.iaea.org/exfor/servlet/X4sGetSubent?subID=O0446002}{O0446002} \\ 
 & 15.4 & \href{https://www-nds.iaea.org/exfor/servlet/X4sGetSubent?subID=O0446002}{O0446002} \\ 
 & 16.0 & \href{https://www-nds.iaea.org/exfor/servlet/X4sGetSubent?subID=C0893012}{C0893012} \\ 
 & 17.0 & \href{https://www-nds.iaea.org/exfor/servlet/X4sGetSubent?subID=O0262007}{O0262007} \\ 
 & 18.6 & \href{https://www-nds.iaea.org/exfor/servlet/X4sGetSubent?subID=D0286017}{D0286017} \\ 
 & 21.3 & \href{https://www-nds.iaea.org/exfor/servlet/X4sGetSubent?subID=O0434008}{O0434008} \\ 
 & 22.2 & \href{https://www-nds.iaea.org/exfor/servlet/X4sGetSubent?subID=C1019017}{C1019017} \\ 
 & 30.3 & \href{https://www-nds.iaea.org/exfor/servlet/X4sGetSubent?subID=O0142007}{O0142007} \\ 
 & 35.2 & \href{https://www-nds.iaea.org/exfor/servlet/X4sGetSubent?subID=O1198021}{O1198021} \\ 
 & 39.6 & \href{https://www-nds.iaea.org/exfor/servlet/X4sGetSubent?subID=O04360031}{O04360031} \\ 
 & 40.0 & \href{https://www-nds.iaea.org/exfor/servlet/X4sGetSubent?subID=O02080071}{O02080071} \\ 
 & 61.4 & \href{https://www-nds.iaea.org/exfor/servlet/X4sGetSubent?subID=O0211013}{O0211013} \\ 
 & 65.0 & \href{https://www-nds.iaea.org/exfor/servlet/X4sGetSubent?subID=O0032008S}{O0032008S} \\ 
 & 100.4 & \href{https://www-nds.iaea.org/exfor/servlet/X4sGetSubent?subID=O0300003}{O0300003} \\ 
 & 160.0 & \href{https://www-nds.iaea.org/exfor/servlet/X4sGetSubent?subID=O0302005}{O0302005} \\ 
 & 178.0 & \href{https://www-nds.iaea.org/exfor/servlet/X4sGetSubent?subID=D0189002}{D0189002} \\ 
 & 192.0 & \href{https://www-nds.iaea.org/exfor/servlet/X4sGetSubent?subID=E1704002}{E1704002} \\ 
\hline 
\niSixty & 14.4 & \href{https://www-nds.iaea.org/exfor/servlet/X4sGetSubent?subID=O0446003}{O0446003} \\ 
 & 15.4 & \href{https://www-nds.iaea.org/exfor/servlet/X4sGetSubent?subID=O0446003}{O0446003} \\ 
 & 16.0 & \href{https://www-nds.iaea.org/exfor/servlet/X4sGetSubent?subID=C0893014}{C0893014} \\ 
 & 18.6 & \href{https://www-nds.iaea.org/exfor/servlet/X4sGetSubent?subID=D0286018}{D0286018} \\ 
 & 30.3 & \href{https://www-nds.iaea.org/exfor/servlet/X4sGetSubent?subID=O0142008}{O0142008} \\ 
 & 30.8 & \href{https://www-nds.iaea.org/exfor/servlet/X4sGetSubent?subID=O0157002}{O0157002} \\ 
 & 39.6 & \href{https://www-nds.iaea.org/exfor/servlet/X4sGetSubent?subID=O04360021}{O04360021} \\ 
 & 65.0 & \href{https://www-nds.iaea.org/exfor/servlet/X4sGetSubent?subID=O0032063S}{O0032063S} \\ 
\hline 
\zrNinety & 9.7 & \href{https://www-nds.iaea.org/exfor/servlet/X4sGetSubent?subID=O03930071}{O03930071} \\ 
 & 12.7 & \href{https://www-nds.iaea.org/exfor/servlet/X4sGetSubent?subID=O03890021}{O03890021} \\ 
 & 14.71 & \href{https://www-nds.iaea.org/exfor/servlet/X4sGetSubent?subID=O0372013}{O0372013} \\ 
 & 16.0 & \href{https://www-nds.iaea.org/exfor/servlet/X4sGetSubent?subID=C0893028}{C0893028} \\ 
 & 18.8 & \href{https://www-nds.iaea.org/exfor/servlet/X4sGetSubent?subID=O0370002}{O0370002} \\ 
 & 22.5 & \href{https://www-nds.iaea.org/exfor/servlet/X4sGetSubent?subID=C0085002}{C0085002} \\ 
 & 30.0 & \href{https://www-nds.iaea.org/exfor/servlet/X4sGetSubent?subID=D0295002}{D0295002} \\ 
 & 40.0 & \href{https://www-nds.iaea.org/exfor/servlet/X4sGetSubent?subID=O02080081}{O02080081} \\ 
 & 49.35 & \href{https://www-nds.iaea.org/exfor/servlet/X4sGetSubent?subID=O0788017}{O0788017} \\ 
 & 61.4 & \href{https://www-nds.iaea.org/exfor/servlet/X4sGetSubent?subID=O0211015}{O0211015} \\ 
 & 65.0 & \href{https://www-nds.iaea.org/exfor/servlet/X4sGetSubent?subID=O0032021S}{O0032021S} \\ 
 & 80.0 & \href{https://www-nds.iaea.org/exfor/servlet/X4sGetSubent?subID=T0101003}{T0101003} \\ 
 & 100.4 & \href{https://www-nds.iaea.org/exfor/servlet/X4sGetSubent?subID=O0300004}{O0300004} \\ 
 & 135.0 & \href{https://www-nds.iaea.org/exfor/servlet/X4sGetSubent?subID=T0101003}{T0101003} \\ 
 & 156.0 & \href{https://www-nds.iaea.org/exfor/servlet/X4sGetSubent?subID=O0049009}{O0049009} \\ 
 & 160.0 & \href{https://www-nds.iaea.org/exfor/servlet/X4sGetSubent?subID=T0101003}{T0101003} \\ 
\hline 
\snOneHundredTwenty & 9.7 & \href{https://www-nds.iaea.org/exfor/servlet/X4sGetSubent?subID=O03930081}{O03930081} \\ 
 & 16.0 & \href{https://www-nds.iaea.org/exfor/servlet/X4sGetSubent?subID=C0893032}{C0893032} \\ 
 & 20.4 & \href{https://www-nds.iaea.org/exfor/servlet/X4sGetSubent?subID=O0169020}{O0169020} \\ 
 & 24.6 & \href{https://www-nds.iaea.org/exfor/servlet/X4sGetSubent?subID=O0169026}{O0169026} \\ 
 & 30.3 & \href{https://www-nds.iaea.org/exfor/servlet/X4sGetSubent?subID=O0142010}{O0142010} \\ 
 & 40.0 & \href{https://www-nds.iaea.org/exfor/servlet/X4sGetSubent?subID=O0328007}{O0328007} \\ 
 & 100.4 & \href{https://www-nds.iaea.org/exfor/servlet/X4sGetSubent?subID=O0300005}{O0300005} \\ 
 & 156.0 & \href{https://www-nds.iaea.org/exfor/servlet/X4sGetSubent?subID=O0049010}{O0049010} \\ 
 & 160.0 & \href{https://www-nds.iaea.org/exfor/servlet/X4sGetSubent?subID=O0479004}{O0479004} \\ 
\hline 
\pbTwoHundredEight & 16.0 & \href{https://www-nds.iaea.org/exfor/servlet/X4sGetSubent?subID=C0893042}{C0893042} \\ 
 & 21.0 & \href{https://www-nds.iaea.org/exfor/servlet/X4sGetSubent?subID=O0287011}{O0287011} \\ 
 & 24.1 & \href{https://www-nds.iaea.org/exfor/servlet/X4sGetSubent?subID=O0287012}{O0287012} \\ 
 & 26.3 & \href{https://www-nds.iaea.org/exfor/servlet/X4sGetSubent?subID=O0287013}{O0287013} \\ 
 & 30.3 & \href{https://www-nds.iaea.org/exfor/servlet/X4sGetSubent?subID=O0142011}{O0142011} \\ 
 & 35.0 & \href{https://www-nds.iaea.org/exfor/servlet/X4sGetSubent?subID=O0287015}{O0287015} \\ 
 & 40.0 & \href{https://www-nds.iaea.org/exfor/servlet/X4sGetSubent?subID=O02080091}{O02080091} \\ 
 & 45.0 & \href{https://www-nds.iaea.org/exfor/servlet/X4sGetSubent?subID=O0287016}{O0287016} \\ 
 & 47.3 & \href{https://www-nds.iaea.org/exfor/servlet/X4sGetSubent?subID=O0287017}{O0287017} \\ 
 & 49.35 & \href{https://www-nds.iaea.org/exfor/servlet/X4sGetSubent?subID=O0788009}{O0788009} \\ 
 & 61.4 & \href{https://www-nds.iaea.org/exfor/servlet/X4sGetSubent?subID=O0211017}{O0211017} \\ 
 & 65.0 & \href{https://www-nds.iaea.org/exfor/servlet/X4sGetSubent?subID=O0032025S}{O0032025S} \\ 
 & 80.0 & \href{https://www-nds.iaea.org/exfor/servlet/X4sGetSubent?subID=T0101004}{T0101004} \\ 
 & 121.0 & \href{https://www-nds.iaea.org/exfor/servlet/X4sGetSubent?subID=T0101004}{T0101004} \\ 
 & 156.0 & \href{https://www-nds.iaea.org/exfor/servlet/X4sGetSubent?subID=O0049012}{O0049012} \\ 
 & 160.0 & \href{https://www-nds.iaea.org/exfor/servlet/X4sGetSubent?subID=O0302002}{O0302002} \\ 
 & 182.0 & \href{https://www-nds.iaea.org/exfor/servlet/X4sGetSubent?subID=T0101004}{T0101004} \\ 
 & 185.0 & \href{https://www-nds.iaea.org/exfor/servlet/X4sGetSubent?subID=O0287002}{O0287002} \\ 
 & 200.0 & \href{https://www-nds.iaea.org/exfor/servlet/X4sGetSubent?subID=C0081005}{C0081005} \\ 
\hline 
\biTwoHundredNine & 16.0 & \href{https://www-nds.iaea.org/exfor/servlet/X4sGetSubent?subID=C0893044}{C0893044} \\ 
 & 65.0 & \href{https://www-nds.iaea.org/exfor/servlet/X4sGetSubent?subID=E0773008}{E0773008} \\ 
 & 78.0 & \href{https://www-nds.iaea.org/exfor/servlet/X4sGetSubent?subID=O0553014}{O0553014} \\ 
 & 153.0 & \href{https://www-nds.iaea.org/exfor/servlet/X4sGetSubent?subID=O0553006}{O0553006} \\ 
 & 156.0 & \href{https://www-nds.iaea.org/exfor/servlet/X4sGetSubent?subID=O0049013}{O0049013} \\ 
\hline 

        \end{longtable}
        \label{ECS_p_bi209}

\subsection*{Comments}
\begin{itemize}
    \item[] \siTwentyEight: The data set listed in the KD corpus as having a scattering energy of 198.1 MeV and being from Olmer et al. (PRC 29 (1984) 361) does not appear in the referenced publication; instead, there are data sets at 179 MeV for elastic scattering and 175 MeV for analyzing powers that are listed in EXFOR for this reference. We used these latter data sets in the KDUQ corpus. The data set at 134.2 MeV and listed in the KD corpus as being from Schwandt et al. (PRC 26 (1982) 55) is instead listed in EXFOR as being from Olmer et al. (PRC 29 (1984) 361). We used this latter reference for the KDUQ corpus.
    \item[] \caForty: The data sets at 14.5, 18.6 and 21 MeV from Boschitz, Bercaw, and Vincent (Phys. Lett. 13 (1964) 322) were not available in EXFOR and apparently have not been digitized. We omitted these data sets from the KDUQ corpus.
    \item[] \feFiftyFour: The data set at 11 MeV from Beneviste, Mitchel, and Fulmer (Phys. Rev. 133 (1964) B317) listed in the KD corpus as being from \feFiftyFour\ is actually from \feFiftySix; it already is referenced under \feFiftySix\ in the KD corpus. We included it as being from \feFiftySix\ in the KDUQ corpus. The data set at 35.0 MeV from Colombo et al. (J. Phys. Soc. Jpn. 44 (1978) 543) was not available in EXFOR; for the KDUQ corpus, we used the EXFOR data set at 35.2 MeV associated with most of the same authors (PRC 21 844 (1980)).
    \item[] \feFiftySix: The data sets at 17.2, 20.4, and 24.6 listed in the KD corpus as being from J.P.M.G. Melssen's PhD thesis (Technische Hogeschool Eindhoven, 1978) are instead listed in EXFOR as being from Van Hall et al. (NP/A 291 63 (1977)); we used this latter reference and its data sets for the KDUQ corpus. The data set listed in the KD corpus as having a scattering energy of 15.3 MeV was not available in the listed Van Hall et al. reference; as such, we omitted it from the KDUQ corpus. The data set at 14.5 MeV from Rosen et al. (Ann. Phys. (N.Y.) 34 (1965) 96) was not available in EXFOR; for the KDUQ corpus, we used the data set at 14.0 MeV from Rosen et al. (PRL 10 246 (1963)) from EXFOR. We were unable to locate the data sets at 10.93 and 11.7 \MeV\ in EXFOR. For these data we used the same reference as listed in the KD corpus: Benveniste, Mitchell, and Fulmer (Phys. Rev. 133 B317). 
    \item[] \niFiftyEight: The data set listed in the KD corpus as having a scattering energy of 17.8 MeV from Payton and Schrank (PR 101 1358 (1956)) was apparently collected at 17.0 MeV, per the original publication and EXFOR. We used the latter energy in the KDUQ corpus. The data set at 35.2 MeV listed in the KD corpus as being from Eliyakut-Roshko et al. (PRC 51 1295 (1995)) is apparently from Fabrici et al. (PRC 21 844 (1980)) and is available in EXFOR; we used the latter reference and EXFOR data sets for the KDUQ corpus. The data set at 200 MeV from Sakaguchi et al. (RCNP Annual Report, 1993, p.4) was not available in EXFOR; for the KDUQ corpus, we used EXFOR data at 192 MeV from Sakaguchi et al. (PRC 57 1749 (1998)). We were unable to locate the data set 11.7 \MeV\ in EXFOR. For these data we used the same reference as listed in the KD corpus: Benveniste, Mitchell, and Fulmer (Phys. Rev. 133 B317). 
    \item[] \niNat: The data set listed in the KD corpus as having scattering energy of 17 MeV and being from Devins et al. (Nucl. Phys. 35 (1962) 617) was apparently actually collected at a scattering energy of 30.8 MeV, and using a \niSixty\ target. For the KDUQ corpus, we assigned this latter energy and target to this data set.
    \item[] \niSixty: The data set at 40 MeV from Blumberg et al. (PR 147 812 (1966)) listed in the KD corpus as being from \niSixty\ are actually from \niFiftyEight; we assigned it to \niFiftyEight\ for the KDUQ corpus.
    \item[] \zrNinety: The data sets listed in the KD corpus as having scattering energies of 12.7 and 14.7 \MeV\ do not appear in Greenlees et al. (PRC 3 1231 (1971)); for the KDUQ corpus, we used the data set at 12.7 MeV from Dickens et al. (PR 168 1355 (1968)) and at 14.7 MeV from Matsuda et al. (JPJ 22 1311 (1967)), as listed in EXFOR. The data set at 180 MeV from Nadasen et al. (PRC 23 1023 (1981)) is not available on EXFOR, only in a difficult-to-parse figure from the original paper. As such, we omitted this dataset from the KDUQ corpus.
    \item[] \pbTwoHundredEight: The data sets at 11, 12, and 13 MeV from Kretschmer et al. (PLB 87 343 (1979)) are not available in EXFOR, only in a difficult-to-parse figure in the original paper. As such, we omitted these data sets from the KDUQ corpus. The data set at 100 MeV doesn't appear to originate from Nadasen et al. (PRC 23 (1981) 1023); there is a data set at 98 MeV from a related publication Schwandt et al. (PRC 26 55 (1982)) for analyzing powers. We included the latter data set in the KDUQ corpus.
    \item[] \biTwoHundredNine: The data set at 57 MeV from Yamabe et al. (JPJ 17 729 (1962)) does not include errors; we omitted it from the KDUQ corpus.
\end{itemize}

\clearpage 
\subsection*{Proton analyzing powers}
\twocolumngrid 
\begin{longtable}{c c c}
    Isotope & Energy & EXFOR Acc. \\ 
\siTwentyEight & 17.8 & \href{https://www-nds.iaea.org/exfor/servlet/X4sGetSubent?subID=O0254003}{O0254003} \\ 
 & 65 & \href{https://www-nds.iaea.org/exfor/servlet/X4sGetSubent?subID=E0166009}{E0166009} \\ 
 & 80 & \href{https://www-nds.iaea.org/exfor/servlet/X4sGetSubent?subID=C0084003}{C0084003} \\ 
 & 100 & \href{https://www-nds.iaea.org/exfor/servlet/X4sGetSubent?subID=C0084003}{C0084003} \\ 
 & 135 & \href{https://www-nds.iaea.org/exfor/servlet/X4sGetSubent?subID=C0084003}{C0084003} \\ 
 & 175 & \href{https://www-nds.iaea.org/exfor/servlet/X4sGetSubent?subID=C0084003}{C0084003} \\ 
 & 180 & \href{https://www-nds.iaea.org/exfor/servlet/X4sGetSubent?subID=C0152004}{C0152004} \\ 
\hline 
\caForty & 16 & \href{https://www-nds.iaea.org/exfor/servlet/X4sGetSubent?subID=C0893003}{C0893003} \\ 
 & 26.3 & \href{https://www-nds.iaea.org/exfor/servlet/X4sGetSubent?subID=O0490006}{O0490006} \\ 
 & 30.05 & \href{https://www-nds.iaea.org/exfor/servlet/X4sGetSubent?subID=O0490003}{O0490003} \\ 
 & 40 & \href{https://www-nds.iaea.org/exfor/servlet/X4sGetSubent?subID=O02080062}{O02080062} \\ 
 & 45.5 & \href{https://www-nds.iaea.org/exfor/servlet/X4sGetSubent?subID=C0076003}{C0076003} \\ 
 & 65 & \href{https://www-nds.iaea.org/exfor/servlet/X4sGetSubent?subID=E0166013}{E0166013} \\ 
 & 75 & \href{https://www-nds.iaea.org/exfor/servlet/X4sGetSubent?subID=O0553011}{O0553011} \\ 
 & 80.2 & \href{https://www-nds.iaea.org/exfor/servlet/X4sGetSubent?subID=T0108004}{T0108004} \\ 
 & 152 & \href{https://www-nds.iaea.org/exfor/servlet/X4sGetSubent?subID=O0553005}{O0553005} \\ 
 & 181.5 & \href{https://www-nds.iaea.org/exfor/servlet/X4sGetSubent?subID=T0108004}{T0108004} \\ 
 & 201.4 & \href{https://www-nds.iaea.org/exfor/servlet/X4sGetSubent?subID=C0148008}{C0148008} \\ 
\hline 
\feFiftyFour & 9.69 & \href{https://www-nds.iaea.org/exfor/servlet/X4sGetSubent?subID=O03930022}{O03930022} \\ 
 & 16 & \href{https://www-nds.iaea.org/exfor/servlet/X4sGetSubent?subID=C0893009}{C0893009} \\ 
 & 17.2 & \href{https://www-nds.iaea.org/exfor/servlet/X4sGetSubent?subID=O0091003}{O0091003} \\ 
 & 18.6 & \href{https://www-nds.iaea.org/exfor/servlet/X4sGetSubent?subID=D0286005}{D0286005} \\ 
 & 20.4 & \href{https://www-nds.iaea.org/exfor/servlet/X4sGetSubent?subID=O0091005}{O0091005} \\ 
 & 24.6 & \href{https://www-nds.iaea.org/exfor/servlet/X4sGetSubent?subID=O0091007}{O0091007} \\ 
\hline 
\feNat & 155 & \href{https://www-nds.iaea.org/exfor/servlet/X4sGetSubent?subID=O0218002}{O0218002} \\ 
 & 179 & \href{https://www-nds.iaea.org/exfor/servlet/X4sGetSubent?subID=D0285008}{D0285008} \\ 
\hline 
\feFiftySix & 14 & \href{https://www-nds.iaea.org/exfor/servlet/X4sGetSubent?subID=C1100005}{C1100005} \\ 
 & 17.2 & \href{https://www-nds.iaea.org/exfor/servlet/X4sGetSubent?subID=O0091009}{O0091009} \\ 
 & 18.6 & \href{https://www-nds.iaea.org/exfor/servlet/X4sGetSubent?subID=D0286006}{D0286006} \\ 
 & 20.4 & \href{https://www-nds.iaea.org/exfor/servlet/X4sGetSubent?subID=O0091011}{O0091011} \\ 
 & 24.6 & \href{https://www-nds.iaea.org/exfor/servlet/X4sGetSubent?subID=O0091013}{O0091013} \\ 
 & 65 & \href{https://www-nds.iaea.org/exfor/servlet/X4sGetSubent?subID=E1201003}{E1201003} \\ 
\hline 
\niFiftyEight & 18.6 & \href{https://www-nds.iaea.org/exfor/servlet/X4sGetSubent?subID=D0286007}{D0286007} \\ 
 & 20.4 & \href{https://www-nds.iaea.org/exfor/servlet/X4sGetSubent?subID=O0091015}{O0091015} \\ 
 & 20.9 & \href{https://www-nds.iaea.org/exfor/servlet/X4sGetSubent?subID=O0434009}{O0434009} \\ 
 & 24.6 & \href{https://www-nds.iaea.org/exfor/servlet/X4sGetSubent?subID=O0091017}{O0091017} \\ 
 & 30.04 & \href{https://www-nds.iaea.org/exfor/servlet/X4sGetSubent?subID=O0490004}{O0490004} \\ 
 & 40 & \href{https://www-nds.iaea.org/exfor/servlet/X4sGetSubent?subID=O02080072}{O02080072} \\ 
 & 65 & \href{https://www-nds.iaea.org/exfor/servlet/X4sGetSubent?subID=E1201007}{E1201007} \\ 
 & 178 & \href{https://www-nds.iaea.org/exfor/servlet/X4sGetSubent?subID=D0189003}{D0189003} \\ 
 & 192 & \href{https://www-nds.iaea.org/exfor/servlet/X4sGetSubent?subID=E1704005}{E1704005} \\ 
\hline 
\niNat & 155 & \href{https://www-nds.iaea.org/exfor/servlet/X4sGetSubent?subID=O0218003}{O0218003} \\ 
\hline 
\niSixty & 20.4 & \href{https://www-nds.iaea.org/exfor/servlet/X4sGetSubent?subID=O0091019}{O0091019} \\ 
 & 24.6 & \href{https://www-nds.iaea.org/exfor/servlet/X4sGetSubent?subID=O0091021}{O0091021} \\ 
 & 65 & \href{https://www-nds.iaea.org/exfor/servlet/X4sGetSubent?subID=E1201009}{E1201009} \\ 
\hline 
\zrNinety & 9.7 & \href{https://www-nds.iaea.org/exfor/servlet/X4sGetSubent?subID=O03930072}{O03930072} \\ 
 & 20.25 & \href{https://www-nds.iaea.org/exfor/servlet/X4sGetSubent?subID=O0392007}{O0392007} \\ 
 & 30 & \href{https://www-nds.iaea.org/exfor/servlet/X4sGetSubent?subID=D0295003}{D0295003} \\ 
 & 40 & \href{https://www-nds.iaea.org/exfor/servlet/X4sGetSubent?subID=O02080082}{O02080082} \\ 
\hline 
\snOneHundredTwenty & 9.7 & \href{https://www-nds.iaea.org/exfor/servlet/X4sGetSubent?subID=O03930082}{O03930082} \\ 
 & 20.4 & \href{https://www-nds.iaea.org/exfor/servlet/X4sGetSubent?subID=O0169021}{O0169021} \\ 
 & 24.6 & \href{https://www-nds.iaea.org/exfor/servlet/X4sGetSubent?subID=O0169027}{O0169027} \\ 
 & 40 & \href{https://www-nds.iaea.org/exfor/servlet/X4sGetSubent?subID=O0328011}{O0328011} \\ 
\hline 
\pbTwoHundredEight & 26.3 & \href{https://www-nds.iaea.org/exfor/servlet/X4sGetSubent?subID=O0490008}{O0490008} \\ 
 & 40 & \href{https://www-nds.iaea.org/exfor/servlet/X4sGetSubent?subID=O02080092}{O02080092} \\ 
 & 49.35 & \href{https://www-nds.iaea.org/exfor/servlet/X4sGetSubent?subID=O0788025}{O0788025} \\ 
 & 65 & \href{https://www-nds.iaea.org/exfor/servlet/X4sGetSubent?subID=E1201015}{E1201015} \\ 
 & 79.8 & \href{https://www-nds.iaea.org/exfor/servlet/X4sGetSubent?subID=T0108010}{T0108010} \\ 
 & 98 & \href{https://www-nds.iaea.org/exfor/servlet/X4sGetSubent?subID=T0108010}{T0108010} \\ 
 & 182 & \href{https://www-nds.iaea.org/exfor/servlet/X4sGetSubent?subID=T0108010}{T0108010} \\ 
 & 200 & \href{https://www-nds.iaea.org/exfor/servlet/X4sGetSubent?subID=C0081006}{C0081006} \\ 
\hline 

        \end{longtable}
        \label{APower_p_pb208}

\subsection*{Comments}
No discrepancies were identified between the KD and KDUQ corpora for proton
analyzing powers. For related information, see the Comments for proton
differential elastic cross sections above (note that the formatting of the
original KD corpus did not explicitly distinguish between inclusion of
differential elastic cross section data sets and analyzing power data sets).

\clearpage 
\subsection*{Proton reaction cross sections}
\twocolumngrid 
\begin{longtable}{c c c}
    Isotope & Energy & EXFOR Acc. \\ 
\alTwentySeven
 & 8.87-10.39 & \href{https://www-nds.iaea.org/exfor/servlet/X4sGetSubent?subID=D0314002}{D0314002} \\ 
 & 8.8 & \href{https://www-nds.iaea.org/exfor/servlet/X4sGetSubent?subID=D0535002}{D0535002} \\ 
 & 9.89 & \href{https://www-nds.iaea.org/exfor/servlet/X4sGetSubent?subID=C1042003}{C1042003} \\ 
 & 9.9-10.12 & \href{https://www-nds.iaea.org/exfor/servlet/X4sGetSubent?subID=O0741004}{O0741004} \\ 
 & 16.29 & \href{https://www-nds.iaea.org/exfor/servlet/X4sGetSubent?subID=O0368004}{O0368004} \\ 
 & 24.8-46.3 & \href{https://www-nds.iaea.org/exfor/servlet/X4sGetSubent?subID=O0325003}{O0325003} \\ 
 & 29 & \href{https://www-nds.iaea.org/exfor/servlet/X4sGetSubent?subID=O0369003}{O0369003} \\ 
 & 34 & \href{https://www-nds.iaea.org/exfor/servlet/X4sGetSubent?subID=O0732003}{O0732003} \\ 
 & 40-60.8 & \href{https://www-nds.iaea.org/exfor/servlet/X4sGetSubent?subID=O0081008}{O0081008} \\ 
 & 61 & \href{https://www-nds.iaea.org/exfor/servlet/X4sGetSubent?subID=O0475003}{O0475003} \\ 
 & 77-133 & \href{https://www-nds.iaea.org/exfor/servlet/X4sGetSubent?subID=C1211003}{C1211003} \\ 
 & 99.7 & \href{https://www-nds.iaea.org/exfor/servlet/X4sGetSubent?subID=O0340006}{O0340006} \\ 
 & 134 & \href{https://www-nds.iaea.org/exfor/servlet/X4sGetSubent?subID=O1947003}{O1947003} \\ 
 & 179.6 & \href{https://www-nds.iaea.org/exfor/servlet/X4sGetSubent?subID=D0533005}{D0533005} \\ 
 & 185 & - \\ 
 & 234 & \href{https://www-nds.iaea.org/exfor/servlet/X4sGetSubent?subID=O0213004}{O0213004} \\ 
\hline 
\siNat & 20.7-38.1 & \href{https://www-nds.iaea.org/exfor/servlet/X4sGetSubent?subID=C1862005}{C1862005} \\ 
 & 24.7-47.8 & \href{https://www-nds.iaea.org/exfor/servlet/X4sGetSubent?subID=O0325006}{O0325006} \\ 
 & 65.5 & \href{https://www-nds.iaea.org/exfor/servlet/X4sGetSubent?subID=O0579005}{O0579005} \\ 
\hline 
\caForty
 & 10.34-21.59 & \href{https://www-nds.iaea.org/exfor/servlet/X4sGetSubent?subID=O0341003}{O0341003} \\ 
 & 24.9-48 & \href{https://www-nds.iaea.org/exfor/servlet/X4sGetSubent?subID=O0330003}{O0330003} \\ 
 & 28.5 & \href{https://www-nds.iaea.org/exfor/servlet/X4sGetSubent?subID=O0150002}{O0150002} \\ 
 & 65.5 & \href{https://www-nds.iaea.org/exfor/servlet/X4sGetSubent?subID=O0579006}{O0579006} \\ 
\hline 
\caNat
 & 99.3 & \href{https://www-nds.iaea.org/exfor/servlet/X4sGetSubent?subID=O0340007}{O0340007} \\ 
 & 179.6 & \href{https://www-nds.iaea.org/exfor/servlet/X4sGetSubent?subID=D0533006}{D0533006} \\ 
\hline 
\feNat
 & 8.8 & \href{https://www-nds.iaea.org/exfor/servlet/X4sGetSubent?subID=D0535004}{D0535004} \\ 
 & 9.21-11.25 & \href{https://www-nds.iaea.org/exfor/servlet/X4sGetSubent?subID=D0314003}{D0314003} \\ 
 & 9.89 & \href{https://www-nds.iaea.org/exfor/servlet/X4sGetSubent?subID=C1042006}{C1042006} \\ 
 & 9.97-10.2 & \href{https://www-nds.iaea.org/exfor/servlet/X4sGetSubent?subID=O0741013}{O0741013} \\ 
 & 15.8 & \href{https://www-nds.iaea.org/exfor/servlet/X4sGetSubent?subID=C1864004}{C1864004} \\ 
 & 28 & \href{https://www-nds.iaea.org/exfor/servlet/X4sGetSubent?subID=C1212007}{C1212007} \\ 
 & 34 & \href{https://www-nds.iaea.org/exfor/servlet/X4sGetSubent?subID=O0732004}{O0732004} \\ 
 & 61 & \href{https://www-nds.iaea.org/exfor/servlet/X4sGetSubent?subID=O0475004}{O0475004} \\ 
 & 98.7 & \href{https://www-nds.iaea.org/exfor/servlet/X4sGetSubent?subID=O0340011}{O0340011} \\ 
 & 179.6 & \href{https://www-nds.iaea.org/exfor/servlet/X4sGetSubent?subID=D0533007}{D0533007} \\ 
 & 230 & \href{https://www-nds.iaea.org/exfor/servlet/X4sGetSubent?subID=O0213005}{O0213005} \\ 
\hline 
\feFiftySix & 14.5 & \href{https://www-nds.iaea.org/exfor/servlet/X4sGetSubent?subID=C1217004}{C1217004} \\ 
 & 20.8-47.8 & \href{https://www-nds.iaea.org/exfor/servlet/X4sGetSubent?subID=T0100004}{T0100004} \\ 
 & 28.5 & \href{https://www-nds.iaea.org/exfor/servlet/X4sGetSubent?subID=O0150003}{O0150003} \\ 
 & 40-60.8 & \href{https://www-nds.iaea.org/exfor/servlet/X4sGetSubent?subID=O0081013}{O0081013} \\ 
\hline 
\cuSixtyThree
 & 6.75 & - \\ 
 & 8.7 & \href{https://www-nds.iaea.org/exfor/servlet/X4sGetSubent?subID=D0535009}{D0535009} \\ 
 & 9.1 & \href{https://www-nds.iaea.org/exfor/servlet/X4sGetSubent?subID=D0534004}{D0534004} \\ 
 & 9.11-11.18 & \href{https://www-nds.iaea.org/exfor/servlet/X4sGetSubent?subID=D0314010}{D0314010} \\ 
 & 9.85 & - \\ 
 & 9.89 & \href{https://www-nds.iaea.org/exfor/servlet/X4sGetSubent?subID=C1042009}{C1042009} \\ 
 & 14.5 & \href{https://www-nds.iaea.org/exfor/servlet/X4sGetSubent?subID=C1217010}{C1217010} \\ 
\hline 
\cuNat
 & 8.78-11.21 & \href{https://www-nds.iaea.org/exfor/servlet/X4sGetSubent?subID=D0337003}{D0337003} \\ 
 & 8.8 & \href{https://www-nds.iaea.org/exfor/servlet/X4sGetSubent?subID=D0535008}{D0535008} \\ 
 & 8.9 & \href{https://www-nds.iaea.org/exfor/servlet/X4sGetSubent?subID=C0067002}{C0067002} \\ 
 & 9.05 & \href{https://www-nds.iaea.org/exfor/servlet/X4sGetSubent?subID=D0534006}{D0534006} \\ 
 & 9.3 & \href{https://www-nds.iaea.org/exfor/servlet/X4sGetSubent?subID=O1948003}{O1948003} \\ 
 & 9.9-10.12 & \href{https://www-nds.iaea.org/exfor/servlet/X4sGetSubent?subID=O0741006}{O0741006} \\ 
 & 15.8 & \href{https://www-nds.iaea.org/exfor/servlet/X4sGetSubent?subID=C1864003}{C1864003} \\ 
 & 16.37 & \href{https://www-nds.iaea.org/exfor/servlet/X4sGetSubent?subID=O0368006}{O0368006} \\ 
 & 28 & \href{https://www-nds.iaea.org/exfor/servlet/X4sGetSubent?subID=C1212016}{C1212016} \\ 
 & 77-133 & \href{https://www-nds.iaea.org/exfor/servlet/X4sGetSubent?subID=C1211004}{C1211004} \\ 
 & 99 & \href{https://www-nds.iaea.org/exfor/servlet/X4sGetSubent?subID=O0340014}{O0340014} \\ 
 & 134 & \href{https://www-nds.iaea.org/exfor/servlet/X4sGetSubent?subID=O1947004}{O1947004} \\ 
 & 185 & - \\ 
 & 225 & \href{https://www-nds.iaea.org/exfor/servlet/X4sGetSubent?subID=O0213006}{O0213006} \\ 
\hline 
\zrNinety
 & 14.5 & \href{https://www-nds.iaea.org/exfor/servlet/X4sGetSubent?subID=C1217014}{C1217014} \\ 
 & 30-60.8 & \href{https://www-nds.iaea.org/exfor/servlet/X4sGetSubent?subID=O0081004}{O0081004} \\ 
\hline 
\zrNat
 & 9.2 & \href{https://www-nds.iaea.org/exfor/servlet/X4sGetSubent?subID=D0534007}{D0534007} \\ 
 & 10.03-10.25 & \href{https://www-nds.iaea.org/exfor/servlet/X4sGetSubent?subID=O0741016}{O0741016} \\ 
 & 98.8 & \href{https://www-nds.iaea.org/exfor/servlet/X4sGetSubent?subID=O0340016}{O0340016} \\ 
\hline 
\snNat
 & 9.99-10.21 & \href{https://www-nds.iaea.org/exfor/servlet/X4sGetSubent?subID=O0741015}{O0741015} \\ 
 & 34 & \href{https://www-nds.iaea.org/exfor/servlet/X4sGetSubent?subID=O0732005}{O0732005} \\ 
 & 61 & \href{https://www-nds.iaea.org/exfor/servlet/X4sGetSubent?subID=O0475005}{O0475005} \\ 
 & 99.1 & \href{https://www-nds.iaea.org/exfor/servlet/X4sGetSubent?subID=O0340023}{O0340023} \\ 
 & 221 & \href{https://www-nds.iaea.org/exfor/servlet/X4sGetSubent?subID=O0213009}{O0213009} \\ 
\hline 
\snOneHundredTwenty
 & 14.5 & \href{https://www-nds.iaea.org/exfor/servlet/X4sGetSubent?subID=C1217022}{C1217022} \\ 
 & 22.8-47.9 & \href{https://www-nds.iaea.org/exfor/servlet/X4sGetSubent?subID=C0424006}{C0424006} \\ 
 & 28.5 & \href{https://www-nds.iaea.org/exfor/servlet/X4sGetSubent?subID=O0150007}{O0150007} \\ 
 & 30-49.5 & \href{https://www-nds.iaea.org/exfor/servlet/X4sGetSubent?subID=O0081005}{O0081005} \\ 
 & 65.5 & \href{https://www-nds.iaea.org/exfor/servlet/X4sGetSubent?subID=O0579012}{O0579012} \\ 
\hline 
\pbNat
 & 9.92 & \href{https://www-nds.iaea.org/exfor/servlet/X4sGetSubent?subID=O0741019}{O0741019} \\ 
 & 16.31 & \href{https://www-nds.iaea.org/exfor/servlet/X4sGetSubent?subID=O0368007}{O0368007} \\ 
 & 34 & \href{https://www-nds.iaea.org/exfor/servlet/X4sGetSubent?subID=O0732006}{O0732006} \\ 
 & 61 & \href{https://www-nds.iaea.org/exfor/servlet/X4sGetSubent?subID=O0475006}{O0475006} \\ 
 & 77-133 & \href{https://www-nds.iaea.org/exfor/servlet/X4sGetSubent?subID=C1211006}{C1211006} \\ 
 & 99.2 & \href{https://www-nds.iaea.org/exfor/servlet/X4sGetSubent?subID=O0340040}{O0340040} \\ 
 & 134 & \href{https://www-nds.iaea.org/exfor/servlet/X4sGetSubent?subID=O1947006}{O1947006} \\ 
 & 185 & - \\ 
 & 226 & \href{https://www-nds.iaea.org/exfor/servlet/X4sGetSubent?subID=O0213010}{O0213010} \\ 
\hline 
\pbTwoHundredEight
 & 21.1-48 & \href{https://www-nds.iaea.org/exfor/servlet/X4sGetSubent?subID=O0330004}{O0330004} \\ 
 & 28.5 & \href{https://www-nds.iaea.org/exfor/servlet/X4sGetSubent?subID=O0150008}{O0150008} \\ 
 & 30-60.8 & \href{https://www-nds.iaea.org/exfor/servlet/X4sGetSubent?subID=O0081007}{O0081007} \\ 
 & 65.5 & \href{https://www-nds.iaea.org/exfor/servlet/X4sGetSubent?subID=O0579014}{O0579014} \\ 
\hline 

\end{longtable}

\subsection*{Comments}
\begin{itemize}
    \item[] \alTwentySeven: we were unable to locate the datum at 185 \MeV\ in EXFOR. For this datum we used the same reference as listed in the KD corpus: Millburn et al. (Phys. Rev. 95 1268 (1954)).
    \item[] \caForty: the data from 99.3 and 179.6 \MeV\ listed as being for \caForty\ in the KD corpus are actually from \caNat; we assigned them to \caNat\ for the KDUQ corpus.
    \item[] \feFiftySix: many of the data listed as being for \feFiftySix\ in the KD Corpus are actually for \feNat; we assigned them to \feNat\ for the KDUQ corpus.
    \item[] \cuSixtyThree: many of the data listed as being for \cuSixtyThree\ in the KD Corpus are actually for \cuNat; we assigned them to \cuNat\ for the KDUQ corpus. We were unable to locate the data at 6.75 and 9.85 \MeV\ in EXFOR. For these data we used the same references as listed in the KD corpus: Dell, Ploughe, and Hausman (Nucl. Phys. 64 (1965) p. 513), and Albert and Hansen (PRL 6 (1961) p. 13), respectively.
    \item[] \cuNat: we were unable to locate the datum at 185 \MeV\ in EXFOR. For this datum we used the same reference as listed in the KD corpus: Millburn et al. (Phys. Rev. 95 1268 (1954)).
    \item[] \zrNinety: several of the data listed as being for \zrNinety\ in the KD Corpus are actually for \zrNat; we assigned them to \zrNat\ for the KDUQ corpus.
    \item[] \snOneHundredTwenty: several of the data listed as being for \snOneHundredTwenty\ in the KD Corpus are actually for \snNat; we assigned them to \snNat\ for the KDUQ corpus.
    \item[] \pbNat: we were unable to locate the datum at 185 \MeV\ in EXFOR. For this datum we used the same reference as listed in the KD corpus: Millburn et al. (Phys. Rev. 95 1268 (1954)).
    \item[] \pbTwoHundredEight: several of the data listed as being for \pbTwoHundredEight\ in the KD Corpus are actually for \pbNat; we assigned them to \pbNat\ for the KDUQ corpus.
\end{itemize}

\clearpage 
\section*{CHUQ corpus}
\subsection*{Neutron differential elastic cross sections}
\begin{longtable}{c c c}
                Isotope & Energies & EXFOR Acc. \# \\ 
\caForty & 13.905 & \href{https://www-nds.iaea.org/exfor/servlet/X4sGetSubent?subID=12996002}{12996002} \\ 
 & 16.916 & \href{https://www-nds.iaea.org/exfor/servlet/X4sGetSubent?subID=12996002}{12996002} \\ 
\hline 
\caNat & 11.01 & \href{https://www-nds.iaea.org/exfor/servlet/X4sGetSubent?subID=10633005}{10633005} \\ 
\hline 
\vFiftyOne & 11.01 & \href{https://www-nds.iaea.org/exfor/servlet/X4sGetSubent?subID=10633006}{10633006} \\ 
\hline 
\feFiftyFour & 9.94 & \href{https://www-nds.iaea.org/exfor/servlet/X4sGetSubent?subID=10958002}{10958002} \\ 
 & 11.0 & \href{https://www-nds.iaea.org/exfor/servlet/X4sGetSubent?subID=12862002}{12862002} \\ 
 & 13.92 & \href{https://www-nds.iaea.org/exfor/servlet/X4sGetSubent?subID=10958002}{10958002} \\ 
 & 20.0 & \href{https://www-nds.iaea.org/exfor/servlet/X4sGetSubent?subID=12862002}{12862002} \\ 
 & 22.0 & \href{https://www-nds.iaea.org/exfor/servlet/X4sGetSubent?subID=12862002}{12862002} \\ 
 & 24.0 & \href{https://www-nds.iaea.org/exfor/servlet/X4sGetSubent?subID=12862002}{12862002} \\ 
 & 26.0 & \href{https://www-nds.iaea.org/exfor/servlet/X4sGetSubent?subID=12862002}{12862002} \\ 
\hline 
\mnFiftyFive & 11.01 & \href{https://www-nds.iaea.org/exfor/servlet/X4sGetSubent?subID=10633007}{10633007} \\ 
\hline 
\feFiftySix & 11.0 & \href{https://www-nds.iaea.org/exfor/servlet/X4sGetSubent?subID=12862003}{12862003} \\ 
 & 20.0 & \href{https://www-nds.iaea.org/exfor/servlet/X4sGetSubent?subID=12862003}{12862003} \\ 
 & 26.0 & \href{https://www-nds.iaea.org/exfor/servlet/X4sGetSubent?subID=12862003}{12862003} \\ 
\hline 
\niFiftyEight & 9.958 & \href{https://www-nds.iaea.org/exfor/servlet/X4sGetSubent?subID=12930002}{12930002} \\ 
 & 13.941 & \href{https://www-nds.iaea.org/exfor/servlet/X4sGetSubent?subID=12930002}{12930002} \\ 
\hline 
\coFiftyNine & 11.01 & \href{https://www-nds.iaea.org/exfor/servlet/X4sGetSubent?subID=10633009}{10633009} \\ 
\hline 
\niSixty & 9.958 & \href{https://www-nds.iaea.org/exfor/servlet/X4sGetSubent?subID=12930004}{12930004} \\ 
 & 13.941 & \href{https://www-nds.iaea.org/exfor/servlet/X4sGetSubent?subID=12930004}{12930004} \\ 
\hline 
\cuSixtyFive & 9.94 & \href{https://www-nds.iaea.org/exfor/servlet/X4sGetSubent?subID=10958008}{10958008} \\ 
 & 13.92 & \href{https://www-nds.iaea.org/exfor/servlet/X4sGetSubent?subID=10958008}{10958008} \\ 
\hline 
\srEightyEight & 11.0 & \href{https://www-nds.iaea.org/exfor/servlet/X4sGetSubent?subID=10729004}{10729004} \\ 
\hline 
\yEightyNine & 11.0 & \href{https://www-nds.iaea.org/exfor/servlet/X4sGetSubent?subID=12774002}{12774002} \\ 
 & 11.0 & \href{https://www-nds.iaea.org/exfor/servlet/X4sGetSubent?subID=12790002}{12790002} \\ 
\hline 
\zrNinety & 11.0 & \href{https://www-nds.iaea.org/exfor/servlet/X4sGetSubent?subID=10729002}{10729002} \\ 
\hline 
\moNinetyTwo & 11.0 & \href{https://www-nds.iaea.org/exfor/servlet/X4sGetSubent?subID=10729007}{10729007} \\ 
 & 11.0 & \href{https://www-nds.iaea.org/exfor/servlet/X4sGetSubent?subID=10867002}{10867002} \\ 
 & 11.01 & \href{https://www-nds.iaea.org/exfor/servlet/X4sGetSubent?subID=10633012}{10633012} \\ 
 & 20.0 & \href{https://www-nds.iaea.org/exfor/servlet/X4sGetSubent?subID=10867002}{10867002} \\ 
 & 26.0 & \href{https://www-nds.iaea.org/exfor/servlet/X4sGetSubent?subID=10867002}{10867002} \\ 
\hline 
\nbNinetyThree & 11.01 & \href{https://www-nds.iaea.org/exfor/servlet/X4sGetSubent?subID=10633011}{10633011} \\ 
\hline 
\moNinetySix & 11.0 & \href{https://www-nds.iaea.org/exfor/servlet/X4sGetSubent?subID=10867003}{10867003} \\ 
 & 11.01 & \href{https://www-nds.iaea.org/exfor/servlet/X4sGetSubent?subID=10633013}{10633013} \\ 
 & 20.0 & \href{https://www-nds.iaea.org/exfor/servlet/X4sGetSubent?subID=10867003}{10867003} \\ 
 & 26.0 & \href{https://www-nds.iaea.org/exfor/servlet/X4sGetSubent?subID=10867003}{10867003} \\ 
\hline 
\moNinetyEight & 11.0 & \href{https://www-nds.iaea.org/exfor/servlet/X4sGetSubent?subID=10867004}{10867004} \\ 
 & 11.01 & \href{https://www-nds.iaea.org/exfor/servlet/X4sGetSubent?subID=10633014}{10633014} \\ 
 & 20.0 & \href{https://www-nds.iaea.org/exfor/servlet/X4sGetSubent?subID=10867004}{10867004} \\ 
 & 26.0 & \href{https://www-nds.iaea.org/exfor/servlet/X4sGetSubent?subID=10867004}{10867004} \\ 
\hline 
\moOneHundred & 11.0 & \href{https://www-nds.iaea.org/exfor/servlet/X4sGetSubent?subID=10867005}{10867005} \\ 
 & 11.01 & \href{https://www-nds.iaea.org/exfor/servlet/X4sGetSubent?subID=10633015}{10633015} \\ 
 & 20.0 & \href{https://www-nds.iaea.org/exfor/servlet/X4sGetSubent?subID=10867005}{10867005} \\ 
 & 26.0 & \href{https://www-nds.iaea.org/exfor/servlet/X4sGetSubent?subID=10867005}{10867005} \\ 
\hline 
\snOneHundredSixteen & 9.945 & \href{https://www-nds.iaea.org/exfor/servlet/X4sGetSubent?subID=13158005}{13158005} \\ 
 & 11.0 & \href{https://www-nds.iaea.org/exfor/servlet/X4sGetSubent?subID=10817006}{10817006} \\ 
 & 13.925 & \href{https://www-nds.iaea.org/exfor/servlet/X4sGetSubent?subID=13158005}{13158005} \\ 
 & 24.0 & \href{https://www-nds.iaea.org/exfor/servlet/X4sGetSubent?subID=10817006}{10817006} \\ 
\hline 
\snOneHundredEighteen & 11.0 & \href{https://www-nds.iaea.org/exfor/servlet/X4sGetSubent?subID=10817007}{10817007} \\ 
 & 24.0 & \href{https://www-nds.iaea.org/exfor/servlet/X4sGetSubent?subID=10817007}{10817007} \\ 
\hline 
\snOneHundredTwenty & 9.943 & \href{https://www-nds.iaea.org/exfor/servlet/X4sGetSubent?subID=13158007}{13158007} \\ 
 & 11.0 & \href{https://www-nds.iaea.org/exfor/servlet/X4sGetSubent?subID=10817008}{10817008} \\ 
 & 11.01 & \href{https://www-nds.iaea.org/exfor/servlet/X4sGetSubent?subID=10633017}{10633017} \\ 
 & 13.923 & \href{https://www-nds.iaea.org/exfor/servlet/X4sGetSubent?subID=13158007}{13158007} \\ 
 & 16.905 & \href{https://www-nds.iaea.org/exfor/servlet/X4sGetSubent?subID=13158007}{13158007} \\ 
\hline 
\snOneHundredTwentyTwo & 11.0 & \href{https://www-nds.iaea.org/exfor/servlet/X4sGetSubent?subID=10817009}{10817009} \\ 
\hline 
\snOneHundredTwentyFour & 11.0 & \href{https://www-nds.iaea.org/exfor/servlet/X4sGetSubent?subID=10817010}{10817010} \\ 
 & 24.0 & \href{https://www-nds.iaea.org/exfor/servlet/X4sGetSubent?subID=10817010}{10817010} \\ 
\hline 
\hoOneHundredSixtyFive & 11.01 & \href{https://www-nds.iaea.org/exfor/servlet/X4sGetSubent?subID=10633018}{10633018} \\ 
\hline 
\pbTwoHundredSix & 11.01 & \href{https://www-nds.iaea.org/exfor/servlet/X4sGetSubent?subID=10633020}{10633020} \\ 
\hline 
\pbTwoHundredEight & 11.0 & \href{https://www-nds.iaea.org/exfor/servlet/X4sGetSubent?subID=10871002}{10871002} \\ 
 & 13.9 & \href{https://www-nds.iaea.org/exfor/servlet/X4sGetSubent?subID=13685002}{13685002} \\ 
 & 16.9 & \href{https://www-nds.iaea.org/exfor/servlet/X4sGetSubent?subID=13685002}{13685002} \\ 
 & 20.0 & \href{https://www-nds.iaea.org/exfor/servlet/X4sGetSubent?subID=10871002}{10871002} \\ 
 & 26.0 & \href{https://www-nds.iaea.org/exfor/servlet/X4sGetSubent?subID=10871002}{10871002} \\ 
\hline 
\biTwoHundredNine & 11.01 & \href{https://www-nds.iaea.org/exfor/servlet/X4sGetSubent?subID=10633022}{10633022} \\ 
\hline 

        \end{longtable}
        \label{ECS_n_bi209}

\subsection*{Comments}
\begin{itemize}
    \item[] \caNat, \vFiftyOne, \mnFiftyFive, \coFiftyNine, \nbNinetyThree, \moNinetyTwo, \moNinetySix, \moNinetyEight, \moOneHundred, \snOneHundredTwenty, \hoOneHundredSixtyFive, \pbTwoHundredSix, \biTwoHundredNine: as listed in EXFOR, the data sets for these nuclei from the publication of Ferrer et al. (Nucl. Phys. A 275, p. 325 (1977)) had both a DATA-ERR1 column containing a partial error in absolute units, and a DATA-ERR2 column containing a separate partial error in percent. It was unclear (to us) the relationship between these sources of error, and we were unable to determine how these data sets were treated in the canonical CH89 fit. For these data sets, we converted the percent error to absolute units and summed both sources of partial error in quadrature to yield an overall error, which was then used in the CHUQ corpus. 
    \item[] \snOneHundredEighteen: For the data sets at 11 and 24 \MeV\ Rapaport et al. (Nucl. Phys. A 341 p. 56 (1980)) listed on EXFOR, a ``null" value was listed in the DATA-ERR column; we elected to use the following column ERR-T, which listed a 5\% relative uncertainty for all data points, for the error of these data in the CHUQ corpus.
    \item[] \feFiftyFour, \feFiftySix: For the data sets at 11, 20, 22, 24, and 26 \MeV\ by Mellema et al. (Phys. Rev. C 28 p. 2267 (1983)) listed in EXFOR, the first and second DATA-ERR columns listed errors in percent and in absolute units, respectively. The first column contained mostly ``null" values. To combine the data from both columns for use in the CHUQ corpus, we converted the percent errors that did exist in the first DATA-ERR column into absolute units, then merged the two columns into one absolute DATA-ERR column.
\end{itemize}

\clearpage 
\subsection*{Neutron analyzing powers}
\begin{longtable}{c c c}
                Isotope & Energies & EXFOR Acc. \# \\ 
\caForty & 10.935 & \href{https://www-nds.iaea.org/exfor/servlet/X4sGetSubent?subID=12996004}{12996004} \\ 
 & 13.904 & \href{https://www-nds.iaea.org/exfor/servlet/X4sGetSubent?subID=12996004}{12996004} \\ 
 & 16.923 & \href{https://www-nds.iaea.org/exfor/servlet/X4sGetSubent?subID=12996004}{12996004} \\ 
\hline 
\niFiftyEight & 9.92 & \href{https://www-nds.iaea.org/exfor/servlet/X4sGetSubent?subID=12930010}{12930010} \\ 
 & 13.91 & \href{https://www-nds.iaea.org/exfor/servlet/X4sGetSubent?subID=12930010}{12930010} \\ 
\hline 
\snOneHundredSixteen & 9.907 & \href{https://www-nds.iaea.org/exfor/servlet/X4sGetSubent?subID=13158002}{13158002} \\ 
 & 13.894 & \href{https://www-nds.iaea.org/exfor/servlet/X4sGetSubent?subID=13158002}{13158002} \\ 
\hline 
\snOneHundredTwenty & 9.906 & \href{https://www-nds.iaea.org/exfor/servlet/X4sGetSubent?subID=13158003}{13158003} \\ 
 & 13.894 & \href{https://www-nds.iaea.org/exfor/servlet/X4sGetSubent?subID=13158003}{13158003} \\ 
\hline 
\pbTwoHundredEight & 9.97 & \href{https://www-nds.iaea.org/exfor/servlet/X4sGetSubent?subID=12844007}{12844007} \\ 
 & 13.9 & \href{https://www-nds.iaea.org/exfor/servlet/X4sGetSubent?subID=12844007}{12844007} \\ 
\hline 

        \end{longtable}
        \label{APower_n_pb208}

\subsection*{Comments}
\begin{itemize}
    \item[] \niSixtyFour: In the data set at 65 \MeV\ by Sakaguchi et al. (Memoirs Faculty of Sci., Kyoto Univ. Ser. Phys. 36 p.305 (1983)) as listed in EXFOR, the first datum possessed an unphysical error of 3.0034 (the analyzing power can only assume a value between -1 and 1). As this is likely a transcription or tabulation error, for the CHUQ corpus we changed this to 0.0034, which is consistent with the other listed errors from this data set.
\end{itemize}

\clearpage 
\subsection*{Proton differential elastic cross sections}
\begin{longtable}{c c c}
                Isotope & Energies & EXFOR Acc. \# \\ 
\caForty & 40.0 & \href{https://www-nds.iaea.org/exfor/servlet/X4sGetSubent?subID=O0328003}{O0328003} \\ 
 & 65.0 & \href{https://www-nds.iaea.org/exfor/servlet/X4sGetSubent?subID=O0032002S}{O0032002S} \\ 
\hline 
\caFortyFour & 65.0 & \href{https://www-nds.iaea.org/exfor/servlet/X4sGetSubent?subID=O0032056S}{O0032056S} \\ 
\hline 
\tiFortySix & 65.0 & \href{https://www-nds.iaea.org/exfor/servlet/X4sGetSubent?subID=O0032058S}{O0032058S} \\ 
\hline 
\caFortyEight & 65.0 & \href{https://www-nds.iaea.org/exfor/servlet/X4sGetSubent?subID=O0032057S}{O0032057S} \\ 
\hline 
\tiFortyEight & 16.0 & \href{https://www-nds.iaea.org/exfor/servlet/X4sGetSubent?subID=C0893004}{C0893004} \\ 
 & 65.0 & \href{https://www-nds.iaea.org/exfor/servlet/X4sGetSubent?subID=O0032059S}{O0032059S} \\ 
\hline 
\tiFifty & 16.0 & \href{https://www-nds.iaea.org/exfor/servlet/X4sGetSubent?subID=C0893006}{C0893006} \\ 
 & 65.0 & \href{https://www-nds.iaea.org/exfor/servlet/X4sGetSubent?subID=O0032060S}{O0032060S} \\ 
\hline 
\feFiftyFour & 16.0 & \href{https://www-nds.iaea.org/exfor/servlet/X4sGetSubent?subID=C0893008}{C0893008} \\ 
 & 17.2 & \href{https://www-nds.iaea.org/exfor/servlet/X4sGetSubent?subID=O0091002}{O0091002} \\ 
 & 20.4 & \href{https://www-nds.iaea.org/exfor/servlet/X4sGetSubent?subID=O0091004}{O0091004} \\ 
 & 24.6 & \href{https://www-nds.iaea.org/exfor/servlet/X4sGetSubent?subID=O0091006}{O0091006} \\ 
 & 40.0 & \href{https://www-nds.iaea.org/exfor/servlet/X4sGetSubent?subID=O0162023}{O0162023} \\ 
 & 65.0 & \href{https://www-nds.iaea.org/exfor/servlet/X4sGetSubent?subID=O0032061S}{O0032061S} \\ 
\hline 
\feFiftySix & 17.2 & \href{https://www-nds.iaea.org/exfor/servlet/X4sGetSubent?subID=O0091008}{O0091008} \\ 
 & 20.4 & \href{https://www-nds.iaea.org/exfor/servlet/X4sGetSubent?subID=O0091010}{O0091010} \\ 
 & 24.6 & \href{https://www-nds.iaea.org/exfor/servlet/X4sGetSubent?subID=O0091012}{O0091012} \\ 
 & 65.0 & \href{https://www-nds.iaea.org/exfor/servlet/X4sGetSubent?subID=O0032062S}{O0032062S} \\ 
\hline 
\niFiftyEight & 16.0 & \href{https://www-nds.iaea.org/exfor/servlet/X4sGetSubent?subID=C0893012}{C0893012} \\ 
 & 20.4 & \href{https://www-nds.iaea.org/exfor/servlet/X4sGetSubent?subID=O0091014}{O0091014} \\ 
 & 24.6 & \href{https://www-nds.iaea.org/exfor/servlet/X4sGetSubent?subID=O0091016}{O0091016} \\ 
 & 40.0 & \href{https://www-nds.iaea.org/exfor/servlet/X4sGetSubent?subID=O0162025}{O0162025} \\ 
 & 65.0 & \href{https://www-nds.iaea.org/exfor/servlet/X4sGetSubent?subID=O0032008S}{O0032008S} \\ 
\hline 
\coFiftyNine & 40.0 & \href{https://www-nds.iaea.org/exfor/servlet/X4sGetSubent?subID=O0328004}{O0328004} \\ 
 & 65.0 & \href{https://www-nds.iaea.org/exfor/servlet/X4sGetSubent?subID=O0032015S}{O0032015S} \\ 
\hline 
\niSixty & 16.0 & \href{https://www-nds.iaea.org/exfor/servlet/X4sGetSubent?subID=C0893014}{C0893014} \\ 
 & 20.4 & \href{https://www-nds.iaea.org/exfor/servlet/X4sGetSubent?subID=O0091018}{O0091018} \\ 
 & 24.6 & \href{https://www-nds.iaea.org/exfor/servlet/X4sGetSubent?subID=O0091020}{O0091020} \\ 
 & 40.0 & \href{https://www-nds.iaea.org/exfor/servlet/X4sGetSubent?subID=O0162027}{O0162027} \\ 
 & 65.0 & \href{https://www-nds.iaea.org/exfor/servlet/X4sGetSubent?subID=O0032063S}{O0032063S} \\ 
\hline 
\niSixtyTwo & 20.4 & \href{https://www-nds.iaea.org/exfor/servlet/X4sGetSubent?subID=O0091022}{O0091022} \\ 
 & 24.6 & \href{https://www-nds.iaea.org/exfor/servlet/X4sGetSubent?subID=O0091024}{O0091024} \\ 
 & 65.0 & \href{https://www-nds.iaea.org/exfor/servlet/X4sGetSubent?subID=O0032064S}{O0032064S} \\ 
\hline 
\cuSixtyThree & 16.0 & \href{https://www-nds.iaea.org/exfor/servlet/X4sGetSubent?subID=C0893016}{C0893016} \\ 
\hline 
\niSixtyFour & 20.4 & \href{https://www-nds.iaea.org/exfor/servlet/X4sGetSubent?subID=O0169004}{O0169004} \\ 
 & 65.0 & \href{https://www-nds.iaea.org/exfor/servlet/X4sGetSubent?subID=O0032065S}{O0032065S} \\ 
\hline 
\znSixtyFour & 20.4 & \href{https://www-nds.iaea.org/exfor/servlet/X4sGetSubent?subID=O1109002}{O1109002} \\ 
\hline 
\cuSixtyFive & 16.0 & \href{https://www-nds.iaea.org/exfor/servlet/X4sGetSubent?subID=C0893018}{C0893018} \\ 
\hline 
\znSixtySix & 20.4 & \href{https://www-nds.iaea.org/exfor/servlet/X4sGetSubent?subID=O1109004}{O1109004} \\ 
\hline 
\znSixtyEight & 20.4 & \href{https://www-nds.iaea.org/exfor/servlet/X4sGetSubent?subID=O1109006}{O1109006} \\ 
 & 40.0 & \href{https://www-nds.iaea.org/exfor/servlet/X4sGetSubent?subID=O0328005}{O0328005} \\ 
\hline 
\znSeventy & 20.4 & \href{https://www-nds.iaea.org/exfor/servlet/X4sGetSubent?subID=O1109008}{O1109008} \\ 
\hline 
\geSeventyTwo & 22.3 & \href{https://www-nds.iaea.org/exfor/servlet/X4sGetSubent?subID=O1103004}{O1103004} \\ 
\hline 
\geSeventyFour & 22.3 & \href{https://www-nds.iaea.org/exfor/servlet/X4sGetSubent?subID=O1103006}{O1103006} \\ 
\hline 
\seSeventySix & 16.0 & \href{https://www-nds.iaea.org/exfor/servlet/X4sGetSubent?subID=C0893020}{C0893020} \\ 
\hline 
\seSeventyEight & 16.0 & \href{https://www-nds.iaea.org/exfor/servlet/X4sGetSubent?subID=C0893022}{C0893022} \\ 
 & 22.3 & \href{https://www-nds.iaea.org/exfor/servlet/X4sGetSubent?subID=O1103012}{O1103012} \\ 
\hline 
\seEighty & 16.0 & \href{https://www-nds.iaea.org/exfor/servlet/X4sGetSubent?subID=C0893024}{C0893024} \\ 
 & 22.3 & \href{https://www-nds.iaea.org/exfor/servlet/X4sGetSubent?subID=O1103014}{O1103014} \\ 
\hline 
\seEightyTwo & 16.0 & \href{https://www-nds.iaea.org/exfor/servlet/X4sGetSubent?subID=C0893026}{C0893026} \\ 
\hline 
\srEightySix & 24.6 & \href{https://www-nds.iaea.org/exfor/servlet/X4sGetSubent?subID=O0169006}{O0169006} \\ 
\hline 
\srEightyEight & 24.6 & \href{https://www-nds.iaea.org/exfor/servlet/X4sGetSubent?subID=O0169008}{O0169008} \\ 
\hline 
\yEightyNine & 65.0 & \href{https://www-nds.iaea.org/exfor/servlet/X4sGetSubent?subID=O0032066S}{O0032066S} \\ 
\hline 
\zrNinety & 16.0 & \href{https://www-nds.iaea.org/exfor/servlet/X4sGetSubent?subID=C0893028}{C0893028} \\ 
 & 40.0 & \href{https://www-nds.iaea.org/exfor/servlet/X4sGetSubent?subID=O0328006}{O0328006} \\ 
 & 65.0 & \href{https://www-nds.iaea.org/exfor/servlet/X4sGetSubent?subID=O0032021S}{O0032021S} \\ 
\hline 
\moNinetyEight & 65.0 & \href{https://www-nds.iaea.org/exfor/servlet/X4sGetSubent?subID=O0032067S}{O0032067S} \\ 
\hline 
\moOneHundred & 65.0 & \href{https://www-nds.iaea.org/exfor/servlet/X4sGetSubent?subID=O0032068S}{O0032068S} \\ 
\hline 
\cdOneHundredSix & 22.3 & \href{https://www-nds.iaea.org/exfor/servlet/X4sGetSubent?subID=O1104002}{O1104002} \\ 
\hline 
\cdOneHundredEight & 22.3 & \href{https://www-nds.iaea.org/exfor/servlet/X4sGetSubent?subID=O1104004}{O1104004} \\ 
\hline 
\cdOneHundredTen & 20.4 & \href{https://www-nds.iaea.org/exfor/servlet/X4sGetSubent?subID=O0169010}{O0169010} \\ 
 & 22.3 & \href{https://www-nds.iaea.org/exfor/servlet/X4sGetSubent?subID=O1104006}{O1104006} \\ 
\hline 
\cdOneHundredTwelve & 20.4 & \href{https://www-nds.iaea.org/exfor/servlet/X4sGetSubent?subID=O0169012}{O0169012} \\ 
 & 22.3 & \href{https://www-nds.iaea.org/exfor/servlet/X4sGetSubent?subID=O1104008}{O1104008} \\ 
\hline 
\cdOneHundredFourteen & 20.4 & \href{https://www-nds.iaea.org/exfor/servlet/X4sGetSubent?subID=O0169014}{O0169014} \\ 
 & 22.3 & \href{https://www-nds.iaea.org/exfor/servlet/X4sGetSubent?subID=O1104010}{O1104010} \\ 
\hline 
\cdOneHundredSixteen & 22.3 & \href{https://www-nds.iaea.org/exfor/servlet/X4sGetSubent?subID=O1104012}{O1104012} \\ 
\hline 
\snOneHundredSixteen & 16.0 & \href{https://www-nds.iaea.org/exfor/servlet/X4sGetSubent?subID=C0893030}{C0893030} \\ 
 & 20.4 & \href{https://www-nds.iaea.org/exfor/servlet/X4sGetSubent?subID=O0169016}{O0169016} \\ 
\hline 
\snOneHundredEighteen & 20.4 & \href{https://www-nds.iaea.org/exfor/servlet/X4sGetSubent?subID=O0169018}{O0169018} \\ 
\hline 
\snOneHundredTwenty & 16.0 & \href{https://www-nds.iaea.org/exfor/servlet/X4sGetSubent?subID=C0893032}{C0893032} \\ 
 & 20.4 & \href{https://www-nds.iaea.org/exfor/servlet/X4sGetSubent?subID=O0169020}{O0169020} \\ 
 & 24.6 & \href{https://www-nds.iaea.org/exfor/servlet/X4sGetSubent?subID=O0169026}{O0169026} \\ 
 & 40.0 & \href{https://www-nds.iaea.org/exfor/servlet/X4sGetSubent?subID=O0328007}{O0328007} \\ 
\hline 
\snOneHundredTwentyTwo & 20.4 & \href{https://www-nds.iaea.org/exfor/servlet/X4sGetSubent?subID=O0169022}{O0169022} \\ 
\hline 
\snOneHundredTwentyFour & 16.0 & \href{https://www-nds.iaea.org/exfor/servlet/X4sGetSubent?subID=C0893034}{C0893034} \\ 
 & 20.4 & \href{https://www-nds.iaea.org/exfor/servlet/X4sGetSubent?subID=O0169024}{O0169024} \\ 
\hline 
\baOneHundredThirtyFour & 16.0 & \href{https://www-nds.iaea.org/exfor/servlet/X4sGetSubent?subID=C0893036}{C0893036} \\ 
\hline 
\baOneHundredThirtySix & 16.0 & \href{https://www-nds.iaea.org/exfor/servlet/X4sGetSubent?subID=C0893038}{C0893038} \\ 
\hline 
\baOneHundredThirtyEight & 16.0 & \href{https://www-nds.iaea.org/exfor/servlet/X4sGetSubent?subID=C0893040}{C0893040} \\ 
\hline 
\smOneHundredFortyFour & 65.0 & \href{https://www-nds.iaea.org/exfor/servlet/X4sGetSubent?subID=O0032069S}{O0032069S} \\ 
\hline 
\pbTwoHundredEight & 40.0 & \href{https://www-nds.iaea.org/exfor/servlet/X4sGetSubent?subID=O0328008}{O0328008} \\ 
 & 65.0 & \href{https://www-nds.iaea.org/exfor/servlet/X4sGetSubent?subID=O0032025S}{O0032025S} \\ 
\hline 
\biTwoHundredNine & 65.0 & \href{https://www-nds.iaea.org/exfor/servlet/X4sGetSubent?subID=O0032070S}{O0032070S} \\ 
\hline 

        \end{longtable}
        \label{ECS_p_bi209}

\subsection*{Comments}
\begin{itemize}
    \item[] \baOneHundredThirtyFour, \baOneHundredThirtySix: the data sets for these nuclei from Varner, PhD thesis (1986) are listed in the CH89 Corpus as being from \baOneHundredThirtyTwo\ and \baOneHundredThirtyFour, but instead listed in EXFOR as being from \baOneHundredThirtyFour\ and \baOneHundredThirtySix. As \baOneHundredThirtyTwo\ is 0.1\% naturally abundant, it is highly unlikely this was actually used as a target. For the CHUQ corpus, we assigned the data sets to \baOneHundredThirtyFour\ and \baOneHundredThirtySix, respectively.
    \item[] \niFiftyEight: we were unable to locate the data set listed in the CH89 corpus as having a scattering energy of 27.2 \MeV. This scattering energy was not reported (to our knowledge) in any of the references cited in the CH89 corpus for the Eindhoven (p,p) scattering data sets.
    \item[] \moNinetyEight: in the data set from Sakaguchi et al. (Memoirs Faculty of Sci., Kyoto Univ. Ser. Phys. 36 p.305 (1983)), the cross section value recorded in EXFOR at an angle of 58.03 degrees is anomalously lower than the adjacent cross sections reported at 55.52 and 60.54 degrees, possibly due to a tabulation error. For the CHUQ corpus we retained the anomalous value.
    \item[] \ndOneHundredFortyTwo: we were unable to locate the data set listed in the CH89 corpus with a scattering energy of 17.2 \MeV. To our knowledge, no data for proton scattering on \ndOneHundredFortyTwo\ are available in EXFOR, nor is \ndOneHundredFortyTwo\ listed as a target under any of the references cited in the CH89 corpus for the Eindhoven (p,p) scattering data sets.
    \item[] \smOneHundredFortyEight, \smOneHundredFifty: we were unable to locate the data sets listed in the CH89 corpus with a scattering energy of 20.4 \MeV. To our knowledge, no data for proton scattering on Sm isotopes at or around 20 MeV are available in EXFOR, nor are any Sm isotopes listed as a target under any of the references cited in the CH89 corpus for the Eindhoven (p,p) scattering data sets.
    \item[] \caForty, \feFiftyFour, \niFiftyEight, \niSixty, \coFiftyNine, \znSixtyEight, \zrNinety, \snOneHundredTwenty, \pbTwoHundredEight: as listed in EXFOR, the data sets for these nuclei, associated with two publications by Fricke et al. (Phys. Rev. 163 p.1153 (1967); Phys. Rev. 156 p.1207 (1967)) have mostly null values in the ERR-T column. It was unclear to us how to combine or assess the few given ERR-T data with the error data in the other provided error columns. As such, we ignored the ERR-T column in these cases and took the DATA-ERR column for the total error.
\end{itemize}

\clearpage 
\subsection*{Proton analyzing powers}
\begin{longtable}{c c c}
                Isotope & Energies & EXFOR Acc. \# \\ 
\caForty & 65.0 & \href{https://www-nds.iaea.org/exfor/servlet/X4sGetSubent?subID=O0032002A}{O0032002A} \\ 
\hline 
\caFortyFour & 65.0 & \href{https://www-nds.iaea.org/exfor/servlet/X4sGetSubent?subID=O0032056A}{O0032056A} \\ 
\hline 
\tiFortySix & 65.0 & \href{https://www-nds.iaea.org/exfor/servlet/X4sGetSubent?subID=O0032058A}{O0032058A} \\ 
\hline 
\caFortyEight & 65.0 & \href{https://www-nds.iaea.org/exfor/servlet/X4sGetSubent?subID=O0032057A}{O0032057A} \\ 
\hline 
\tiFortyEight & 16.0 & \href{https://www-nds.iaea.org/exfor/servlet/X4sGetSubent?subID=C0893005}{C0893005} \\ 
 & 65.0 & \href{https://www-nds.iaea.org/exfor/servlet/X4sGetSubent?subID=O0032059A}{O0032059A} \\ 
\hline 
\tiFifty & 16.0 & \href{https://www-nds.iaea.org/exfor/servlet/X4sGetSubent?subID=C0893007}{C0893007} \\ 
 & 65.0 & \href{https://www-nds.iaea.org/exfor/servlet/X4sGetSubent?subID=O0032060A}{O0032060A} \\ 
\hline 
\feFiftyFour & 16.0 & \href{https://www-nds.iaea.org/exfor/servlet/X4sGetSubent?subID=C0893009}{C0893009} \\ 
 & 17.2 & \href{https://www-nds.iaea.org/exfor/servlet/X4sGetSubent?subID=O0091003}{O0091003} \\ 
 & 20.4 & \href{https://www-nds.iaea.org/exfor/servlet/X4sGetSubent?subID=O0091005}{O0091005} \\ 
 & 24.6 & \href{https://www-nds.iaea.org/exfor/servlet/X4sGetSubent?subID=O0091007}{O0091007} \\ 
 & 40.0 & \href{https://www-nds.iaea.org/exfor/servlet/X4sGetSubent?subID=O0162024}{O0162024} \\ 
 & 65.0 & \href{https://www-nds.iaea.org/exfor/servlet/X4sGetSubent?subID=O0032061A}{O0032061A} \\ 
\hline 
\feFiftySix & 17.2 & \href{https://www-nds.iaea.org/exfor/servlet/X4sGetSubent?subID=O0091009}{O0091009} \\ 
 & 20.4 & \href{https://www-nds.iaea.org/exfor/servlet/X4sGetSubent?subID=O0091011}{O0091011} \\ 
 & 24.6 & \href{https://www-nds.iaea.org/exfor/servlet/X4sGetSubent?subID=O0091013}{O0091013} \\ 
 & 65.0 & \href{https://www-nds.iaea.org/exfor/servlet/X4sGetSubent?subID=O0032062A}{O0032062A} \\ 
\hline 
\niFiftyEight & 16.0 & \href{https://www-nds.iaea.org/exfor/servlet/X4sGetSubent?subID=C0893013}{C0893013} \\ 
 & 20.4 & \href{https://www-nds.iaea.org/exfor/servlet/X4sGetSubent?subID=O0091015}{O0091015} \\ 
 & 24.6 & \href{https://www-nds.iaea.org/exfor/servlet/X4sGetSubent?subID=O0091017}{O0091017} \\ 
 & 40.0 & \href{https://www-nds.iaea.org/exfor/servlet/X4sGetSubent?subID=O0162026}{O0162026} \\ 
 & 65.0 & \href{https://www-nds.iaea.org/exfor/servlet/X4sGetSubent?subID=O0032008A}{O0032008A} \\ 
\hline 
\coFiftyNine & 40.0 & \href{https://www-nds.iaea.org/exfor/servlet/X4sGetSubent?subID=O0328009}{O0328009} \\ 
 & 65.0 & \href{https://www-nds.iaea.org/exfor/servlet/X4sGetSubent?subID=O0032015A}{O0032015A} \\ 
\hline 
\niSixty & 16.0 & \href{https://www-nds.iaea.org/exfor/servlet/X4sGetSubent?subID=C0893015}{C0893015} \\ 
 & 20.4 & \href{https://www-nds.iaea.org/exfor/servlet/X4sGetSubent?subID=O0091019}{O0091019} \\ 
 & 24.6 & \href{https://www-nds.iaea.org/exfor/servlet/X4sGetSubent?subID=O0091021}{O0091021} \\ 
 & 40.0 & \href{https://www-nds.iaea.org/exfor/servlet/X4sGetSubent?subID=O0162028}{O0162028} \\ 
 & 65.0 & \href{https://www-nds.iaea.org/exfor/servlet/X4sGetSubent?subID=O0032063A}{O0032063A} \\ 
\hline 
\niSixtyTwo & 20.4 & \href{https://www-nds.iaea.org/exfor/servlet/X4sGetSubent?subID=O0091023}{O0091023} \\ 
 & 24.6 & \href{https://www-nds.iaea.org/exfor/servlet/X4sGetSubent?subID=O0091025}{O0091025} \\ 
 & 65.0 & \href{https://www-nds.iaea.org/exfor/servlet/X4sGetSubent?subID=O0032064A}{O0032064A} \\ 
\hline 
\cuSixtyThree & 16.0 & \href{https://www-nds.iaea.org/exfor/servlet/X4sGetSubent?subID=C0893017}{C0893017} \\ 
\hline 
\niSixtyFour & 20.4 & \href{https://www-nds.iaea.org/exfor/servlet/X4sGetSubent?subID=O0169005}{O0169005} \\ 
 & 65.0 & \href{https://www-nds.iaea.org/exfor/servlet/X4sGetSubent?subID=O0032065A}{O0032065A} \\ 
\hline 
\znSixtyFour & 20.4 & \href{https://www-nds.iaea.org/exfor/servlet/X4sGetSubent?subID=O1109003}{O1109003} \\ 
\hline 
\cuSixtyFive & 16.0 & \href{https://www-nds.iaea.org/exfor/servlet/X4sGetSubent?subID=C0893019}{C0893019} \\ 
\hline 
\znSixtySix & 20.4 & \href{https://www-nds.iaea.org/exfor/servlet/X4sGetSubent?subID=O1109005}{O1109005} \\ 
\hline 
\znSixtyEight & 20.4 & \href{https://www-nds.iaea.org/exfor/servlet/X4sGetSubent?subID=O1109007}{O1109007} \\ 
 & 40.0 & \href{https://www-nds.iaea.org/exfor/servlet/X4sGetSubent?subID=O0328010}{O0328010} \\ 
\hline 
\znSeventy & 20.4 & \href{https://www-nds.iaea.org/exfor/servlet/X4sGetSubent?subID=O1109009}{O1109009} \\ 
\hline 
\geSeventyTwo & 22.3 & \href{https://www-nds.iaea.org/exfor/servlet/X4sGetSubent?subID=O1103005}{O1103005} \\ 
\hline 
\geSeventyFour & 22.3 & \href{https://www-nds.iaea.org/exfor/servlet/X4sGetSubent?subID=O1103007}{O1103007} \\ 
\hline 
\seSeventySix & 16.0 & \href{https://www-nds.iaea.org/exfor/servlet/X4sGetSubent?subID=C0893021}{C0893021} \\ 
\hline 
\seSeventyEight & 16.0 & \href{https://www-nds.iaea.org/exfor/servlet/X4sGetSubent?subID=C0893023}{C0893023} \\ 
 & 22.3 & \href{https://www-nds.iaea.org/exfor/servlet/X4sGetSubent?subID=O1103013}{O1103013} \\ 
\hline 
\seEighty & 16.0 & \href{https://www-nds.iaea.org/exfor/servlet/X4sGetSubent?subID=C0893025}{C0893025} \\ 
 & 22.3 & \href{https://www-nds.iaea.org/exfor/servlet/X4sGetSubent?subID=O1103015}{O1103015} \\ 
\hline 
\seEightyTwo & 16.0 & \href{https://www-nds.iaea.org/exfor/servlet/X4sGetSubent?subID=C0893027}{C0893027} \\ 
\hline 
\srEightySix & 24.6 & \href{https://www-nds.iaea.org/exfor/servlet/X4sGetSubent?subID=O0169007}{O0169007} \\ 
\hline 
\srEightyEight & 24.6 & \href{https://www-nds.iaea.org/exfor/servlet/X4sGetSubent?subID=O0169009}{O0169009} \\ 
\hline 
\yEightyNine & 65.0 & \href{https://www-nds.iaea.org/exfor/servlet/X4sGetSubent?subID=O0032066A}{O0032066A} \\ 
\hline 
\zrNinety & 16.0 & \href{https://www-nds.iaea.org/exfor/servlet/X4sGetSubent?subID=C0893029}{C0893029} \\ 
 & 65.0 & \href{https://www-nds.iaea.org/exfor/servlet/X4sGetSubent?subID=O0032021A}{O0032021A} \\ 
\hline 
\moNinetyEight & 65.0 & \href{https://www-nds.iaea.org/exfor/servlet/X4sGetSubent?subID=O0032067A}{O0032067A} \\ 
\hline 
\moOneHundred & 65.0 & \href{https://www-nds.iaea.org/exfor/servlet/X4sGetSubent?subID=O0032068A}{O0032068A} \\ 
\hline 
\cdOneHundredSix & 22.3 & \href{https://www-nds.iaea.org/exfor/servlet/X4sGetSubent?subID=O1104003}{O1104003} \\ 
\hline 
\cdOneHundredEight & 22.3 & \href{https://www-nds.iaea.org/exfor/servlet/X4sGetSubent?subID=O1104005}{O1104005} \\ 
\hline 
\cdOneHundredTen & 20.4 & \href{https://www-nds.iaea.org/exfor/servlet/X4sGetSubent?subID=O0169011}{O0169011} \\ 
 & 22.3 & \href{https://www-nds.iaea.org/exfor/servlet/X4sGetSubent?subID=O1104007}{O1104007} \\ 
\hline 
\cdOneHundredTwelve & 20.4 & \href{https://www-nds.iaea.org/exfor/servlet/X4sGetSubent?subID=O0169013}{O0169013} \\ 
 & 22.3 & \href{https://www-nds.iaea.org/exfor/servlet/X4sGetSubent?subID=O1104009}{O1104009} \\ 
\hline 
\cdOneHundredFourteen & 20.4 & \href{https://www-nds.iaea.org/exfor/servlet/X4sGetSubent?subID=O0169015}{O0169015} \\ 
 & 22.3 & \href{https://www-nds.iaea.org/exfor/servlet/X4sGetSubent?subID=O1104011}{O1104011} \\ 
\hline 
\cdOneHundredSixteen & 22.3 & \href{https://www-nds.iaea.org/exfor/servlet/X4sGetSubent?subID=O1104013}{O1104013} \\ 
\hline 
\snOneHundredSixteen & 16.0 & \href{https://www-nds.iaea.org/exfor/servlet/X4sGetSubent?subID=C0893031}{C0893031} \\ 
 & 20.4 & \href{https://www-nds.iaea.org/exfor/servlet/X4sGetSubent?subID=O0169017}{O0169017} \\ 
\hline 
\snOneHundredEighteen & 20.4 & \href{https://www-nds.iaea.org/exfor/servlet/X4sGetSubent?subID=O0169019}{O0169019} \\ 
\hline 
\snOneHundredTwenty & 16.0 & \href{https://www-nds.iaea.org/exfor/servlet/X4sGetSubent?subID=C0893033}{C0893033} \\ 
 & 20.4 & \href{https://www-nds.iaea.org/exfor/servlet/X4sGetSubent?subID=O0169021}{O0169021} \\ 
 & 24.6 & \href{https://www-nds.iaea.org/exfor/servlet/X4sGetSubent?subID=O0169027}{O0169027} \\ 
 & 40.0 & \href{https://www-nds.iaea.org/exfor/servlet/X4sGetSubent?subID=O0328011}{O0328011} \\ 
\hline 
\snOneHundredTwentyTwo & 20.4 & \href{https://www-nds.iaea.org/exfor/servlet/X4sGetSubent?subID=O0169023}{O0169023} \\ 
\hline 
\snOneHundredTwentyFour & 16.0 & \href{https://www-nds.iaea.org/exfor/servlet/X4sGetSubent?subID=C0893035}{C0893035} \\ 
 & 20.4 & \href{https://www-nds.iaea.org/exfor/servlet/X4sGetSubent?subID=O0169025}{O0169025} \\ 
\hline 
\baOneHundredThirtyFour & 16.0 & \href{https://www-nds.iaea.org/exfor/servlet/X4sGetSubent?subID=C0893037}{C0893037} \\ 
\hline 
\baOneHundredThirtySix & 16.0 & \href{https://www-nds.iaea.org/exfor/servlet/X4sGetSubent?subID=C0893039}{C0893039} \\ 
\hline 
\baOneHundredThirtyEight & 16.0 & \href{https://www-nds.iaea.org/exfor/servlet/X4sGetSubent?subID=C0893041}{C0893041} \\ 
\hline 
\smOneHundredFortyFour & 65.0 & \href{https://www-nds.iaea.org/exfor/servlet/X4sGetSubent?subID=O0032069A}{O0032069A} \\ 
\hline 
\pbTwoHundredEight & 65.0 & \href{https://www-nds.iaea.org/exfor/servlet/X4sGetSubent?subID=O0032025A}{O0032025A} \\ 
\hline 
\biTwoHundredNine & 65.0 & \href{https://www-nds.iaea.org/exfor/servlet/X4sGetSubent?subID=O0032070A}{O0032070A} \\ 
\hline 

        \end{longtable}
        \label{APower_p_bi209}

\subsection*{Comments}
\begin{itemize}
    \item[] \baOneHundredThirtyFour, \baOneHundredThirtySix: the data sets for these nuclei from Varner, PhD thesis (1986) are listed in the CH89 Corpus as being from \baOneHundredThirtyTwo\ and \baOneHundredThirtyFour, but instead listed in EXFOR as being from \baOneHundredThirtyFour\ and \baOneHundredThirtySix. As \baOneHundredThirtyTwo\ is 0.1\% naturally abundant, it is highly unlikely this was actually used as a target. For the CHUQ corpus, we assigned the data sets to \baOneHundredThirtyFour\ and \baOneHundredThirtySix, respectively.
    \item[] \niFiftyEight: we were unable to locate the data set listed in the CH89 corpus with a scattering energy of 27.2 \MeV. This scattering energy was not reported (to our knowledge) in any of the references cited in the CH89 corpus for the Eindhoven (p,p) scattering data sets.
    \item[] \ndOneHundredFortyTwo: we were unable to locate the data set listed in the CH89 corpus with a scattering energy of 17.2 \MeV. To our knowledge, no data for proton scattering on \ndOneHundredFortyTwo\ are available in EXFOR, nor is \ndOneHundredFortyTwo\ listed as a target under any of the references cited in the CH89 corpus for the Eindhoven (p,p) scattering data sets.
    \item[] \smOneHundredFortyEight, \smOneHundredFifty: we were unable to locate the data sets listed in the CH89 corpus with a scattering energy of 20.4 \MeV. To our knowledge, no data for proton scattering on Sm isotopes at or around 20 MeV are available in EXFOR, nor are any Sm isotopes listed as a target under any of the references cited in the CH89 corpus for the Eindhoven (p,p) scattering data sets.
    \item[] \feFiftyFour, \niFiftyEight, \niSixty, \niSixtyFour, \coFiftyNine, \znSixtyFour, \znSixtySix, \znSixtyEight, \znSeventy, \geSeventyTwo, \geSeventyFour, \seSeventyEight, \seEighty, \srEightySix, \srEightyEight, \cdOneHundredTen, \cdOneHundredTwelve, \cdOneHundredFourteen, \snOneHundredSixteen, \snOneHundredEighteen, \snOneHundredTwenty, \snOneHundredTwentyTwo, \snOneHundredTwentyFour: as listed in EXFOR, the data sets for these nuclei, associated with two publications by Fricke et al. (Phys. Rev. 163 p.1153 (1967); Phys. Rev. 156 p.1207 (1967)), one publication by Moonen et al. (J. Phys. G 19, p.635 (1993)), and one publication by Wassenaar et al. (J. Phys. G 15, p.181 (1989)) have mostly (or many) null values in the ERR-T column. It was unclear to us how to combine or assess the given ERR-T data with the error data in the other provided error columns. As such, we ignored the ERR-T column in these cases and took the DATA-ERR column for the total error.
\end{itemize}

\clearpage 
\section*{Test corpus}
\subsection*{Neutron differential elastic cross sections}
\twocolumngrid 
\begin{longtable}{c c c}
                Isotope & Energies & EXFOR Acc. \\ 
\alTwentySeven & 15.431 & \href{https://www-nds.iaea.org/exfor/servlet/X4sGetSubent?subID=13903002}{13903002} \\ 
\hline 
\siTwentyEight & 15.43 & \href{https://www-nds.iaea.org/exfor/servlet/X4sGetSubent?subID=14345002}{14345002} \\ 
 & 18.9 & \href{https://www-nds.iaea.org/exfor/servlet/X4sGetSubent?subID=14345002}{14345002} \\ 
\hline 
\sThirtyTwo & 7.96 & \href{https://www-nds.iaea.org/exfor/servlet/X4sGetSubent?subID=14345004}{14345004} \\ 
 & 9.95 & \href{https://www-nds.iaea.org/exfor/servlet/X4sGetSubent?subID=14345004}{14345004} \\ 
 & 11.93 & \href{https://www-nds.iaea.org/exfor/servlet/X4sGetSubent?subID=14345004}{14345004} \\ 
 & 13.92 & \href{https://www-nds.iaea.org/exfor/servlet/X4sGetSubent?subID=14345004}{14345004} \\ 
 & 15.44 & \href{https://www-nds.iaea.org/exfor/servlet/X4sGetSubent?subID=14345004}{14345004} \\ 
 & 16.92 & \href{https://www-nds.iaea.org/exfor/servlet/X4sGetSubent?subID=14345004}{14345004} \\ 
 & 18.9 & \href{https://www-nds.iaea.org/exfor/servlet/X4sGetSubent?subID=14345004}{14345004} \\ 
\hline 
\arNat & 6.0 & \href{https://www-nds.iaea.org/exfor/servlet/X4sGetSubent?subID=14371004}{14371004} \\ 
\hline 
\caForty & 11.9 & \href{https://www-nds.iaea.org/exfor/servlet/X4sGetSubent?subID=14303002}{14303002} \\ 
 & 16.9 & \href{https://www-nds.iaea.org/exfor/servlet/X4sGetSubent?subID=14303003}{14303003} \\ 
 & 65.0 & \href{https://www-nds.iaea.org/exfor/servlet/X4sGetSubent?subID=13946003}{13946003} \\ 
 & 75.0 & \href{https://www-nds.iaea.org/exfor/servlet/X4sGetSubent?subID=13946003}{13946003} \\ 
 & 85.0 & \href{https://www-nds.iaea.org/exfor/servlet/X4sGetSubent?subID=13946003}{13946003} \\ 
 & 95.0 & \href{https://www-nds.iaea.org/exfor/servlet/X4sGetSubent?subID=13946003}{13946003} \\ 
 & 107.5 & \href{https://www-nds.iaea.org/exfor/servlet/X4sGetSubent?subID=13946003}{13946003} \\ 
 & 127.5 & \href{https://www-nds.iaea.org/exfor/servlet/X4sGetSubent?subID=13946003}{13946003} \\ 
 & 155.0 & \href{https://www-nds.iaea.org/exfor/servlet/X4sGetSubent?subID=13946003}{13946003} \\ 
 & 185.0 & \href{https://www-nds.iaea.org/exfor/servlet/X4sGetSubent?subID=13946003}{13946003} \\ 
 & 225.0 & \href{https://www-nds.iaea.org/exfor/servlet/X4sGetSubent?subID=13946003}{13946003} \\ 
\hline 
\caFortyEight & 11.9 & \href{https://www-nds.iaea.org/exfor/servlet/X4sGetSubent?subID=14303004}{14303004} \\ 
 & 16.8 & \href{https://www-nds.iaea.org/exfor/servlet/X4sGetSubent?subID=14303005}{14303005} \\ 
\hline 
\feFiftyFour & 2.0 & \href{https://www-nds.iaea.org/exfor/servlet/X4sGetSubent?subID=14451008}{14451008} \\ 
 & 2.25 & \href{https://www-nds.iaea.org/exfor/servlet/X4sGetSubent?subID=14451008}{14451008} \\ 
 & 2.5 & \href{https://www-nds.iaea.org/exfor/servlet/X4sGetSubent?subID=14451008}{14451008} \\ 
 & 2.75 & \href{https://www-nds.iaea.org/exfor/servlet/X4sGetSubent?subID=14451008}{14451008} \\ 
 & 3.0 & \href{https://www-nds.iaea.org/exfor/servlet/X4sGetSubent?subID=14451008}{14451008} \\ 
 & 3.5 & \href{https://www-nds.iaea.org/exfor/servlet/X4sGetSubent?subID=14451008}{14451008} \\ 
 & 4.0 & \href{https://www-nds.iaea.org/exfor/servlet/X4sGetSubent?subID=14451008}{14451008} \\ 
 & 5.0 & \href{https://www-nds.iaea.org/exfor/servlet/X4sGetSubent?subID=14451008}{14451008} \\ 
 & 6.0 & \href{https://www-nds.iaea.org/exfor/servlet/X4sGetSubent?subID=14451008}{14451008} \\ 
\hline 
\feNat & 1.75 & \href{https://www-nds.iaea.org/exfor/servlet/X4sGetSubent?subID=14451003}{14451003} \\ 
 & 8.17 & \href{https://www-nds.iaea.org/exfor/servlet/X4sGetSubent?subID=32673002}{32673002} \\ 
\hline 
\feFiftySix & 1.3 & \href{https://www-nds.iaea.org/exfor/servlet/X4sGetSubent?subID=14462002}{14462002} \\ 
 & 1.5 & \href{https://www-nds.iaea.org/exfor/servlet/X4sGetSubent?subID=14462002}{14462002} \\ 
 & 1.8 & \href{https://www-nds.iaea.org/exfor/servlet/X4sGetSubent?subID=14462002}{14462002} \\ 
 & 2.0 & \href{https://www-nds.iaea.org/exfor/servlet/X4sGetSubent?subID=14462002}{14462002} \\ 
 & 2.25 & \href{https://www-nds.iaea.org/exfor/servlet/X4sGetSubent?subID=14462002}{14462002} \\ 
 & 2.5 & \href{https://www-nds.iaea.org/exfor/servlet/X4sGetSubent?subID=14462002}{14462002} \\ 
 & 2.75 & \href{https://www-nds.iaea.org/exfor/servlet/X4sGetSubent?subID=14462002}{14462002} \\ 
 & 3.0 & \href{https://www-nds.iaea.org/exfor/servlet/X4sGetSubent?subID=14462002}{14462002} \\ 
 & 3.5 & \href{https://www-nds.iaea.org/exfor/servlet/X4sGetSubent?subID=14462002}{14462002} \\ 
 & 4.0 & \href{https://www-nds.iaea.org/exfor/servlet/X4sGetSubent?subID=14462002}{14462002} \\ 
 & 4.5 & \href{https://www-nds.iaea.org/exfor/servlet/X4sGetSubent?subID=14462002}{14462002} \\ 
 & 4.9 & \href{https://www-nds.iaea.org/exfor/servlet/X4sGetSubent?subID=14462002}{14462002} \\ 
 & 5.94 & \href{https://www-nds.iaea.org/exfor/servlet/X4sGetSubent?subID=14462002}{14462002} \\ 
 & 6.96 & \href{https://www-nds.iaea.org/exfor/servlet/X4sGetSubent?subID=14462002}{14462002} \\ 
 & 7.96 & \href{https://www-nds.iaea.org/exfor/servlet/X4sGetSubent?subID=14462002}{14462002} \\ 
 & 96.0 & \href{https://www-nds.iaea.org/exfor/servlet/X4sGetSubent?subID=22987002}{22987002} \\ 
 & 96.0 & \href{https://www-nds.iaea.org/exfor/servlet/X4sGetSubent?subID=22987005}{22987005} \\ 
 & 96.0 & \href{https://www-nds.iaea.org/exfor/servlet/X4sGetSubent?subID=23059003}{23059003} \\ 
\hline 
\coFiftyNine & 9.953 & \href{https://www-nds.iaea.org/exfor/servlet/X4sGetSubent?subID=13903004}{13903004} \\ 
 & 11.944 & \href{https://www-nds.iaea.org/exfor/servlet/X4sGetSubent?subID=13903004}{13903004} \\ 
 & 15.425 & \href{https://www-nds.iaea.org/exfor/servlet/X4sGetSubent?subID=13903004}{13903004} \\ 
 & 16.879 & \href{https://www-nds.iaea.org/exfor/servlet/X4sGetSubent?subID=13903004}{13903004} \\ 
 & 18.862 & \href{https://www-nds.iaea.org/exfor/servlet/X4sGetSubent?subID=13903004}{13903004} \\ 
\hline 
\cuNat & 6.95 & \href{https://www-nds.iaea.org/exfor/servlet/X4sGetSubent?subID=22974002}{22974002} \\ 
 & 8.07 & \href{https://www-nds.iaea.org/exfor/servlet/X4sGetSubent?subID=22974002}{22974002} \\ 
 & 8.96 & \href{https://www-nds.iaea.org/exfor/servlet/X4sGetSubent?subID=22974002}{22974002} \\ 
 & 9.9 & \href{https://www-nds.iaea.org/exfor/servlet/X4sGetSubent?subID=22974002}{22974002} \\ 
 & 10.86 & \href{https://www-nds.iaea.org/exfor/servlet/X4sGetSubent?subID=22974002}{22974002} \\ 
 & 11.9 & \href{https://www-nds.iaea.org/exfor/servlet/X4sGetSubent?subID=22974002}{22974002} \\ 
 & 12.93 & \href{https://www-nds.iaea.org/exfor/servlet/X4sGetSubent?subID=22974002}{22974002} \\ 
 & 14.18 & \href{https://www-nds.iaea.org/exfor/servlet/X4sGetSubent?subID=22974002}{22974002} \\ 
\hline 
\yEightyNine & 96.0 & \href{https://www-nds.iaea.org/exfor/servlet/X4sGetSubent?subID=22987003}{22987003} \\ 
\hline 
\snOneHundredTwelve & 11.0 & \href{https://www-nds.iaea.org/exfor/servlet/X4sGetSubent?subID=14662002}{14662002} \\ 
 & 17.0 & \href{https://www-nds.iaea.org/exfor/servlet/X4sGetSubent?subID=14662003}{14662003} \\ 
\hline 
\snOneHundredTwentyFour & 11.0 & \href{https://www-nds.iaea.org/exfor/servlet/X4sGetSubent?subID=14662004}{14662004} \\ 
 & 17.0 & \href{https://www-nds.iaea.org/exfor/servlet/X4sGetSubent?subID=14662005}{14662005} \\ 
\hline 
\gdNat & 0.334 & \href{https://www-nds.iaea.org/exfor/servlet/X4sGetSubent?subID=13894002}{13894002} \\ 
 & 0.456 & \href{https://www-nds.iaea.org/exfor/servlet/X4sGetSubent?subID=13894002}{13894002} \\ 
 & 0.55 & \href{https://www-nds.iaea.org/exfor/servlet/X4sGetSubent?subID=13894002}{13894002} \\ 
 & 0.649 & \href{https://www-nds.iaea.org/exfor/servlet/X4sGetSubent?subID=13894002}{13894002} \\ 
 & 0.785 & \href{https://www-nds.iaea.org/exfor/servlet/X4sGetSubent?subID=13894002}{13894002} \\ 
 & 0.919 & \href{https://www-nds.iaea.org/exfor/servlet/X4sGetSubent?subID=13894002}{13894002} \\ 
 & 1.08 & \href{https://www-nds.iaea.org/exfor/servlet/X4sGetSubent?subID=13894002}{13894002} \\ 
 & 1.264 & \href{https://www-nds.iaea.org/exfor/servlet/X4sGetSubent?subID=13894002}{13894002} \\ 
 & 1.432 & \href{https://www-nds.iaea.org/exfor/servlet/X4sGetSubent?subID=13894002}{13894002} \\ 
 & 4.51 & \href{https://www-nds.iaea.org/exfor/servlet/X4sGetSubent?subID=13894004}{13894004} \\ 
 & 5.01 & \href{https://www-nds.iaea.org/exfor/servlet/X4sGetSubent?subID=13894004}{13894004} \\ 
 & 5.51 & \href{https://www-nds.iaea.org/exfor/servlet/X4sGetSubent?subID=13894004}{13894004} \\ 
 & 5.91 & \href{https://www-nds.iaea.org/exfor/servlet/X4sGetSubent?subID=13894004}{13894004} \\ 
 & 6.51 & \href{https://www-nds.iaea.org/exfor/servlet/X4sGetSubent?subID=13894004}{13894004} \\ 
 & 7.14 & \href{https://www-nds.iaea.org/exfor/servlet/X4sGetSubent?subID=13894004}{13894004} \\ 
 & 7.51 & \href{https://www-nds.iaea.org/exfor/servlet/X4sGetSubent?subID=13894004}{13894004} \\ 
 & 8.03 & \href{https://www-nds.iaea.org/exfor/servlet/X4sGetSubent?subID=13894004}{13894004} \\ 
 & 8.41 & \href{https://www-nds.iaea.org/exfor/servlet/X4sGetSubent?subID=13894004}{13894004} \\ 
 & 9.06 & \href{https://www-nds.iaea.org/exfor/servlet/X4sGetSubent?subID=13894004}{13894004} \\ 
 & 9.51 & \href{https://www-nds.iaea.org/exfor/servlet/X4sGetSubent?subID=13894004}{13894004} \\ 
 & 9.99 & \href{https://www-nds.iaea.org/exfor/servlet/X4sGetSubent?subID=13894004}{13894004} \\ 
\hline 
\taOneHundredEightyOne & 0.323 & \href{https://www-nds.iaea.org/exfor/servlet/X4sGetSubent?subID=13965002}{13965002} \\ 
 & 0.37 & \href{https://www-nds.iaea.org/exfor/servlet/X4sGetSubent?subID=13965002}{13965002} \\ 
 & 0.414 & \href{https://www-nds.iaea.org/exfor/servlet/X4sGetSubent?subID=13965002}{13965002} \\ 
 & 0.469 & \href{https://www-nds.iaea.org/exfor/servlet/X4sGetSubent?subID=13965002}{13965002} \\ 
 & 0.522 & \href{https://www-nds.iaea.org/exfor/servlet/X4sGetSubent?subID=13965002}{13965002} \\ 
 & 0.568 & \href{https://www-nds.iaea.org/exfor/servlet/X4sGetSubent?subID=13965002}{13965002} \\ 
 & 0.615 & \href{https://www-nds.iaea.org/exfor/servlet/X4sGetSubent?subID=13965002}{13965002} \\ 
 & 0.666 & \href{https://www-nds.iaea.org/exfor/servlet/X4sGetSubent?subID=13965002}{13965002} \\ 
 & 0.719 & \href{https://www-nds.iaea.org/exfor/servlet/X4sGetSubent?subID=13965002}{13965002} \\ 
 & 0.767 & \href{https://www-nds.iaea.org/exfor/servlet/X4sGetSubent?subID=13965002}{13965002} \\ 
 & 0.815 & \href{https://www-nds.iaea.org/exfor/servlet/X4sGetSubent?subID=13965002}{13965002} \\ 
 & 0.867 & \href{https://www-nds.iaea.org/exfor/servlet/X4sGetSubent?subID=13965002}{13965002} \\ 
 & 0.911 & \href{https://www-nds.iaea.org/exfor/servlet/X4sGetSubent?subID=13965002}{13965002} \\ 
 & 0.967 & \href{https://www-nds.iaea.org/exfor/servlet/X4sGetSubent?subID=13965002}{13965002} \\ 
 & 1.015 & \href{https://www-nds.iaea.org/exfor/servlet/X4sGetSubent?subID=13965002}{13965002} \\ 
 & 1.068 & \href{https://www-nds.iaea.org/exfor/servlet/X4sGetSubent?subID=13965002}{13965002} \\ 
 & 1.11 & \href{https://www-nds.iaea.org/exfor/servlet/X4sGetSubent?subID=13965002}{13965002} \\ 
 & 1.163 & \href{https://www-nds.iaea.org/exfor/servlet/X4sGetSubent?subID=13965002}{13965002} \\ 
 & 1.213 & \href{https://www-nds.iaea.org/exfor/servlet/X4sGetSubent?subID=13965002}{13965002} \\ 
 & 1.261 & \href{https://www-nds.iaea.org/exfor/servlet/X4sGetSubent?subID=13965002}{13965002} \\ 
 & 1.309 & \href{https://www-nds.iaea.org/exfor/servlet/X4sGetSubent?subID=13965002}{13965002} \\ 
 & 1.359 & \href{https://www-nds.iaea.org/exfor/servlet/X4sGetSubent?subID=13965002}{13965002} \\ 
 & 1.415 & \href{https://www-nds.iaea.org/exfor/servlet/X4sGetSubent?subID=13965002}{13965002} \\ 
 & 1.465 & \href{https://www-nds.iaea.org/exfor/servlet/X4sGetSubent?subID=13965002}{13965002} \\ 
 & 4.51 & \href{https://www-nds.iaea.org/exfor/servlet/X4sGetSubent?subID=13965002}{13965002} \\ 
 & 5.01 & \href{https://www-nds.iaea.org/exfor/servlet/X4sGetSubent?subID=13965002}{13965002} \\ 
 & 5.51 & \href{https://www-nds.iaea.org/exfor/servlet/X4sGetSubent?subID=13965002}{13965002} \\ 
 & 5.91 & \href{https://www-nds.iaea.org/exfor/servlet/X4sGetSubent?subID=13965002}{13965002} \\ 
 & 6.51 & \href{https://www-nds.iaea.org/exfor/servlet/X4sGetSubent?subID=13965002}{13965002} \\ 
 & 7.14 & \href{https://www-nds.iaea.org/exfor/servlet/X4sGetSubent?subID=13965002}{13965002} \\ 
 & 7.51 & \href{https://www-nds.iaea.org/exfor/servlet/X4sGetSubent?subID=13965002}{13965002} \\ 
 & 8.03 & \href{https://www-nds.iaea.org/exfor/servlet/X4sGetSubent?subID=13965002}{13965002} \\ 
 & 8.41 & \href{https://www-nds.iaea.org/exfor/servlet/X4sGetSubent?subID=13965002}{13965002} \\ 
 & 9.06 & \href{https://www-nds.iaea.org/exfor/servlet/X4sGetSubent?subID=13965002}{13965002} \\ 
 & 9.51 & \href{https://www-nds.iaea.org/exfor/servlet/X4sGetSubent?subID=13965002}{13965002} \\ 
 & 9.99 & \href{https://www-nds.iaea.org/exfor/servlet/X4sGetSubent?subID=13965002}{13965002} \\ 
\hline 
\wNat & 7.19 & \href{https://www-nds.iaea.org/exfor/servlet/X4sGetSubent?subID=22962002}{22962002} \\ 
 & 8.08 & \href{https://www-nds.iaea.org/exfor/servlet/X4sGetSubent?subID=22962002}{22962002} \\ 
 & 9.08 & \href{https://www-nds.iaea.org/exfor/servlet/X4sGetSubent?subID=22962002}{22962002} \\ 
 & 10.11 & \href{https://www-nds.iaea.org/exfor/servlet/X4sGetSubent?subID=22962002}{22962002} \\ 
 & 11.04 & \href{https://www-nds.iaea.org/exfor/servlet/X4sGetSubent?subID=22962002}{22962002} \\ 
 & 12.12 & \href{https://www-nds.iaea.org/exfor/servlet/X4sGetSubent?subID=22962002}{22962002} \\ 
 & 13.1 & \href{https://www-nds.iaea.org/exfor/servlet/X4sGetSubent?subID=22962002}{22962002} \\ 
 & 14.1 & \href{https://www-nds.iaea.org/exfor/servlet/X4sGetSubent?subID=22962002}{22962002} \\ 
\hline 
\reNat & 0.352 & \href{https://www-nds.iaea.org/exfor/servlet/X4sGetSubent?subID=13878002}{13878002} \\ 
 & 0.446 & \href{https://www-nds.iaea.org/exfor/servlet/X4sGetSubent?subID=13878002}{13878002} \\ 
 & 0.544 & \href{https://www-nds.iaea.org/exfor/servlet/X4sGetSubent?subID=13878002}{13878002} \\ 
 & 0.643 & \href{https://www-nds.iaea.org/exfor/servlet/X4sGetSubent?subID=13878002}{13878002} \\ 
 & 0.748 & \href{https://www-nds.iaea.org/exfor/servlet/X4sGetSubent?subID=13878002}{13878002} \\ 
 & 0.846 & \href{https://www-nds.iaea.org/exfor/servlet/X4sGetSubent?subID=13878002}{13878002} \\ 
 & 0.944 & \href{https://www-nds.iaea.org/exfor/servlet/X4sGetSubent?subID=13878002}{13878002} \\ 
 & 1.034 & \href{https://www-nds.iaea.org/exfor/servlet/X4sGetSubent?subID=13878002}{13878002} \\ 
 & 1.144 & \href{https://www-nds.iaea.org/exfor/servlet/X4sGetSubent?subID=13878002}{13878002} \\ 
 & 1.241 & \href{https://www-nds.iaea.org/exfor/servlet/X4sGetSubent?subID=13878002}{13878002} \\ 
 & 1.345 & \href{https://www-nds.iaea.org/exfor/servlet/X4sGetSubent?subID=13878002}{13878002} \\ 
 & 1.44 & \href{https://www-nds.iaea.org/exfor/servlet/X4sGetSubent?subID=13878002}{13878002} \\ 
 & 4.51 & \href{https://www-nds.iaea.org/exfor/servlet/X4sGetSubent?subID=13878004}{13878004} \\ 
 & 5.01 & \href{https://www-nds.iaea.org/exfor/servlet/X4sGetSubent?subID=13878004}{13878004} \\ 
 & 5.51 & \href{https://www-nds.iaea.org/exfor/servlet/X4sGetSubent?subID=13878004}{13878004} \\ 
 & 5.91 & \href{https://www-nds.iaea.org/exfor/servlet/X4sGetSubent?subID=13878004}{13878004} \\ 
 & 6.51 & \href{https://www-nds.iaea.org/exfor/servlet/X4sGetSubent?subID=13878004}{13878004} \\ 
 & 7.14 & \href{https://www-nds.iaea.org/exfor/servlet/X4sGetSubent?subID=13878004}{13878004} \\ 
 & 7.51 & \href{https://www-nds.iaea.org/exfor/servlet/X4sGetSubent?subID=13878004}{13878004} \\ 
 & 8.03 & \href{https://www-nds.iaea.org/exfor/servlet/X4sGetSubent?subID=13878004}{13878004} \\ 
 & 8.41 & \href{https://www-nds.iaea.org/exfor/servlet/X4sGetSubent?subID=13878004}{13878004} \\ 
 & 9.06 & \href{https://www-nds.iaea.org/exfor/servlet/X4sGetSubent?subID=13878004}{13878004} \\ 
 & 9.51 & \href{https://www-nds.iaea.org/exfor/servlet/X4sGetSubent?subID=13878004}{13878004} \\ 
 & 9.99 & \href{https://www-nds.iaea.org/exfor/servlet/X4sGetSubent?subID=13878004}{13878004} \\ 
\hline 
\auOneHundredNinetySeven & 4.51 & \href{https://www-nds.iaea.org/exfor/servlet/X4sGetSubent?subID=14033002}{14033002} \\ 
 & 5.01 & \href{https://www-nds.iaea.org/exfor/servlet/X4sGetSubent?subID=14033002}{14033002} \\ 
 & 5.51 & \href{https://www-nds.iaea.org/exfor/servlet/X4sGetSubent?subID=14033002}{14033002} \\ 
 & 5.91 & \href{https://www-nds.iaea.org/exfor/servlet/X4sGetSubent?subID=14033002}{14033002} \\ 
 & 6.51 & \href{https://www-nds.iaea.org/exfor/servlet/X4sGetSubent?subID=14033002}{14033002} \\ 
 & 7.14 & \href{https://www-nds.iaea.org/exfor/servlet/X4sGetSubent?subID=14033002}{14033002} \\ 
 & 7.51 & \href{https://www-nds.iaea.org/exfor/servlet/X4sGetSubent?subID=14033002}{14033002} \\ 
 & 8.03 & \href{https://www-nds.iaea.org/exfor/servlet/X4sGetSubent?subID=14033002}{14033002} \\ 
 & 8.41 & \href{https://www-nds.iaea.org/exfor/servlet/X4sGetSubent?subID=14033002}{14033002} \\ 
 & 9.06 & \href{https://www-nds.iaea.org/exfor/servlet/X4sGetSubent?subID=14033002}{14033002} \\ 
 & 9.51 & \href{https://www-nds.iaea.org/exfor/servlet/X4sGetSubent?subID=14033002}{14033002} \\ 
 & 9.99 & \href{https://www-nds.iaea.org/exfor/servlet/X4sGetSubent?subID=14033002}{14033002} \\ 
\hline 
\pbNat & 2.24 & \href{https://www-nds.iaea.org/exfor/servlet/X4sGetSubent?subID=23156002}{23156002} \\ 
 & 2.71 & \href{https://www-nds.iaea.org/exfor/servlet/X4sGetSubent?subID=23156003}{23156003} \\ 
 & 2.9 & \href{https://www-nds.iaea.org/exfor/servlet/X4sGetSubent?subID=31687002}{31687002} \\ 
 & 2.94 & \href{https://www-nds.iaea.org/exfor/servlet/X4sGetSubent?subID=23156004}{23156004} \\ 
 & 3.0 & \href{https://www-nds.iaea.org/exfor/servlet/X4sGetSubent?subID=31687002}{31687002} \\ 
 & 3.1 & \href{https://www-nds.iaea.org/exfor/servlet/X4sGetSubent?subID=31687002}{31687002} \\ 
 & 3.2 & \href{https://www-nds.iaea.org/exfor/servlet/X4sGetSubent?subID=31687002}{31687002} \\ 
 & 3.4 & \href{https://www-nds.iaea.org/exfor/servlet/X4sGetSubent?subID=31687002}{31687002} \\ 
 & 4.02 & \href{https://www-nds.iaea.org/exfor/servlet/X4sGetSubent?subID=23156005}{23156005} \\ 
\hline 
\pbTwoHundredEight & 30.4 & \href{https://www-nds.iaea.org/exfor/servlet/X4sGetSubent?subID=14317002}{14317002} \\ 
 & 40.0 & \href{https://www-nds.iaea.org/exfor/servlet/X4sGetSubent?subID=14317003}{14317003} \\ 
 & 65.0 & \href{https://www-nds.iaea.org/exfor/servlet/X4sGetSubent?subID=13946004}{13946004} \\ 
 & 75.0 & \href{https://www-nds.iaea.org/exfor/servlet/X4sGetSubent?subID=13946004}{13946004} \\ 
 & 85.0 & \href{https://www-nds.iaea.org/exfor/servlet/X4sGetSubent?subID=13946004}{13946004} \\ 
 & 95.0 & \href{https://www-nds.iaea.org/exfor/servlet/X4sGetSubent?subID=13946004}{13946004} \\ 
 & 96.0 & \href{https://www-nds.iaea.org/exfor/servlet/X4sGetSubent?subID=22987004}{22987004} \\ 
 & 96.0 & \href{https://www-nds.iaea.org/exfor/servlet/X4sGetSubent?subID=22987007}{22987007} \\ 
 & 96.0 & \href{https://www-nds.iaea.org/exfor/servlet/X4sGetSubent?subID=23059002}{23059002} \\ 
 & 107.5 & \href{https://www-nds.iaea.org/exfor/servlet/X4sGetSubent?subID=13946004}{13946004} \\ 
 & 127.5 & \href{https://www-nds.iaea.org/exfor/servlet/X4sGetSubent?subID=13946004}{13946004} \\ 
 & 155.0 & \href{https://www-nds.iaea.org/exfor/servlet/X4sGetSubent?subID=13946004}{13946004} \\ 
 & 185.0 & \href{https://www-nds.iaea.org/exfor/servlet/X4sGetSubent?subID=13946004}{13946004} \\ 
 & 225.0 & \href{https://www-nds.iaea.org/exfor/servlet/X4sGetSubent?subID=13946004}{13946004} \\ 
\hline 
\biTwoHundredNine & 3.99 & \href{https://www-nds.iaea.org/exfor/servlet/X4sGetSubent?subID=23156006}{23156006} \\ 
\hline 

        \end{longtable}
        \label{ECS_n_bi209}

\subsection*{Comments}
\begin{itemize}
    \item[] \feNat: We did not include the data set of E. Pirovano et al. (Phys. Rev. C 99 (2019) 024601), as it included over 100 scattering energies but only a few angles for each energy.
    \item[] \feFiftyFour: We reduced the cross section values and reported errors of J. R. Vanhoy et al. (Nucl. Phys. A 972 (2018) 107) by a factor of 1000; the data as listed in EXFOR appear to be 1000 too small, possible due to a units mismatch.
    \item[] \wNat: We combined the data set of 0-degree elastic scattering of Schmidt et al. (EXFOR Acc. No. \href{https://www-nds.iaea.org/exfor/servlet/X4sGetSubent?subID=22962024}{22962024}) with the non-0-degree elastic scattering data at the same energies by the same authors (EXFOR Acc. No. \href{https://www-nds.iaea.org/exfor/servlet/X4sGetSubent?subID=22962002}{22962002}), to avoid duplicate calculations/figures of these data while using the Test corpus.
    \item[] \yEightyNine: We combined the data set of 0-degree elastic scattering of Oehrn et al.  (EXFOR Acc. No. \href{https://www-nds.iaea.org/exfor/servlet/X4sGetSubent?subID=22987006}{22987006}) with the non-0-degree elastic scattering data at the same energies by the same authors (EXFOR Acc. No. \href{https://www-nds.iaea.org/exfor/servlet/X4sGetSubent?subID=22987003}{22987003}), to avoid duplicate calculations/figures of these data while using the Test corpus.
\end{itemize}

\clearpage 
\subsection*{Neutron analyzing powers}
\twocolumngrid 
\begin{longtable}{c c c}
                Isotope & Energies & EXFOR Acc. \\ 
\alTwentySeven & 15.425 & \href{https://www-nds.iaea.org/exfor/servlet/X4sGetSubent?subID=13903003}{13903003} \\ 
\hline 
\siTwentyEight & 15.4 & \href{https://www-nds.iaea.org/exfor/servlet/X4sGetSubent?subID=14345003}{14345003} \\ 
 & 18.6 & \href{https://www-nds.iaea.org/exfor/servlet/X4sGetSubent?subID=14345003}{14345003} \\ 
\hline 
\sThirtyTwo & 9.9 & \href{https://www-nds.iaea.org/exfor/servlet/X4sGetSubent?subID=14345005}{14345005} \\ 
 & 13.9 & \href{https://www-nds.iaea.org/exfor/servlet/X4sGetSubent?subID=14345005}{14345005} \\ 
 & 15.4 & \href{https://www-nds.iaea.org/exfor/servlet/X4sGetSubent?subID=14345005}{14345005} \\ 
 & 16.9 & \href{https://www-nds.iaea.org/exfor/servlet/X4sGetSubent?subID=14345005}{14345005} \\ 
\hline 
\coFiftyNine & 15.273 & \href{https://www-nds.iaea.org/exfor/servlet/X4sGetSubent?subID=13903005}{13903005} \\ 
\hline 

        \end{longtable}
        \label{APower_n_co59}

\subsection*{Comments}
No special cases were encountered in assembling neutron analyzing powers for the
Test corpus.

\clearpage 
\subsection*{Neutron total cross sections}
\twocolumngrid 
\begin{longtable}{c c c}
                Isotope & Energies & EXFOR Acc. \\ 
\sNat & 14.1 & \href{https://www-nds.iaea.org/exfor/servlet/X4sGetSubent?subID=31807004}{31807004} \\ 
\hline 
\caForty & 12.04-276.13 & \href{https://www-nds.iaea.org/exfor/servlet/X4sGetSubent?subID=14269004}{14269004} \\ 
\hline 
\tiNat & 0.4-24.75 & \href{https://www-nds.iaea.org/exfor/servlet/X4sGetSubent?subID=14576004}{14576004} \\ 
 & 0.401-24.69 & \href{https://www-nds.iaea.org/exfor/servlet/X4sGetSubent?subID=14576005}{14576005} \\ 
\hline 
\caFortyEight & 12.04-276.13 & \href{https://www-nds.iaea.org/exfor/servlet/X4sGetSubent?subID=14269003}{14269003} \\ 
\hline 
\niFiftyEight & 2.505-290.209 & \href{https://www-nds.iaea.org/exfor/servlet/X4sGetSubent?subID=14661007}{14661007} \\ 
\hline 
\niNat & 2.505-290.209 & \href{https://www-nds.iaea.org/exfor/servlet/X4sGetSubent?subID=14661008}{14661008} \\ 
\hline 
\niSixtyFour & 2.505-290.209 & \href{https://www-nds.iaea.org/exfor/servlet/X4sGetSubent?subID=14661006}{14661006} \\ 
\hline 
\zrNat & 0.4-24.74 & \href{https://www-nds.iaea.org/exfor/servlet/X4sGetSubent?subID=14576008}{14576008} \\ 
 & 0.4-24.74 & \href{https://www-nds.iaea.org/exfor/servlet/X4sGetSubent?subID=14576009}{14576009} \\ 
\hline 
\rhOneHundredThree & 2.505-290.209 & \href{https://www-nds.iaea.org/exfor/servlet/X4sGetSubent?subID=14661005}{14661005} \\ 
\hline 
\snOneHundredTwelve & 3.006-299.233 & \href{https://www-nds.iaea.org/exfor/servlet/X4sGetSubent?subID=14661004}{14661004} \\ 
\hline 
\inNat & 14.1 & \href{https://www-nds.iaea.org/exfor/servlet/X4sGetSubent?subID=31807005}{31807005} \\ 
\hline 
\snNat & 3.006-299.233 & \href{https://www-nds.iaea.org/exfor/servlet/X4sGetSubent?subID=14661002}{14661002} \\ 
\hline 
\snOneHundredTwentyFour & 3.006-299.233 & \href{https://www-nds.iaea.org/exfor/servlet/X4sGetSubent?subID=14661003}{14661003} \\ 
\hline 
\teNat & 14.1 & \href{https://www-nds.iaea.org/exfor/servlet/X4sGetSubent?subID=31807006}{31807006} \\ 
\hline 
\taNat & 0.2-9.121 & \href{https://www-nds.iaea.org/exfor/servlet/X4sGetSubent?subID=23199004}{23199004} \\ 
 & 0.732-1.853 & \href{https://www-nds.iaea.org/exfor/servlet/X4sGetSubent?subID=30831002}{30831002} \\ 
\hline 
\taOneHundredEightyOne & 0.4-24.72 & \href{https://www-nds.iaea.org/exfor/servlet/X4sGetSubent?subID=14576002}{14576002} \\ 
 & 0.4-24.72 & \href{https://www-nds.iaea.org/exfor/servlet/X4sGetSubent?subID=14576003}{14576003} \\ 
\hline 
\wOneHundredEightyTwo & 5.48-299.34 & \href{https://www-nds.iaea.org/exfor/servlet/X4sGetSubent?subID=13887002}{13887002} \\ 
\hline 
\wOneHundredEightyFour & 5.48-299.34 & \href{https://www-nds.iaea.org/exfor/servlet/X4sGetSubent?subID=13887003}{13887003} \\ 
\hline 
\wOneHundredEightySix & 5.48-299.34 & \href{https://www-nds.iaea.org/exfor/servlet/X4sGetSubent?subID=13887004}{13887004} \\ 
\hline 
\auOneHundredNinetySeven & 0.2-9.121 & \href{https://www-nds.iaea.org/exfor/servlet/X4sGetSubent?subID=23199002}{23199002} \\ 
\hline 
\pbTwoHundredFour & 26.993-26.993 & \href{https://www-nds.iaea.org/exfor/servlet/X4sGetSubent?subID=13907002}{13907002} \\ 
\hline 
\pbNat & 2.505-290.209 & \href{https://www-nds.iaea.org/exfor/servlet/X4sGetSubent?subID=14661015}{14661015} \\ 
 & 14.1 & \href{https://www-nds.iaea.org/exfor/servlet/X4sGetSubent?subID=31807007}{31807007} \\ 
\hline 
\biTwoHundredNine & 0.682 & \href{https://www-nds.iaea.org/exfor/servlet/X4sGetSubent?subID=30831003}{30831003} \\ 
\hline 

        \end{longtable}
        \label{TCS_n_bi209}

\subsection*{Comments}
\begin{itemize}
    \item[] \arForty: We did not include the data set of B. Bhandari et al. (Phys. Rev. Lett. 123 (2019) 042502) because the very large reported energy uncertainties made the data unsuitable for comparison against OMP predictions. For example, the reported scattering energy uncertainty for one cross section datum was approximately $\pm$50 \MeV.
    \item[] \auOneHundredNinetySeven, \wNat, \feNat: We did not include the data sets of R. Beyer et al. (Eur. Phys. J. A 54 (2018) 81), as the scattering energies used were only up to 8 \MeV, making the data only marginally useful for comparison against an optical potential.
    \item[] \feNat: We did not include the data set of G. D. Kim et al. (J. Rad. and Nucl. Chem.271 (2007) 541), as the scattering energies used were from 1.1 to 2 \MeV, making the data only marginally useful for comparison against an optical potential.
    \item[] \iOneHundredTwentySeven: We did not include the data set of G. Noguere et al. (Priv. Comm: Noguere (2005)), as the scattering energies used were from a few keV to less than 2 \MeV, making the data only marginally useful for comparison against an optical potential.
    \item[] \pbNat: We did not include the data set of A. B. Laptev (Thesis: Laptev (2004)); the data appear not to have been corrected for dead time and are inconsistent with many other measurements of the same quantity.
\end{itemize}

\clearpage 
\subsection*{Proton differential elastic cross sections}
\twocolumngrid 
\begin{longtable}{c c c}
                Isotope & Energies & EXFOR Acc. \\ 
\mgTwentyFour & 7.4 & \href{https://www-nds.iaea.org/exfor/servlet/X4sGetSubent?subID=F1277002}{F1277002} \\ 
\hline 
\niFiftyEight & 172.0 & \href{https://www-nds.iaea.org/exfor/servlet/X4sGetSubent?subID=D0350003}{D0350003} \\ 
 & 250.0 & \href{https://www-nds.iaea.org/exfor/servlet/X4sGetSubent?subID=E2042003}{E2042003} \\ 
\hline 
\znSixtyFour & 24.0 & \href{https://www-nds.iaea.org/exfor/servlet/X4sGetSubent?subID=O2165003}{O2165003} \\ 
\hline 
\snOneHundredSixteen & 295.0 & \href{https://www-nds.iaea.org/exfor/servlet/X4sGetSubent?subID=E2098002}{E2098002} \\ 
\hline 
\snOneHundredEighteen & 295.0 & \href{https://www-nds.iaea.org/exfor/servlet/X4sGetSubent?subID=E2098003}{E2098003} \\ 
\hline 
\snOneHundredTwenty & 200.0 & \href{https://www-nds.iaea.org/exfor/servlet/X4sGetSubent?subID=E2042007}{E2042007} \\ 
 & 250.0 & \href{https://www-nds.iaea.org/exfor/servlet/X4sGetSubent?subID=E2042010}{E2042010} \\ 
 & 295.0 & \href{https://www-nds.iaea.org/exfor/servlet/X4sGetSubent?subID=E2098004}{E2098004} \\ 
 & 300.0 & \href{https://www-nds.iaea.org/exfor/servlet/X4sGetSubent?subID=E2042012}{E2042012} \\ 
\hline 
\snOneHundredTwentyTwo & 295.0 & \href{https://www-nds.iaea.org/exfor/servlet/X4sGetSubent?subID=E2098005}{E2098005} \\ 
\hline 
\snOneHundredTwentyFour & 295.0 & \href{https://www-nds.iaea.org/exfor/servlet/X4sGetSubent?subID=E2098006}{E2098006} \\ 
\hline 

        \end{longtable}
        \label{ECS_p_sn124}

\subsection*{Comments}
\begin{itemize}
    \item[] \alTwentySeven: we did not include the data sets of Zdravko Siketic et al. (Nucl. Instrum. Meth. 261 (2007) 414) and K. Shahzad et al. (Nucl. Sci. Techn. (Shanghai) 27 (2016) 33), as they included over 100 scattering energies but only a few angles for each energy. We also did not include the data set of S. T. Pittman et al. (Phys. Rev. C 85 (2012) 065804), as it had arbitrary (non-normalized) units.
    \item[] \caFortyFour: we did not include the data set of S. J. Lokitz et al. (Phys. Lett. B, 599, (2004) 223), as it included over 100 scattering energies but only a few angles for each energy.
    \item[] \kNat: we did not include the data set of M. Kokkoris et al. (Nucl. Instrum. Meth. 268 (2010) 1797), as it included over 100 scattering energies but only a few angles for each energy.
    \item[] \mgNat: we did not include the data set of Xiaodong Zhang et al. (Nucl. Instrum. Meth. 201 (2003) 551), as it included over 100 scattering energies but only a few angles for each energy.
    \item[] \niFiftyEight, \pbTwoHundredFour, \pbTwoHundredSix, \pbTwoHundredEight: we did not include the data sets of J. Zenihiro et al. (Phys. Rev. C 82 (2010) 044611), as they had no associated cross section errors.
    \item[] \scFortyFive: we did not include the data set of G. Provatas et al. (Nucl. Instrum. Meth. 269 (2011) 2994), as it included over 100 scattering energies but only a few angles for each energy.
    \item[] \siNat: we did not include the data set of L. Csedreki et al. (Nucl. Instrum. Meth. 443 (2019) 48), as it included over 100 scattering energies but only a few angles for each energy. We did not include the data sets of Becker (Private communication (2008)), as these elastic scattering data are dominated by resonances, making them categorically distinct from the other cross section and analyzing power data considered in this work.
    \item[] \snOneHundredSixteen: There were several null values in the data set of S. Terashima et al. (Phys. Rev. C 77 (2008) 024317) for the ``flag" column in the EXFOR data file; we mapped these to zero, to silence the parsing error into JSON, and included the full data set for analysis.
    \item[] \tiNat: There were several null values in the data set of P. Hu et al. (Nucl. Instrum. Meth. 217 (2004) 551) for the ``flag" column in the EXFOR data file; we mapped these to zero, to silence the parsing error into JSON, and included the full data set for analysis.
    \item[] \vNat: We did not include the data set of X. Zhang et al. (J. Rad. Nucl. Chem. 266, (2005) 149), as it included over 100 scattering energies but only a few angles for each energy.
    \item[] \crFiftyTwo: We did not include the data set of K. G. Leach et al. (Phys. Rev. C 100 (2019) 014320); the data as listed in EXFOR are likely subject to a tabulation error and do not correspond with the values in the referenced publication.
    \item[] \znSixtyFour: We reduced the cross section values and reported errors of K. G. Leach et al. (Phys. Rev. C 100 (2019) 014320) by a factor of 1000; the data as listed in EXFOR appear to be a factor of 1000 too large, possibly due to a units mismatch.

\end{itemize}

\clearpage 
\subsection*{Proton analyzing powers}
\twocolumngrid 
\begin{longtable}{c c c}
                Isotope & Energies & EXFOR Acc.  \\ 
\niFiftyEight & 172.0 & \href{https://www-nds.iaea.org/exfor/servlet/X4sGetSubent?subID=D0350004}{D0350004} \\ 
 & 250.0 & \href{https://www-nds.iaea.org/exfor/servlet/X4sGetSubent?subID=E2042004}{E2042004} \\ 
 & 295.0 & \href{https://www-nds.iaea.org/exfor/servlet/X4sGetSubent?subID=E2291009}{E2291009} \\ 
\hline 
\snOneHundredSixteen & 295.0 & \href{https://www-nds.iaea.org/exfor/servlet/X4sGetSubent?subID=E2098007}{E2098007} \\ 
\hline 
\snOneHundredEighteen & 295.0 & \href{https://www-nds.iaea.org/exfor/servlet/X4sGetSubent?subID=E2098008}{E2098008} \\ 
\hline 
\snOneHundredTwenty & 200.0 & \href{https://www-nds.iaea.org/exfor/servlet/X4sGetSubent?subID=E2042008}{E2042008} \\ 
 & 250.0 & \href{https://www-nds.iaea.org/exfor/servlet/X4sGetSubent?subID=E2042011}{E2042011} \\ 
 & 295.0 & \href{https://www-nds.iaea.org/exfor/servlet/X4sGetSubent?subID=E2098009}{E2098009} \\ 
 & 300.0 & \href{https://www-nds.iaea.org/exfor/servlet/X4sGetSubent?subID=E2042013}{E2042013} \\ 
\hline 
\snOneHundredTwentyTwo & 295.0 & \href{https://www-nds.iaea.org/exfor/servlet/X4sGetSubent?subID=E2098010}{E2098010} \\ 
\hline 
\snOneHundredTwentyFour & 295.0 & \href{https://www-nds.iaea.org/exfor/servlet/X4sGetSubent?subID=E2098011}{E2098011} \\ 
\hline 
\pbTwoHundredFour & 295.0 & \href{https://www-nds.iaea.org/exfor/servlet/X4sGetSubent?subID=E2291006}{E2291006} \\ 
\hline 
\pbTwoHundredSix & 295.0 & \href{https://www-nds.iaea.org/exfor/servlet/X4sGetSubent?subID=E2291007}{E2291007} \\ 
\hline 
\pbTwoHundredEight & 295.0 & \href{https://www-nds.iaea.org/exfor/servlet/X4sGetSubent?subID=E2291008}{E2291008} \\ 
\hline 

        \end{longtable}
        \label{APower_p_pb208}

\subsection*{Comments}
No special cases were encountered in assembling proton analyzing powers for the
Test corpus.

\clearpage 
\subsection*{Proton reaction cross sections}
\twocolumngrid 
\begin{longtable}{c c c}
                Isotope & Energies & EXFOR Acc. \\ 
\caForty & 81.0-180.0 & \href{https://www-nds.iaea.org/exfor/servlet/X4sGetSubent?subID=D0356003}{D0356003} \\ 
\hline 
\niFiftyEight & 81.0 & \href{https://www-nds.iaea.org/exfor/servlet/X4sGetSubent?subID=D0356004}{D0356004} \\ 
\hline 
\zrNinety & 81.0-180.0 & \href{https://www-nds.iaea.org/exfor/servlet/X4sGetSubent?subID=D0356005}{D0356005} \\ 
\hline 
\pbTwoHundredEight & 81.0-180.0 & \href{https://www-nds.iaea.org/exfor/servlet/X4sGetSubent?subID=D0356006}{D0356006} \\ 
\hline 
        \end{longtable}
        \label{RCS_p_pb208}

\subsection*{Comments}
No special cases were encountered in assembling proton reaction cross sections for the
Test corpus.

\newpage

\section*{Supplemental Material C: OMP performance against experimental data}

This section details how to interpret the attached figures comparing the CH89
and KD OMPs and their UQ counterparts CHUQ and KDUQ against the experimental
training corpora and the Test corpus. All figures are included in the file
\texttt{supplement\_performance.tar}. Figures showing performance of the
CH89/CHUQ and KD/KDUQ OMPs against their respective training data appear in
directories \texttt{CHUQCorpus} and \texttt{KDUQCorpus}, respectively.  Figures
showing performance of the CH89/CHUQ and KD/KDUQ OMPs against the Test corpus
appear in the \texttt{TestCorpus} directory, within which there are two
subdirectories \texttt{CHUQ} and \texttt{KDUQ} containing figures for these
OMPs. Within each of these directories, figures for each target nucleus are
organized into separate subdirectories, inside of which the figures are listed
by data type and projectile. For example, after expanding the tar file
\texttt{supplement\_performance.tar}, the user could locate a figure comparing
the CH89 and CHUQ OMPs against the CHUQ corpus \caNat\ neutron elastic cross
section data at \texttt{CHUQCorpus/ca0/ECS\_n.pdf}. The data plotted in the
figures correspond to the data tabulated in Supplemental Material A.

To eliminate the need for separate captions for each figure, a consistent
format is applied for all figures of each sector of experimental data. Each
figure displays experimental data and predictions only for a single data type
for a given nucleus and projectile. Differential data (proton and neutron
elastic cross sections and analyzing powers) are plotted in columns within
each figure, with up to five scattering energies shown in a single column. The
scattering energy for each experimental data set (and the corresponding OMP
predictions) is listed in \MeV\ immediately above the lowest-angle cross
section data. For legibility, data sets at different energies are offset by a
factor of 100 from the adjacent data sets, as indicated by the labels on the
right-hand-side of each column. Experimental data with their associated
uncertainties are shown as black symbols with error bars. For the training corpora, 
black circles are used, and for the Test corpus, black diamonds are used. For the
data in the training corpora only, any data points that were
identified as outliers with respect to the uncertainty-quantified OMPs are
plotted as white circles, enabling visual identification of points qualifying as outliers 
(see the main text for details on the outlier identification approach).

Predictions made using the CH89 and KD OMPs are indicated with the dashed gray
line. Predictions made using the CHUQDemocratic and KDUQDemocratic OMPs are
indicated with dark and light bands, which represent the 68\% and 95\%
uncertainty intervals. To avoid confusion between CHUQ and KDUQ, calculations
made using CHUQDemocratic are colored in blue and those made using
KDUQDemocratic are colored in red. For protons, the cross sections and cross
section uncertainties have been divided by the Rutherford cross section, to
facilitate visual comparison between neutron and proton data. In the handful of
instances where the reported experimental uncertainty was greater than or equal
to the experimental cross section, the experimental uncertainty plotted in the
figure was reduced to 99\% of the cross section value, to reduce visual clutter
from lines plunging to (unphysical) negative cross sections. Finally, for
differential elastic scattering data sets at energies above roughly 75 \MeV,
cross sections fall so rapidly with angle that high-angle cross sections are
difficult to plot on the same scale as the lower-energy, low-angle cross
sections. (There were no high-angle, high-energy experimental data, so this
issue only impacted calculated values). A similar situation appears in plotting
analyzing powers, which oscillate rapidly at high energies and large angles
($>90$ degrees). To improve legibility, for the highest-energy differential
cross sections and analyzing powers, figures show predicted cross sections only
at forward center-of-momentum angles (typically up to 90 degrees) where
experimental data exist, not to the maximum possible angle of 180 degrees.

For integral quantities (proton reaction cross section and neutron total cross
section), we use the same color conventions and uncertainty intervals for
plotting experimental data and OMP predictions.

Finally, a word of caution about assessing performance against the Test corpus.
The original CH89 OMP was intended to be used only for targets with $A \geq 40$
and for energies $10 \leq E \leq 65$ \MeV. As a significant fraction of the
Test corpus lies at lower and (much) higher energies than this range, it is
unsurprising that extrapolation of CHUQ beyond its intended limits yields very
poor reproduction of certain experimental data in the Test corpus (for example,
for proton scattering on \snOneHundredTwenty\ from 200-300 \MeV). Still, we
include figures depicting CHUQ's performance against the entire Test corpus in
part to illustrate how OMP predictions (even uncertainty-equipped ones)
may rapidly degrade when pushed beyond their limits. This provides some indication
of the peril of extrapolation in other dimensions, for example, away from
$\beta$-stability. For the KD OMP, the nominal validity range is with targets
of $A \geq 27$ and $0.001 \leq E \leq 200$ \MeV, which has much larger overlap
with the Test corpus coverage. As such, the performance of KD and KDUQ against
the Test Corpus is far better than CHUQ in certain regimes, for example, proton
elastic scattering cross sections at 300 \MeV\ even though they are also well
beyond the nominal KD validity range.

\end{document}